\documentclass[12pt, draftclsnofoot, onecolumn]{IEEEtran}

\usepackage[utf8]{inputenc} 
\usepackage[T1]{fontenc}
\usepackage{url}
\usepackage{ifthen}
\usepackage{cite}
\usepackage[cmex10]{amsmath} 
\usepackage{graphicx}
\usepackage{subfig}
\usepackage{booktabs}
\usepackage{soul}

\usepackage{comment}
\usepackage{algorithm,algcompatible}
\algnewcommand\INPUT{\item[\textbf{Input:}]}%
\algnewcommand\OUTPUT{\item[\textbf{Output:}]}%
\usepackage[dvipsnames]{xcolor}
\usepackage{amsfonts}
\usepackage{amssymb}
\usepackage{comment}
\usepackage{latexsym}
\usepackage{bm}
\usepackage{bmpsize}
\usepackage[justification=centering]{caption}

\DeclareMathOperator*{\argmin}{arg\,min}
\DeclareGraphicsExtensions{.eps}
\graphicspath{{../jpg/}{../pdf/}{../jpeg/}{../eps/}}
\newtheorem{prop}{Proposition}

\newtheorem{lemma}{Lemma}

\allowdisplaybreaks

\ifCLASSINFOpdf
\else
\fi
\begin{document}

\title{Energy-Efficient Node Deployment in Heterogeneous Two-Tier Wireless Sensor Networks with Limited Communication Range}

\author{Saeed Karimi-Bidhendi,
        Jun Guo,
        and Hamid Jafarkhani
        \IEEEauthorblockA{}
\thanks{Authors are with the Center for Pervasive Communications \& Computing, University of California, Irvine, Irvine CA, 92697 USA (e-mail: \{skarimib, guoj4, hamidj\}@uci.edu). 
\copyright 2020 IEEE.  Personal use of this material is permitted.  Permission from IEEE must be obtained for all other uses, in any current or future media, including reprinting/republishing this material for advertising or promotional purposes, creating new collective works, for resale or redistribution to servers or lists, or reuse of any copyrighted component of this work in other works.
This work was accepted for publication in IEEE Transactions on Wireless Communications, and presented in part in 2019 IEEE International Symposium on Information Theory \cite{ISIT_paper}. This work was supported in part by the NSF Award CCF-1815339.}
}

\markboth{}%
{Shell \MakeLowercase{\textit{et al.}}: Bare Demo of IEEEtran.cls for IEEE Journals}

\maketitle
\vspace{-9mm}
\begin{abstract}
We study a heterogeneous two-tier wireless sensor network in which $N$ heterogeneous access points (APs) collect sensing data from densely distributed sensors and then forward the data to $M$ heterogeneous fusion centers (FCs). This heterogeneous node deployment problem is modeled as an optimization problem with the total power consumption of the network as its cost function. The necessary conditions of the optimal AP and FC node deployment are explored in this paper. We provide a variation of Voronoi Diagram as the optimal cell partition for this network and show that each AP should be placed between its connected FC and the geometric center of its cell partition. In addition, we propose a heterogeneous two-tier Lloyd algorithm to optimize the node deployment. Furthermore, we study the sensor deployment when the communication range is limited for sensors and APs. Simulation results show that our proposed algorithms outperform the existing clustering methods like Minimum  Energy Routing, Agglomerative Clustering, Divisive  Clustering, Particle Swarm Optimization, Relay Node placement in Double-tiered Wireless Sensor Networks, and Improved Relay Node Placement, on average.
\end{abstract}

\IEEEpeerreviewmaketitle

\section{Introduction}

\IEEEPARstart{W}{ireless} sensor networks (WSNs) have been widely used to gather data from the environment and transfer the sensed information through wireless channels to one or more fusion centers. Based on the network architecture, WSNs can be classified as either hierarchical or non-hierarchical WSNs. In hierarchical WSNs, sensors play different roles as they are often divided into clusters and some of them are selected as cluster heads or relays. In non-hierarchical WSNs every sensor has identical functionality and the connectivity of network is usually maintained by multi-hop wireless communications. WSNs can also be divided into either homogeneous WSNs \cite{ICC_paper,JG,SD,AS,MBMOS}, in which sensors share the same capacity, e.g., storage, computation power, antennas, sensitivity etc., or heterogeneous WSNs where sensors have different capacities \cite{YTH,JG2,JPH,MNMA}.

Energy consumption is a key bottleneck in WSNs due to limited energy resources of sensors, and difficulty or even infeasibility of recharging the batteries of densely deployed sensors. The energy consumption of a sensor node comes from three primary components: communication energy, computation energy \cite{HYHJMM} and sensing energy. The experimental measurements show that, in many applications, the computation energy is negligible compared to communication energy\cite{GMMA, MS}. Furthermore, for passive sensors, such as light sensors and acceleration sensors, the sensing energy is significantly small. Therefore, wireless communication dominates the sensor energy consumption in practice. There are three primary methods to reduce the energy consumption of radio communication in the literature: (i) topology control\cite{XYY,XW}, in which unnecessary energy consumption is avoided by properly switching awake and asleep states,  (ii) energy-efficient routing protocols \cite{MBMOS,JGBC}, that are designed to find an optimal path to transfer data, and (iii) power control protocols \cite{VP,SVRP}, that save communication energy by adjusting the transmitter power at each node while keeping reliable communications. Another widely used method, Clustering \cite{OYSF,VP}, attempts to balance the energy consumption among sensor nodes by iteratively selecting cluster heads. Unfortunately, the above MAC protocols bring about a massive number of message exchanges because the knowledge of geometry and/or energy is required during the operation \cite{OYSF, erdem}. Also, the node deployment is known and fixed in these approaches while it plays an important role in energy consumption of the WSNs. 

While WSNs provide a bridge between the physical and virtual information world, the collected data is not useful if it cannot be transmitted from sensors to access points and eventually to base stations. Connectivity, as a prominent necessity in WSNs, is widely studied under the binary communication model in \cite{SD} and \cite{Coverage_with_connectivity_in_wireless_sensor_networks, Integrated_coverage_and_connectivity_configuration, Deploying_wireless_sensors_to_achieve_both_coverage_and_connectivity, Distributed_algorithms_for_energy_efficient_even, A_geometry_study_on_the_capacity_of, A_study_of_connectivity_in_MIMO}. In the binary communication model, each node can only communicate to other nodes within a certain range due to the limited transmission power. Note that connectivity is guaranteed when nodes are linked by wire lines; however, the same is not true for WSNs due to the limited available power in wireless communication. Many distributed sensor deployment algorithms, such as Lloyd Algorithm, do not take both power consumption and connectivity into account; therefore, they usually converge to a sub-optimal deployment in which nodes are divided into several disconnected components. For a one-tier WSN, the design of optimal deployment algorithms that consider connectivity and coverage is studied in \cite{JG2}. While we consider a 2D deployment in this work, the case of 3D optimal deployment has been studied in \cite{Erdem_UAV,JGPWHJ20}, and the applicability of the evolutionary algorithms to solve UAV deployment problems has been introduced in \cite{Outage-optimized}.

In this paper, we study the node deployment problem in heterogeneous two-tier WSNs consisting of heterogeneous APs and heterogeneous FCs, with and without communication power constraints. We consider the total wireless communication power consumption as the cost function. The optimal energy-efficient sensor deployment in homogeneous WSNs is studied in \cite{JG}. However, the homogeneous two-tier WSNs in \cite{JG} do not address various challenges that exist in the heterogeneous two-tier WSNs, e.g., unlike regular Voronoi diagrams for homogeneous WSNs, the optimal cells in heterogeneous WSNs may be non-convex, not star-shaped or even disconnected, and the cell boundaries may not be hyperplanes. Another challenge in the heterogeneous two-tier networks is that unlike the homogeneous case \cite{JG}, or heterogeneous one-tier case \cite{On the minimum average distortion of quantizers with index-dependent}, some nodes may not contribute to the energy saving. To the best of our knowledge, the optimal node deployment for energy efficiency in heterogeneous WSNs is still an open problem. Our main goal is to find the optimal AP and FC deployment to minimize the total communication power consumption. By deriving the necessary conditions of the optimal deployments in such heterogeneous two-tier WSNs, we design Lloyd-like algorithms to deploy nodes. In addition, we update the designed deployment algorithms to consider the effects of limited communication range. We also study the trade-off between AP and sensor power consumption.

The rest of this paper is organized as follows: In Section \ref{sec:model}, we introduce the system model and problem formulation. In Section \ref{sec:opt}, we study the optimal AP and FC deployment and provide the corresponding necessary conditions. A numerical algorithm is proposed in Section \ref{sec:algorithm} to minimize the energy consumption. An analysis of AP and sensor power trade-off is provided in Section \ref{sec:AP_Sensor_Power_Function}. In Section \ref{sec:limited_communication_range}, 
an algorithm is proposed to maximize the network coverage and minimize the power consumption, simultaneously. Section \ref{sec:simulation} presents the experimental results and Section \ref{sec:conclusion} concludes the paper.

\section{System Model and Problem Formulation}\label{sec:model}

\begin{figure}[!htb]
\centering
\includegraphics[width=11cm]{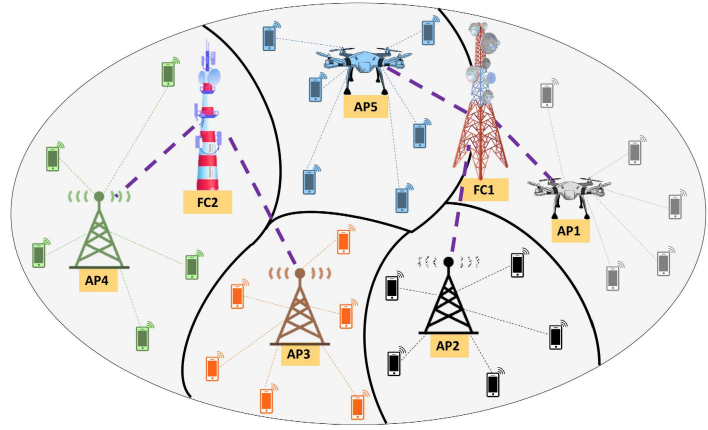}
\caption{System model.}
\label{system-model}
\end{figure}

Here, we study the power consumption of the heterogeneous two-tier WSNs consisting of three types of nodes, i.e., homogeneous sensors, heterogeneous APs and heterogeneous FCs. The power consumption models for homogeneous WSNs are discussed in details in \cite{JG}. The main difference in this work is the heterogeneous characteristics of the APs and FCs. For the sake of completeness, we describe the system model, as shown in Fig. \ref{system-model}, for heterogeneous WSNs here in details. Given the target area $\Omega\subset \mathbb{R}^2$ which is a convex polygon including its interior, $N$ APs and $M$ FCs are deployed to gather data from densely deployed sensors. Throughout this paper, we assume that $N\geq M$. Given the sets of AP and FC indices, i.e., $\mathcal{I_A}=\{1,2,..., N\}$ and $\mathcal{I_B}=\{1,2,..., M\}$, respectively, the index map $T:\mathcal{I_A} \longrightarrow \mathcal{I_B}$ is defined to be $T(n)=m$ if and only if AP $n$ is connected to FC $m$. If AP $n$ has no associated FC, we set $T(n)=-1$. Conversely, $T^{-1}(m)$ is defined to be the set of all AP indices $n$ such that $T(n)=m$, and $\left|T^{-1}(m)\right|$ denotes the cardinality of this set. The AP and FC deployments are then defined by $P=\left(p_1, ..., p_N \right)$ and $Q=\left(q_1, ..., q_M \right)$, where $p_n, q_m\in \mathbb{R}^2$ denote the location of AP $n$ and FC $m$, respectively. Throughout this paper, we assume that each sensor only sends data to one AP. For each $n\in \mathcal{I_A}$, AP $n$ collects data from sensors in the region $\emph{R}_n \subset \Omega$; therefore, for each AP deployment $P$, there exists an AP partition $\mathbf{R}=(\emph{R}_1,...,\emph{R}_N)$ comprised of disjoint subsets of $\mathbb{R}^2$ whose union is $\Omega$.  The density of 
sensors is denoted via a continuous and differentiable function $f: \Omega \longrightarrow \mathbb{R}^+$. The total amount of data gathered from the 
sensors in region $\emph{R}_n$ in one time unit is $g \int_{\emph{R}_n} f(w)dw$,  where $g$ is  the  bit-rate  of  the  sensors. Due to the homogeneity of sensors, $g$ is a constant \cite{JG}. 

We focus on the power consumption of sensors and APs, since FCs usually have reliable energy resources and their energy consumption is not the main concern. First, we discuss the sensors' power consumption. As shown in \cite{JG}, because of the path-loss, the instant transmission power is proportional to the square of the distance between the two nodes and a constant that depends on the characteristics of the two nodes, i.e., $a\times \| p_n - w\|^2$ for a sensor located at $\omega$ that sends its data to AP $n$. 
According to \cite{Heinzelman}, the parameter $a$ is given as $a = \frac{P_{r-th} \left(4\pi\right)^2}{G_tG_r\lambda^2}$, where $P_{r-th}$ is the minimum receiver power threshold, 
$G_t$ and $G_r$ are the transmitter and receiver antenna gains, respectively, and $\lambda$ is the carrier signal wavelength.
For homogeneous WSNs, all nodes in each tier have the same characteristics and therefore, the parameter $a$ is the same and will not affect the optimization. However, in a heterogeneous WSN, the heterogeneity of APs causes nodes to have different antenna gains and SNR thresholds; therefore, the   
 parameter $a$ will be a function of the node index. Hence, the sensors' power consumption can be written as 
\begin{equation}\label{eq2}
    \overline{\mathcal{P}}^{\mathcal{S}}(P, \mathbf{R}) = \sum_{n=1}^N\int_{R_n}a_n\|p_n - w\|^2f(w)dw.
\end{equation}
Similarly, for the AP's power consumption, the instant transmission power between AP $n$ and FC $T(n)$ can be written as $b\times \|p_n - q_{T(n)} \|^2$ where the  parameter $b$ depends on the antenna gain and SNR threshold of FC $T(n)$ and antenna gain of AP $n$ \cite{Heinzelman}. Hence, it is the same for homogeneous WSNs and will not affect the optimization. However, in a heterogeneous WSN, the heterogeneity of APs and FCs causes the parameter $b$ to be a function of the node indices. 
Therefore, the APs' power consumption can be written as

\begin{equation}\label{eq1}
         \overline{\mathcal{P}}^{\mathcal{A}}\left(P, Q, \mathbf{R}, T \right) =  \sum_{n=1}^{N}\int_{R_n}b_{n,T(n)}\|p_n-q_{T(n)}\|^2f(w)dw.  
\end{equation}
Our goal in this work is to minimize the power consumptions in (\ref{eq2}) and (\ref{eq1}). However, as will be shown later, there is a trade-off between the two power consumptions. As such, one objective is to minimize the AP transmission power in (\ref{eq1}) given a constraint on the sensor transmission power in (\ref{eq2}). Mathematically, this results in the AP-Sensor power function defined as 
\begin{equation}\label{AP_Sensor_power_function}
    A(s) \triangleq \inf_{\left(P,Q,\mathbf{R},T \right): \overline{\mathcal{P}}^{\mathcal{S}}\left(P,\mathbf{R} \right)\leq s} \overline{\mathcal{P}}^{\mathcal{A}}\left(P,Q,\mathbf{R},T \right).
\end{equation}
Similarly, one can define the Sensor-AP power function to minimize the sensor power in (\ref{eq2}) given a constraint on the AP transmission power in (\ref{eq1}) as follows:
\begin{equation}\label{Sensor_AP_power_function}
    S(a) \triangleq \inf_{\left(P,Q,\mathbf{R},T \right): \overline{\mathcal{P}}^{\mathcal{A}}\left(P,Q,\mathbf{R},T \right)\leq a}\overline{\mathcal{P}}^{\mathcal{S}}\left(P,\mathbf{R} \right).
\end{equation}
The two-tier power consumption is then defined as the Lagrangian function of (\ref{eq1}) and (\ref{eq2}):
\begin{align}\label{eq4}
        & \overline{\mathcal{P}} \left(P,Q, \mathbf{R}, T \right) = \overline{\mathcal{P}}^\mathcal{S} \left(P, \mathbf{R} \right) + \beta \overline{\mathcal{P}}^\mathcal{A} \left(P,Q, \mathbf{R}, T \right) \\& =  \sum_{n=1}^N\int_{R_n} \left(a_n\|p_n  -  w\|^2  +  \beta b_{n,T(n)}\|p_n  -  q_{T(n)}\|^2 \right) f(w)dw. \nonumber
\end{align}
Our main objective in this paper is to minimize the two-tier power consumption defined in (\ref{eq4}) over the AP deployment $P$, FC deployment $Q$, cell partition $\mathbf{R}$ and index map $T$ and study the behavior of the AP-Sensor power function. 

\section{Optimal Node Deployment in Two-Tier WSNs}\label{sec:opt}

As it is shown in (\ref{eq4}), the two-tier power consumption depends on four variables $P$, $Q$, $\mathbf{R}$ and $T$. Therefore, our goal is to find the optimal AP and FC deployments, cell partitioning and index map, denoted by $P^*=\left(p_1^*,...,p_N^* \right)$, $Q^*=\left(q_1^*,...,q_M^* \right)$, $\mathbf{R}^*=(R^*_1,...,R^*_N)$ and $T^*$, respectively, that minimizes the two-tier power consumption. Note that not only the variables $P$, $Q$, $\mathbf{R}$ and $T$ are intertwined, i.e., the best value for each of them depends on the value of the other three variables, but also this optimization problem is NP-hard. Our approach is to design an iterative algorithm that optimizes the value of one variable while the other three variables are held fixed. To this end, first we derive the necessary conditions for optimal deployment at each step.

Note that the index map only appears in the second term of (\ref{eq4}); thus, for any given AP and FC deployment $P$ and $Q$, the optimal index map is given by:
\begin{equation}\label{eq8}
    T(n) = \argmin_m b_{n,m}\|p_n - q_m\|^2 .
\end{equation}
Ties are broken in favor of the smaller index for a unique mapping. Eq. (\ref{eq8}) implies that an AP may not be connected to its closest FC due to heterogeneity of the APs and FCs, and to minimize the two-tier power consumption, AP $n$ should be connected to FC $m$ that minimizes the weighted distance $b_{n,m}\|p_n - q_m\|^2$. 

Next, we study the properties of optimal cell partitioning. For each $n\in \mathcal{I_A}$, we define the Voronoi cell $V_n$ for AP and FC deployments $P$ and $Q$, and index map $T$ as:
\begin{multline}\label{eq5}
    V_n(P,Q,T) \triangleq \{w: a_n\|p_n-w\|^2 + \beta b_{n, T(n)}\|p_n-q_{T(n)}\|^2 \\ \leq  a_k\|p_k-w\|^2 + \beta b_{k, T(k)}\|p_k-q_{T(k)}\|^2, \forall k \neq n \}.
\end{multline}
Ties are broken in favor of the smaller index to ensure that each Voronoi cell $V_n$ is a Borel set. When it is clear from the context, we write $V_n$ instead of $V_n(P,Q,T)$. The collection
\begin{equation}\label{eq6}
\mathbf{V}(P,Q,T) \triangleq (V_1,V_2,...,V_N)
\end{equation}
is referred to as the generalized Voronoi diagram. Note that unlike the regular Voronoi diagrams, the Voronoi cells defined in (\ref{eq5}) may be non-convex, not star-shaped or even disconnected. The following proposition establishes that the generalized Voronoi diagram in (\ref{eq6}) provides the optimal cell partitions, i.e., $\mathbf{R}^*(P,Q,T) = \mathbf{V}(P,Q,T)$ for a given $P,Q,T$.
\begin{prop}\label{allactive}
For any partition of the target area $\Omega$ such as $U$, and any AP and FC node deployments such as $P$ and $Q$ and each index map $T$ we have:
\begin{equation}\label{eq7}
    \overline{\mathcal{P}}\left(P,Q,U,T \right) \geq \overline{\mathcal{P}}\left(P,Q,\mathbf{V}(P,Q,T), T \right).
\end{equation}
\end{prop}
The proof is provided in Appendix \ref{Appendix_A}. 

Next, we aim to derive the necessary condition for optimal locations of APs and FCs. For this purpose, first we need to show that each FC contributes to the total power consumption in an optimal node deployment, i.e., adding an additional FC results in a strictly lower optimal two-tier power consumption regardless of its weights $b_{n,M+1}$ as long as $M<N$ holds.
\begin{lemma}\label{lemma1}
Let $\left(P^*, Q^*, \mathbf{R}^*, T^* \right)$ be the optimal node deployment for $N$ APs and $M$ FCs. Given an additional FC with parameters $b_{n,M+1}$ for every $n\in \mathcal{I_A}$, the optimal AP and FC deployments, index map and cell partitioning are denoted via $P'=\left(p_1',p_2',...,p_N'\right)$, $Q'=\left(q_1',q_2',...,q_{M+1}'\right)$, $T'$ and $\mathbf{R}'$, respectively. Assuming $M<N$, we have:
\begin{equation}\label{FC_Decrease_Distortion}
\overline{\mathcal{P}}\left(P',Q',\mathbf{R}',T' \right) < \overline{\mathcal{P}}\left(P^*,Q^*,\mathbf{R}^*,T^* \right).
\end{equation}
\end{lemma}
The proof is provided in Appendix \ref{Appendix_B}. 

Let $v_n^*(P,Q,T)=\int_{R_n^*}f(w)dw$ be the Lebesgue measure (volume) of the region $R_n^*$, and $c_n^*(P,Q,T) = \frac{\int_{R_n^*}wf(w)dw}{\int_{R_n^*}f(w)dw}$ be the geometric centroid of the region $R_n^*$. When there is no ambiguity, we write $v_n^*(P,Q,T)$ and $c_n^*(P,Q,T)$ as $v_n^*$ and $c_n^*$, respectively. Lemma \ref{lemma1} immediately leads to the following corollary.

\emph{Corollary 1. }Let $\left(P^*, Q^*, \mathbf{R}^*, T^* \right)$ be the optimal node deployment for $N$ APs and $M$ FCs. If $M\leq N$, then for each $m\in \mathcal{I_B}$, $\sum_{n:T^*(n)=m}v^*_n>0$.

\noindent The proof is provided in Appendix \ref{Appendix_C}. 

Lemma \ref{lemma1} and Corollary 1 are technical results that we need to prove the following proposition that provides the necessary conditions for the optimal AP and FC deployments in the heterogeneous two-tier WSNs.
\begin{prop}\label{necessary}
The necessary conditions for optimal deployments in the heterogeneous two-tier WSNs with power consumption defined in (\ref{eq4}) are:
\begin{equation}\label{eq9}
        p_n^* = \frac{a_n c_n^* + \beta b_{n,T^*(n)}q^*_{T^*(n)}}{a_n + \beta b_{n,T^*(n)}} \qquad , \qquad
        q^*_m  = \frac{\sum_{n:T^*(n)=m}{}b_{n,m}p^*_nv^*_n}{\sum_{n:T^*(n)=m}{}b_{n,m}v^*_n}.
\end{equation}
\end{prop}
The proof is provided in Appendix \ref{Appendix_D}. 

Corollary 1 implies that the denominator of the second equation in (\ref{eq9}) is positive; thus, $q^*_m$ is well-defined. According to (\ref{eq9}), the optimal location of FC $m$ is the linear combination of the locations of its connected APs, and the optimal location of AP $n$ is on the segment $\overline{c^*_nq^*_{T^*(n)}}$. While Lemma \ref{lemma1} indicates that each FC contributes to the power consumption, the same result may not hold for some APs. To show that under certain settings, an AP may not be useful, i.e., no sensor sends data to it, we use the sensor network in the following lemma as an example.

\begin{lemma}\label{AP_usefulness}
Consider two APs and one FC within the target region $\Omega=[0,1]$ with parameters $b_{1,1}=\kappa\times a_1$, $b_{2,1}=\kappa\times a_2$ where $\kappa$ is a positive constant, and a uniform density function. The necessary and sufficient condition for both APs to be useful is
\begin{equation}\label{necesssary_sufficient_condition_for_usefulness}
        \sqrt{\frac{4\beta'+1}{\beta' + 1}} - 1  \leq \sqrt{\frac{a_1}{a_2}} \leq \frac{1}{\sqrt{\frac{4\beta'+1}{\beta' + 1}} - 1},
\end{equation}
where $\beta' = \beta \times \kappa$. If the above condition holds, both APs are useful and the optimal two-tier power consumption is given by:
\begin{equation}\label{optimal_two_tier_distortion_useful_APs}
    \overline{\mathcal{P}} = \frac{\left(4\beta' + 1 \right)}{12\left(\beta' + 1\right)} \times \left( \frac{\sqrt{a_1a_2}}{\sqrt{a_1} + \sqrt{a_2}} \right)^2.
\end{equation}
Otherwise, all sensors send their data to the stronger AP and we have:
\begin{equation}\label{optimal_two_tier_distortion_useless_APs}
    \overline{\mathcal{P}} = \frac{\min\left(a_1, a_2 \right)}{12}.
\end{equation}
The proof is provided in Appendix \ref{Appendix_F}.
\end{lemma}

In the next section, we use the properties derived in Propositions 1 and 2 and in (\ref{eq8}), and design a Lloyd-like algorithm to find the optimal node deployment.

\section{Node Deployment Algorithm}\label{sec:algorithm}

First, we quickly review the conventional Lloyd algorithm. Lloyd Algorithm iterates between two steps: In the first step, the node deployment is optimized while the partitioning is fixed and in the second step, the partitioning is optimized while the node deployment is fixed. Although the conventional Lloyd Algorithm can be used to solve one-tier quantizers or one-tier node deployment problems as shown in \cite{JG2}, it cannot be applied to two-tier WSNs where two kinds of nodes are deployed. Inspired by the properties explored in Section III, we propose a heterogeneous two-tier Lloyd (HTTL) algorithm to solve the optimal deployment problem in heterogeneous two-tier WSNs and minimize the two-tier power consumption defined in (\ref{eq4}). Starting with a random initialization for node deployment $\left(P,Q,\mathbf{R},T \right)$ in the target area $\Omega$, our algorithm iterates between four steps: (i) Update the index map $T$ according to (\ref{eq8}); (ii) Obtain the cell partitioning according to (\ref{eq5}) and update the value of volumes $v_n$ and centroids $c_n$; (iii) For each $m\in \mathcal{I_B}$, if $T^{-1}(m)$ is not empty, update the location of FC $m$ according to (\ref{eq9}); otherwise, randomly select an index $m'\in \mathcal{I_B}$ according to the distribution $ P(m')= \frac{\left|T^{-1}(m')\right|}{N}$, and move FC $m$ to a random location within $\bigcup_{n:T(n)=m'}R_n$; (iv) Update the location of APs according to (\ref{eq9}). The algorithm continues until the stop criterion, $\frac{\overline{\mathcal{P}}_{\textrm{old}} - \overline{\mathcal{P}}_{\textrm{new}}}{\overline{\mathcal{P}}_{\textrm{old}}}\geq \epsilon$ is satisfied ($\overline{\mathcal{P}}_{\textrm{old}}$ and $\overline{\mathcal{P}}_{\textrm{new}}$ are the average powers in the previous and current iterations, respectively.). 

\begin{prop}  HTTL Algorithm is an iterative improvement algorithm, i.e., the Lagrangian function in (\ref{eq4}) is non-increasing and the algorithm converges.
\end{prop}
The proof is provided in Appendix \ref{Appendix_E}.

\section{AP-Sensor Power Function}\label{sec:AP_Sensor_Power_Function}

Note that the Lagrangian two-tier power consumption defined in (\ref{eq4}) is the unconstrained version of the constrained optimization problems defined in (\ref{AP_Sensor_power_function}) and (\ref{Sensor_AP_power_function}). Since the AP-Sensor power function and the Sensor-AP power function are dual of each other, in this section, we only study the properties of the AP-Sensor power function $A(s)$.
An AP-Sensor power pair $(s,a)$ is \emph{achievable} if and only if there is a node deployment $\left(P,Q,\mathbf{R},T \right)$ such that $\overline{\mathcal{P}}^{\mathcal{A}}\left(P,Q,\mathbf{R},T \right)=a$ while $\overline{\mathcal{P}}^{\mathcal{S}}\left(P,\mathbf{R} \right) \leq s$. Moreover, a deployment $\left(P,Q,\mathbf{R},T \right)$ is a \emph{feasible} solution for the power pair $\left(s, a \right)$ if and only if $\overline{\mathcal{P}}^{\mathcal{A}}\left(P,Q,\mathbf{R},T \right)=a$ while $\overline{\mathcal{P}}^{\mathcal{S}}\left(P,\mathbf{R} \right) \leq s$. By definition, it is evident that every point above the curve $A(s)$ is also achievable. In what follows, we analyze the properties of the AP-Sensor power function. Without loss of generality, we assume that $a_1 \leq a_2 \leq \ldots \leq a_N$ holds.  A $K-$level one-tier quantizer is a tuple $(X, \mathbf{R})$, i.e. the location of points $X = \left(x_1, \cdots, x_K  \right)$ and the partitioning $\mathbf{R} = \left(R_1, \cdots, R_K \right)$ of the target region, such that $x_i$ is the quantization point for all $\omega \in R_i$ and $K$ is the number of sub-regions. Let $D_K$ be the minimum distortion of a heterogeneous $K-$level one-tier quantizer in the space $\Omega$ with parameters $a_1, \ldots, a_K$, i.e., we have:
\begin{equation}\label{D_N}
    D_K = \min_{X,\mathbf{R}}\sum_{i=1}^{K}\int_{R_i}a_i \|x_i-w \|^2 f(w)dw,
\end{equation}
where the minimum is over all node deployments $X=\left(x_1, \ldots, x_K \right)$ and partitioning $\mathbf{R} = \left(R_1,\ldots, R_K \right)$ of $\Omega$.
\begin{lemma}\label{AP_Sensor_power_function_is_non_increasing}
Let $N$ and $M$ be the number of APs and FCs where $N>M$. Then, the AP-Sensor power function $A(s)$ is a non-increasing function with the domain $\left[D_N, +\infty \right)$ such that $A(s)>0$ for $s\in \left[D_N, D_M \right)$ and $A(s)=0$ for $s\in \left[D_M, +\infty \right)$.
\end{lemma}
The proof is provided in Appendix \ref{Appendix_H}.

Lemma \ref{AP_Sensor_power_function_is_non_increasing} characterizes the non-increasing property of $A(s)$ in addition to defining its domain based on the properties of a regular quantizer. For a fixed partitioning $\mathbf{R}=\left(R_1, \ldots, R_N \right)$, let $\mathcal{H}(\mathbf{R}) = \sum_{i=1}^{N}\int_{R_i}a_i \|c_i-w \|^2 f(w)dw$ where $c_i$ is the centroid of the region $R_i$, i.e., $\mathcal{H}(\mathbf{R})$ is the minimum one-tier power consumption with parameters $a_1,\ldots,a_N$ for a fixed partitioning $\mathbf{R}$. For the special case of $M=1$, the following lemma derives a closed-form solution for the AP-Sensor power function for any fixed partitioning of $\Omega$.
\begin{lemma}\label{closed_form_fixed_partitioning}
For $Q=(q)$, $P=\left(p_1, \ldots, p_N \right)$, and fixed $\mathbf{R}$, define $A(s,\mathbf{R})$ to be:
\begin{equation}\label{A_(S,R)}
    A(s,\mathbf{R}) \triangleq \inf_{\left(P,Q,T \right): \overline{\mathcal{P}}^{\mathcal{S}}\left(P,\mathbf{R} \right)\leq s} \overline{\mathcal{P}}^{\mathcal{A}}\left(P,Q,\mathbf{R},T \right).
\end{equation}
We have:

(i) The domain of $A(s,\mathbf{R})$ is $\{\left(s,\mathbf{R}\right)\big | s\geq \mathcal{H}(\mathbf{R}) \}$.

(ii) If $b_{i,1} =  \kappa a_i$ for $\kappa  \in  \mathbb{R}^+$ and each $i  \in  \mathcal{I_A}$, when $(s,\mathbf{R})  \in  \{(s,\mathbf{R}) \big | \mathcal{H}(\mathbf{R})  \leq  s  \leq  \mathcal{J}(\mathbf{R})\}$, we have:
\begin{equation}\label{closed_form_formula}
    A\left(s,\mathbf{R} \right)= \kappa\left[\sqrt{\mathcal{J}(\mathbf{R}) - \mathcal{H}(\mathbf{R})} - \sqrt{s - \mathcal{H}(\mathbf{R})}  \right]^2,
\end{equation}
and $A\left(s,\mathbf{R} \right)=0$ for $s\geq \mathcal{J}(\mathbf{R})$ where $\mathcal{J}(\mathbf{R})$ is defined as: 
\begin{equation}\label{J(R)}
    \mathcal{J}(\mathbf{R}) = \sum_{n=1}^{N}\int_{R_n}a_n \bigg | \!  \bigg|\frac{\sum_{i=1}^{N}a_iv_ic_i}{\sum_{i=1}^{N}a_iv_i} - w \bigg | \! \bigg|^2 f(w)dw,
\end{equation}
where $v_i$ and $c_i$ are volume and centroid of the region $R_i$, respectively. 

\noindent The proof is provided in Appendix \ref{Appendix_I}.
\end{lemma}

In Section \ref{sec:simulation}, we experimentally plot the AP-Sensor power function defined in (\ref{AP_Sensor_power_function}) and verify the above properties. We conclude this section by deriving a closed-form formula for the AP-Sensor power function for the same setting used in Lemma \ref{AP_usefulness}.

\begin{lemma}\label{special_case_closed_form_A(s)}
Consider two APs and one FC within the target region $\Omega=[0,1]$ with parameters $b_{1,1}=\kappa\times a_1$, $b_{2,1}=\kappa\times a_2$, and a uniform density function. If (\ref{necesssary_sufficient_condition_for_usefulness}) holds, we have:
\begin{equation}\label{special_case_closed_form_formula_for_A(s)}
    A(s) = \kappa\Bigg[\frac{1}{2}\left(\frac{\sqrt{a_1a_2}}{\sqrt{a_1} + \sqrt{a_2}  }  \right) -\sqrt{s - \frac{1}{12}\left(\frac{\sqrt{a_1a_2}}{\sqrt{a_1} + \sqrt{a_2}  }  \right)^2}    \Bigg]^2,
\end{equation}
for $\frac{1}{12}\left(\frac{\sqrt{a_1a_2}}{\sqrt{a_1} + \sqrt{a_2}  }  \right)^2 \leq s < \frac{\min\left(a_1, a_2\right)}{12}$ and $A(s)=0$ for $s\geq \frac{\min\left(a_1, a_2\right)}{12}$. If (\ref{necesssary_sufficient_condition_for_usefulness}) does not hold, we have $A(s)=0$ for any $s$.

\noindent The proof is provided in Appendix \ref{Appendix_J}.
\end{lemma}
Lemma \ref{special_case_closed_form_A(s)} shows that $A(s)$ is not continuous at $s^* = \frac{\min\left(a_1,a_2 \right)}{12}$ for this example. In addition, $A(s)$ is convex in the intervals $\left[0, s^*\right)$ and $\left[s^*,+\infty \right)$.

\section{Limited Communication Range}\label{sec:limited_communication_range}

Note that when sensors or APs have limited transmission power, not all APs can communicate with FCs. Similarly, only sensors within the sensing range of APs in the set $\left\{n\big | T(n)\neq -1  \right\}$ can transmit their collected information to fusion centers. We consider a common power constraint $\sigma^2$ for homogeneous densely deployed sensors, and power constraints $\sigma_n^2, n\in\mathcal{I_A}$ for the heterogeneous APs. In other words, to maintain the connectivity of the network, a sensor at position $w$ can forward its collected data to AP $n$, and AP $n$ can in turn sends the data to FC $m$ if and only if:
\begin{equation}\label{w_to_Pn_and_Qm_power_constraint}
        a_n\|p_n - w \|^2 \leq \sigma^2 \qquad , \qquad
        b_{n,m}\|p_n - q_m \|^2 \leq \sigma_n^2,
\end{equation}
or equivalently:
\begin{equation}\label{w_to_Pn_and_Qm_radius}
        \|p_n - w \| \leq \frac{\sigma}{\sqrt{a_n}} \qquad , \qquad
        \|p_n - q_m \| \leq \frac{\sigma_n}{\sqrt{b_{n,m}}}.
\end{equation}
Hence, we use the coverage defined by:
\begin{equation}\label{coverage_area_definition}
    C(P, T) = \int_{\bigcup_{n:T(n)\neq -1} \mathcal{B}\left(p_n, \frac{\sigma}{\sqrt{a_n}}\right) \cap \Omega}f(w)dw
\end{equation}
as a performance measure along with the two-tier power consumption in (\ref{eq4}) when communication range is limited, where $\mathcal{B}(c,r)=\{\omega|\|\omega-c\|\leq r\}$ is a disk centered at $c$ with radius $r$. Note that HTTL Algorithm described in Section \ref{sec:algorithm} can converge to a deployment in which (\ref{w_to_Pn_and_Qm_radius}) may not hold. Our main goal in this section is to find a proper deployment that not only minimizes the two-tier power consumption $\overline{\mathcal{P}}(P, Q, \mathbf{R}, T)$ in (\ref{eq4}), but also maximizes the total coverage $C(P, T)$ in (\ref{coverage_area_definition}). In what follows, we describe our approach in details.

Starting with an initial deployment $\left(P,Q,\mathbf{R},T \right)$, if $\left\{m \Big | m\in \mathcal{I_B}, q_m \in \mathcal{B}\left(p_n, \frac{\sigma_n}{\sqrt{b_{n,m}}}\right) \right\}\neq \varnothing$, then the index map $T$ is updated as
\begin{equation}\label{limited_com_range_index_map}
    T(n) = \argmin_{m: q_m \in \mathcal{B}\left(p_n, \frac{\sigma_n}{\sqrt{b_{n,m}}}\right) }b_{n,m}\|p_n - q_m\|^2,
\end{equation}
otherwise, we set $T(n) = -1$, indicating that AP $n$ has no associated FC. Note that although some sensors in the region $R_n, n\in \mathcal{I_A}$, may not be able to transmit their data to AP $n$ due to their limited transmission power, we still partition the target region using the generalized Voronoi diagram in (\ref{eq5}) and (\ref{eq6}) since it minimizes the two-tier power consumption given a fixed node deployment and index map. But instead of using all $N$ APs for generalized Voronoi partitioning, we only use APs in the set $\left\{n\big | T(n)\neq -1 \right\}$.

\begin{figure}[!htb]
\centering
\subfloat[]{\includegraphics[width=39mm]{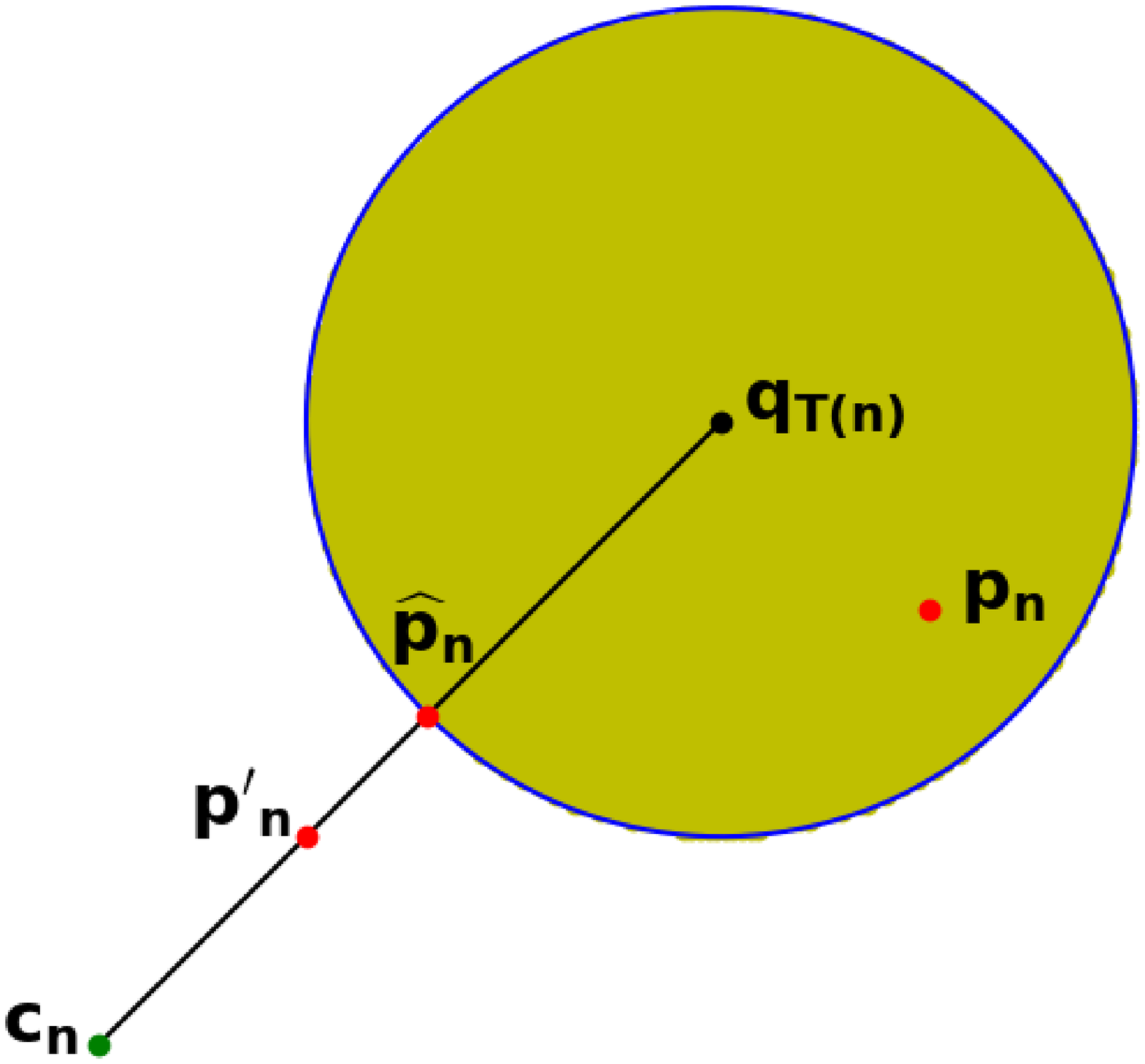}
\label{desired_region_AP}}
\hfil
\subfloat[]{\includegraphics[width=39mm]{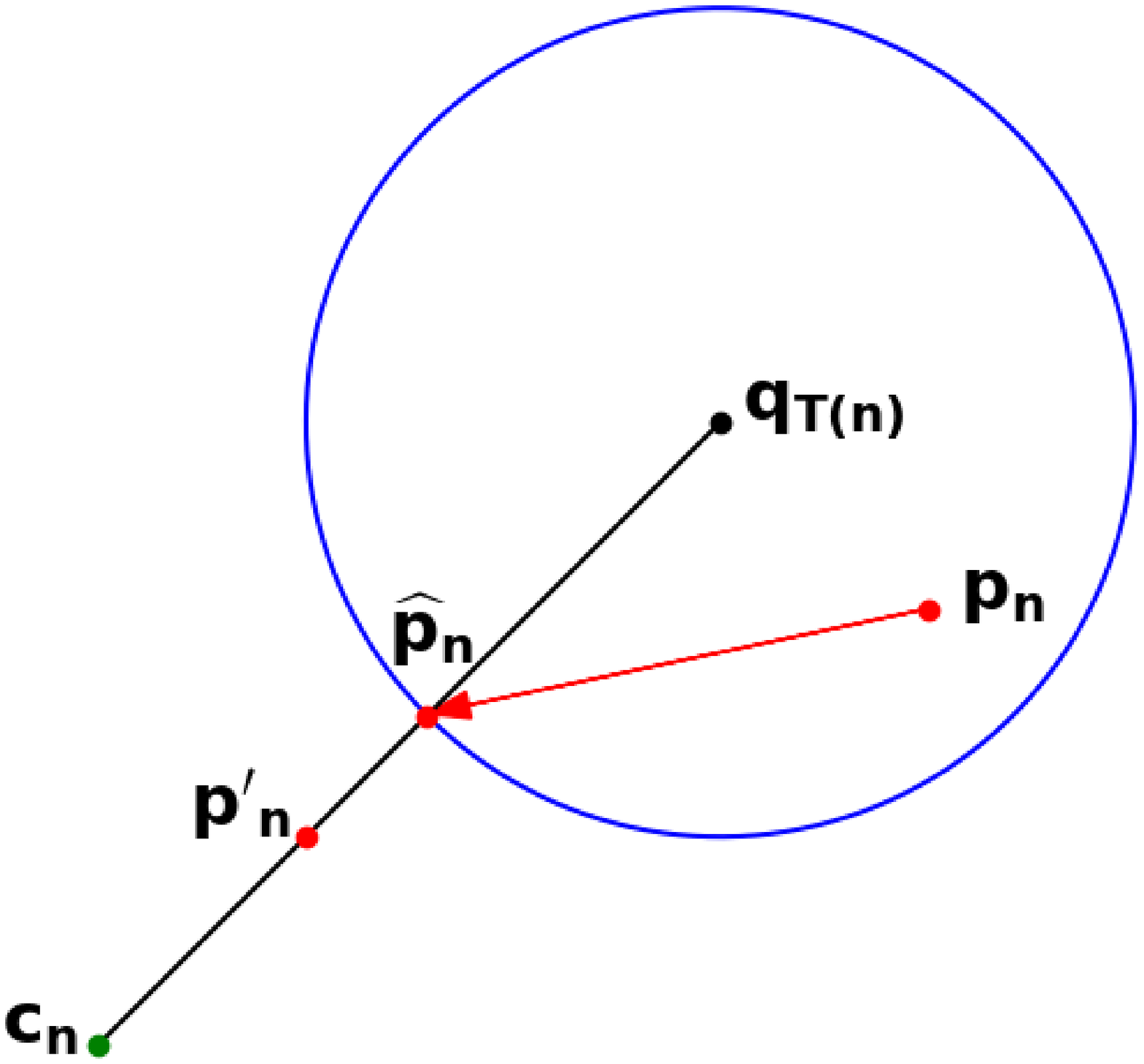}
\label{AP_movement_limited_com_range}}
\hfil
\subfloat[]{\includegraphics[width=39mm]{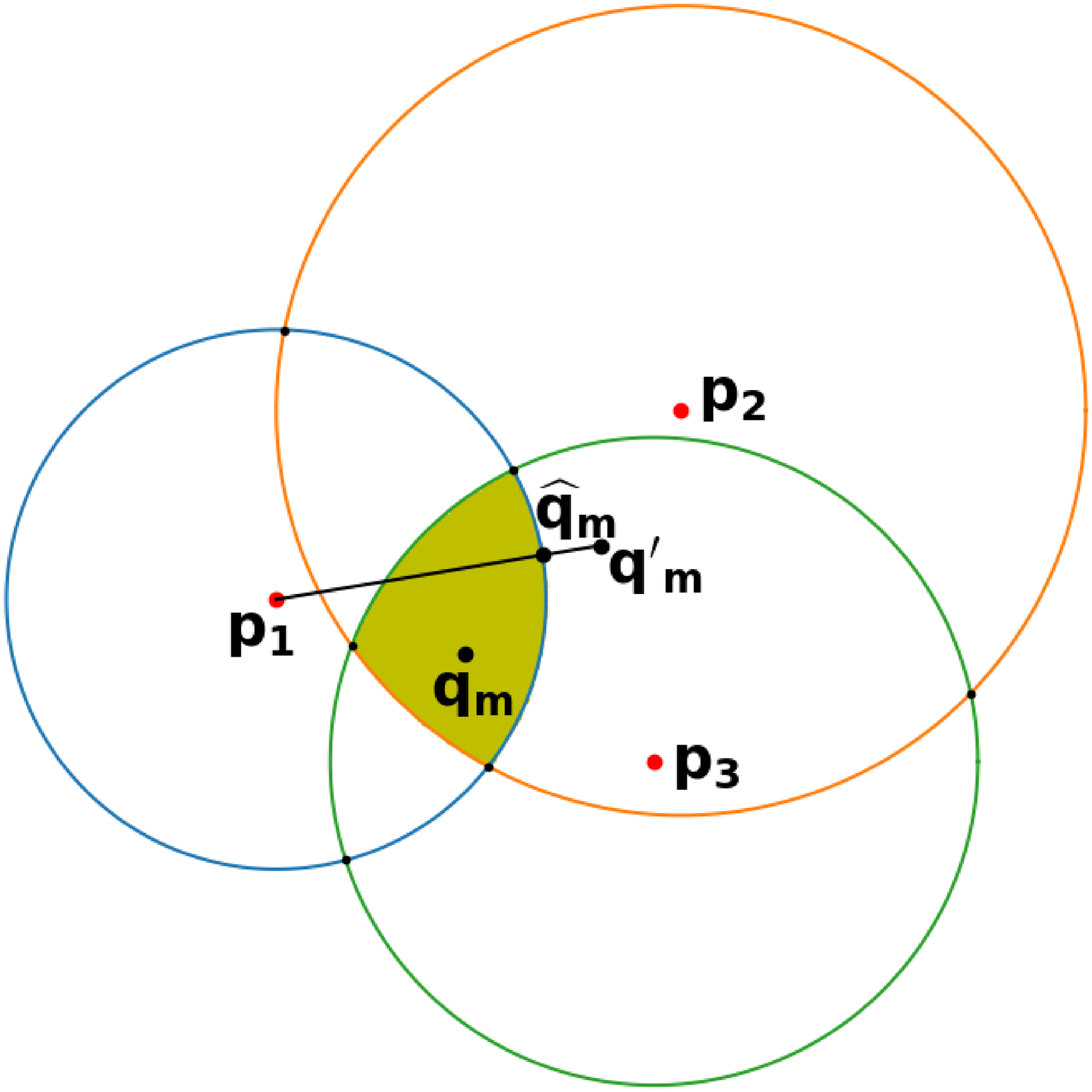}
\label{desired_region_FC}}
\hfil
\subfloat[]{\includegraphics[width=39mm]{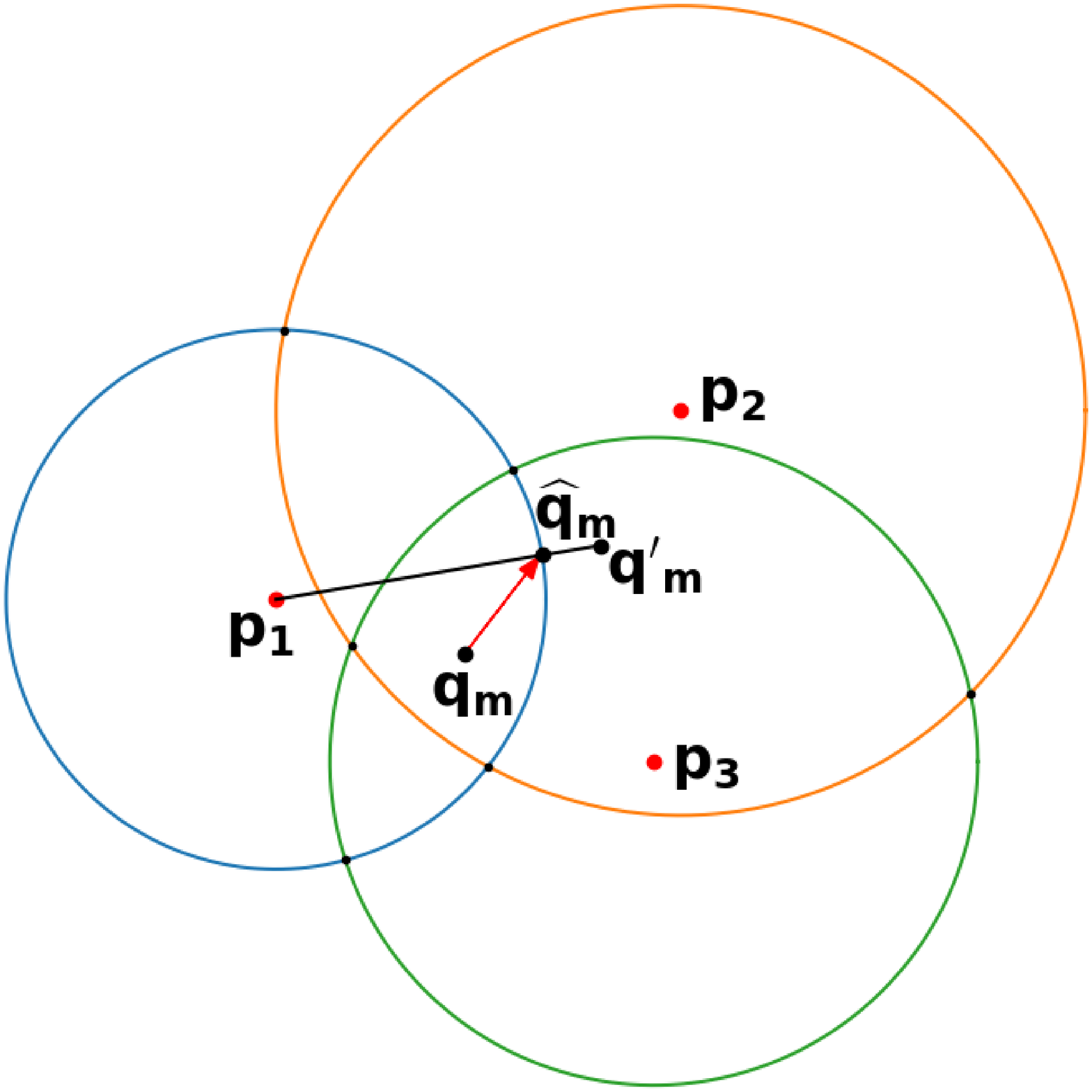}
\label{FC_movement_limited_com_range}}
\vspace{-3mm}
\captionsetup{justification=justified}
\caption{\small{Optimal AP and FC movement. (a) Desired region for AP. (b) Optimal positioning of AP. (c) Desired region for FC. (d) Optimal positioning of FC.}}
\label{desired_and_final_locations}
\end{figure}

For each AP in the set $\left\{n \big| T(n)=-1 \right\}$, we randomly move AP inside the target region. Similarly, for each FC in the set $\left\{m \Big | T^{-1}(m) = \varnothing  \right\}$, we randomly relocate the FC inside $\Omega$. For those APs that have an associated FC, Proposition \ref{necessary} indicates that their current locations should be updated according to (\ref{eq9}), as we did in Step (iv) of the HTTL algorithm; however, as it is illustrated in Fig. \ref{desired_region_AP}, the optimal location $p'_n = \frac{a_n c_n + \beta b_{n,T(n)}q_{T(n)}}{a_n + \beta b_{n,T(n)}}$ for AP $n$ may lie outside the communication range of its corresponding FC, that we refer to as the \emph{desired region} for AP $n$. In that case, AP $n$ is moved to the closest point to $p'_n$ within its desired region, denoted by $\widehat{p}_n$, as it is shown in Fig. \ref{AP_movement_limited_com_range}. Similarly, (\ref{eq9}) implies that FC $m$ should be relocated to the position $q'_m = \frac{\sum_{n:T(n)=m}{}b_{n,m}p_nv_n}{\sum_{n:T(n)=m}{}b_{n,m}v_n}$, as we did in Step (iii) of the HTTL algorithm; however, as it is illustrated in Fig. \ref{desired_region_FC}, $q'_m$ may lie outside the region, that we refer to as the \emph{desired region} for FC $m$, in which all its associated APs can communicate. In that case, we move FC $m$ to the closest point to $q'_m$ within its desired region, denoted by $\widehat{q}_m$, as it is shown in Fig. \ref{FC_movement_limited_com_range}. Note that in order to find $\widehat{q}_m$, we only need to consider a finite number of points. The entire process to optimize the power for a limited communication range is summarized in Algorithm 1. Similar to HTTL Algorithm, each AP lies on the segment connecting its corresponding FC to the centroid of its region once the Limited-HTTL algorithm converges. The following proposition shows that Limited-HTTL Algorithm is an iterative improvement algorithm and converges.

\begin{prop}  Limited-HTTL Algorithm is an iterative improvement algorithm, i.e., the Lagrangian function in (\ref{eq4}) is non-increasing and the algorithm converges.
\end{prop}
The proof is provided in Appendix \ref{Appendix_K}.

\begin{algorithm}
    \caption{Limited-HTTL Algorithm}
  \begin{algorithmic}[1]

    \INPUT Weights $\left\{a_n \right\}_{n\in \mathcal{I_A}}$ and $\left\{b_{n,m} \right\}_{n\in \mathcal{I_A},m\in \mathcal{I_B}}$, $\beta \in \mathbb{R}^+$, powers $\sigma^2$ and $\sigma^2_n, n\in\mathcal{I_A}$ and $\epsilon \in \mathbb{R}^+$.
    \OUTPUT Optimal node deployment $\left(P^*, Q^*, \mathbf{R}^*, T^* \right)$.
    \STATE Randomly initialize the node deployment $\left(P, Q, \mathbf{R}, T \right)$.
    
    \STATE {\bf do}
    \STATE Compute the two-tier power consumption $\overline{\mathcal{P}}_{\textrm{old}}=\overline{\mathcal{P}}(P,Q,\mathbf{R},T)$.
    \STATE Update the index map $T$ according to (\ref{limited_com_range_index_map}).
    \STATE Use APs in the set $\left\{n\big | T(n)\neq -1 \right\}$ for generalized Voronoi partitioning of $\Omega$.
    \STATE Calculate the volumes $\{v_n\}$ and centroids $\{c_n\}$ for each $n\in \left\{n\big | T(n)\neq -1 \right\}$.
    \STATE For each $m\in \mathcal{I_B}$:
    
    \noindent\quad-- if $T^{-1}(m)\neq \varnothing$:

    \noindent\qquad$\bullet$ move FC $m$ to the nearest point to $q'_m =  \frac{\sum_{n:T(n)=m}{}b_{n,m}p_nv_n}{\sum_{n:T(n)=m}{}b_{n,m}v_n}$ inside its desired region.
    
    \noindent\quad-- else:
    
    \noindent\qquad$\bullet$ randomly select an index $m'\in \mathcal{I_B}$ according to the distribution $P(m')=\frac{\left|T^{-1}(m')\right|}{N}$.
    
    \noindent\qquad$\bullet$ move FC $m$ to a random location within the region
    $\bigcup_{n:T(n)=m'}R_n$.
    
    \STATE $\forall n\in \mathcal{I_A}$, move AP $n$ to the nearest point to $p'_n=\frac{a_n c_n + \beta b_{n,T(n)}q_{T(n)}}{a_n + \beta b_{n,T(n)}}$ inside its desired region.
    \STATE Update the two-tier power consumption $\overline{\mathcal{P}}_{\textrm{new}}=\overline{\mathcal{P}}(P,Q,\mathbf{R},T)$.
    \STATE {\bf While }$\frac{\overline{\mathcal{P}}_{\textrm{old}} - \overline{\mathcal{P}}_{\textrm{new}}}{\overline{\mathcal{P}}_{\textrm{old}}}\geq \epsilon$
    \STATE {\bf Return:} The node deployment $\left(P,Q,\mathbf{R},T \right)$.
  \end{algorithmic}
\end{algorithm}

\section{Experiments}\label{sec:simulation}

Simulations are carried out for both synthetic and real-world datasets. For the synthetic data,  we provide the experimental results in two heterogeneous two-tier WSNs: (i) WSN1: A heterogeneous WSN including 1 FC and 20 APs; (ii) WSN2: A heterogeneous WSN including 4 FCs and 20 APs. We consider the same target domain $\Omega$ as in \cite{JG, JPH}, i.e., $\Omega=[0,10]^2$. Simulations are performed for two different sensor density functions, i.e., a uniform distribution $f(\omega) = \frac{1}{\int_{\Omega}dw} = 0.01$, and a mixture of Gaussian distribution:
\begin{equation}\label{Mixture_of_Gaussian_Distribution}
    f(\omega) = \frac{1}{2} \! \times  \mathcal{N}\! \left(\! \begin{bmatrix}
    3 \\
    3 
\end{bmatrix} \! , \! \begin{bmatrix}
    1.5 & 0 \\
    0 & 1.5 
\end{bmatrix} \! \right)\! +  \frac{1}{4} \! \times  \mathcal{N}\! \left(\! \begin{bmatrix}
    6 \\
    7 
\end{bmatrix} \! , \! \begin{bmatrix}
    2 & 0 \\
    0 & 2 
\end{bmatrix}\! \right) \! +  \frac{1}{4} \! \times  \mathcal{N}\! \left(\! \begin{bmatrix}
    7.5 \\
    2.5 
\end{bmatrix} \! , \! \begin{bmatrix}
    1 & 0 \\
    0 & 1
\end{bmatrix} \! \right).
\end{equation}
To evaluate the performance, 10 initial AP and FC deployments on $\Omega$ are generated randomly, i.e, every node location is generated with uniform distribution on $\Omega$. In order to make a fair comparison to prior work, similar to the experimental setting in \cite{JG, JPH}, the maximum number of iterations is set to 100, FCs and APs are denoted, respectively, by black and red circles. Other parameters are provided in Table \ref{SPT}. According to the parameters in Table \ref{SPT}, we divide APs into two groups: strong APs ($n\in\{1,\dots,10\}$) and weak APs ($n\in\{11,\dots,20\}$). Similarly, FCs are divided into strong FCs ($m\in\{1,2\}$) and weak FCs ($m\in\{3,4\}$).


\begin{table}[!bth]
\centering
  \caption{Simulation Parameters}
  \vspace{-3mm}
\begin{tabular}{ccccccccccc}
 \toprule
                   &\multicolumn{2}{c}{{\bf WSN1}} &                     &&                    &                    &      \multicolumn{2}{c}{{\bf WSN2}}  &                       &                       \\\cline{1-4}\cline{6-11}
$\mathbf{a_{1:10}}$&$\mathbf{a_{11:20}}$&$\mathbf{b_{1:4,1}}$&$\mathbf{b_{5:20,1}}$&&$\mathbf{a_{1:10}}$ &$\mathbf{a_{11:20}}$&$\mathbf{b_{1:4,1:2}}$&$\mathbf{b_{1:4,3:4}}$&$\mathbf{b_{5:20,1:2}}$&$\mathbf{b_{5:20,3:4}}$\\
    $\mathbf{1}$   &    $\mathbf{2}$    &   $\mathbf{1}$     &      $\mathbf{2}$   &&   $\mathbf{1}$     &     $\mathbf{2}$   &    $\mathbf{1}$      &     $\mathbf{2}$     &      $\mathbf{2}$     &      $\mathbf{4}$     \\ 
 \bottomrule
\end{tabular}
  \label{SPT}
\end{table}

Like the experiments in \cite{JG}, we compare the weighted power of our proposed algorithm with Minimum Energy Routing (MER) \cite{AG}, Agglomerative Clustering (AC) \cite{DM}, Divisive Clustering (DC) \cite{DM} algorithms, and Particle Swarm Optimization (PSO) \cite{PSO}. PSO is a population-based stochastic algorithm for non-linear optimization. AC and DC are bottom-up and top-down clustering algorithms, respectively. MER is a combination of Multiplicatively weighted Voronoi Partition \cite{VD} and Bellman-Ford algorithms \cite[Section 2.3.4]{BG}. More details about MER, AC, and DC can be found in \cite{JG}. When the communication range is limited, we further compare our method with two other algorithms, i.e., Improved Relay Node Placement (IRNP)\cite{IRNP}, and Relay Node placement in Double tiered Wireless Sensor Network (RNDWSN)\cite{RNDWSN}. IRNP and RNDWSN are node placement algorithms designed to maximize the network coverage. Note that if a small portion of sensors are covered by a particular node placement, since not many sensors will transfer data to fusion centers, the resulting power consumption will be small too. Therefore, our primary goal in node deployment with limited transmission power is to maximize the network coverage and minimize the power consumption, simultaneously.

\begin{figure}[!htb]
\centering
\subfloat[]{\includegraphics[width=42.0mm]{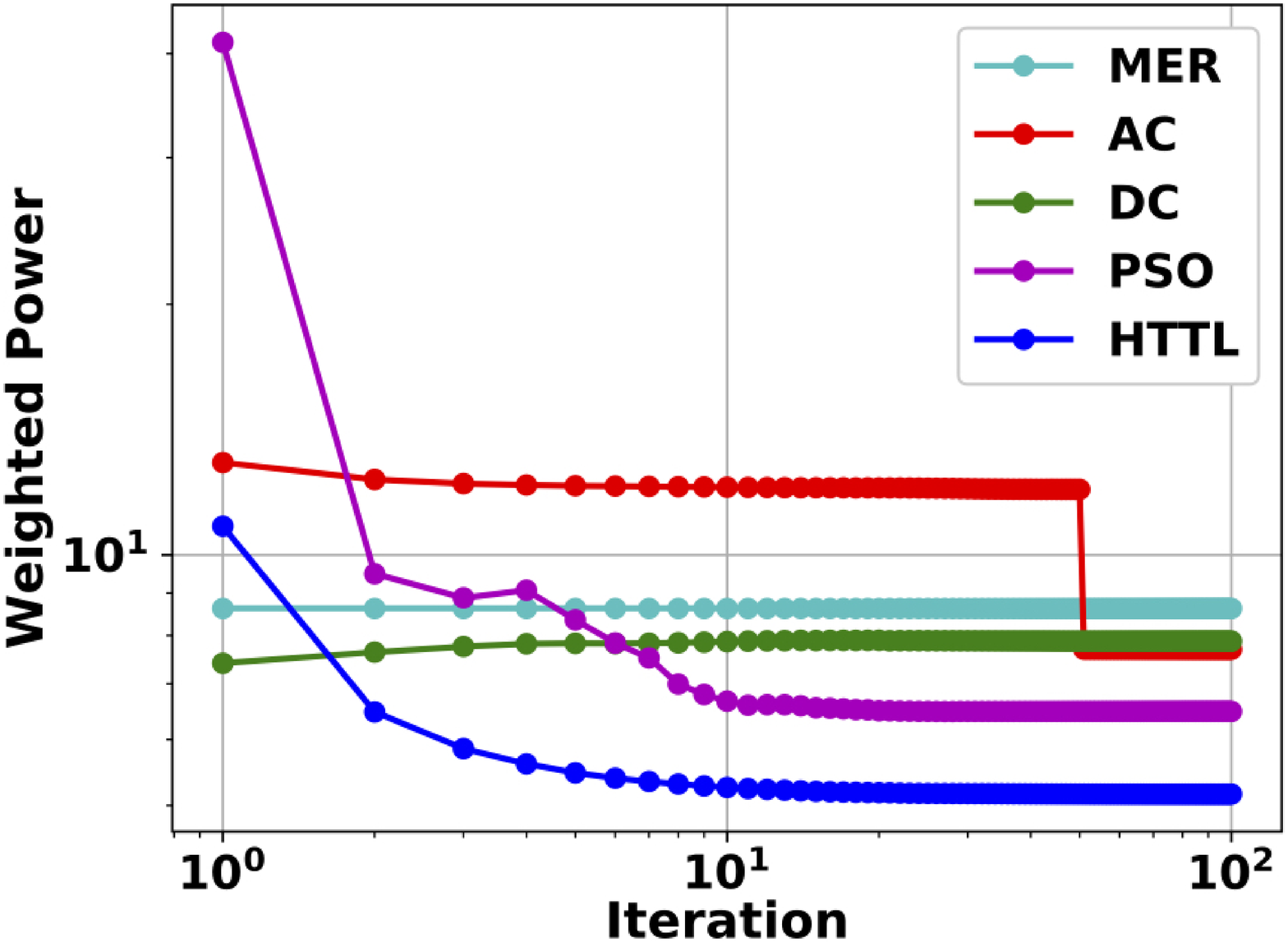}}
\label{distortion-vs-iteration-WSN1-Uniform}
\hspace{-4mm}
\subfloat[]{\includegraphics[width=42.0mm]{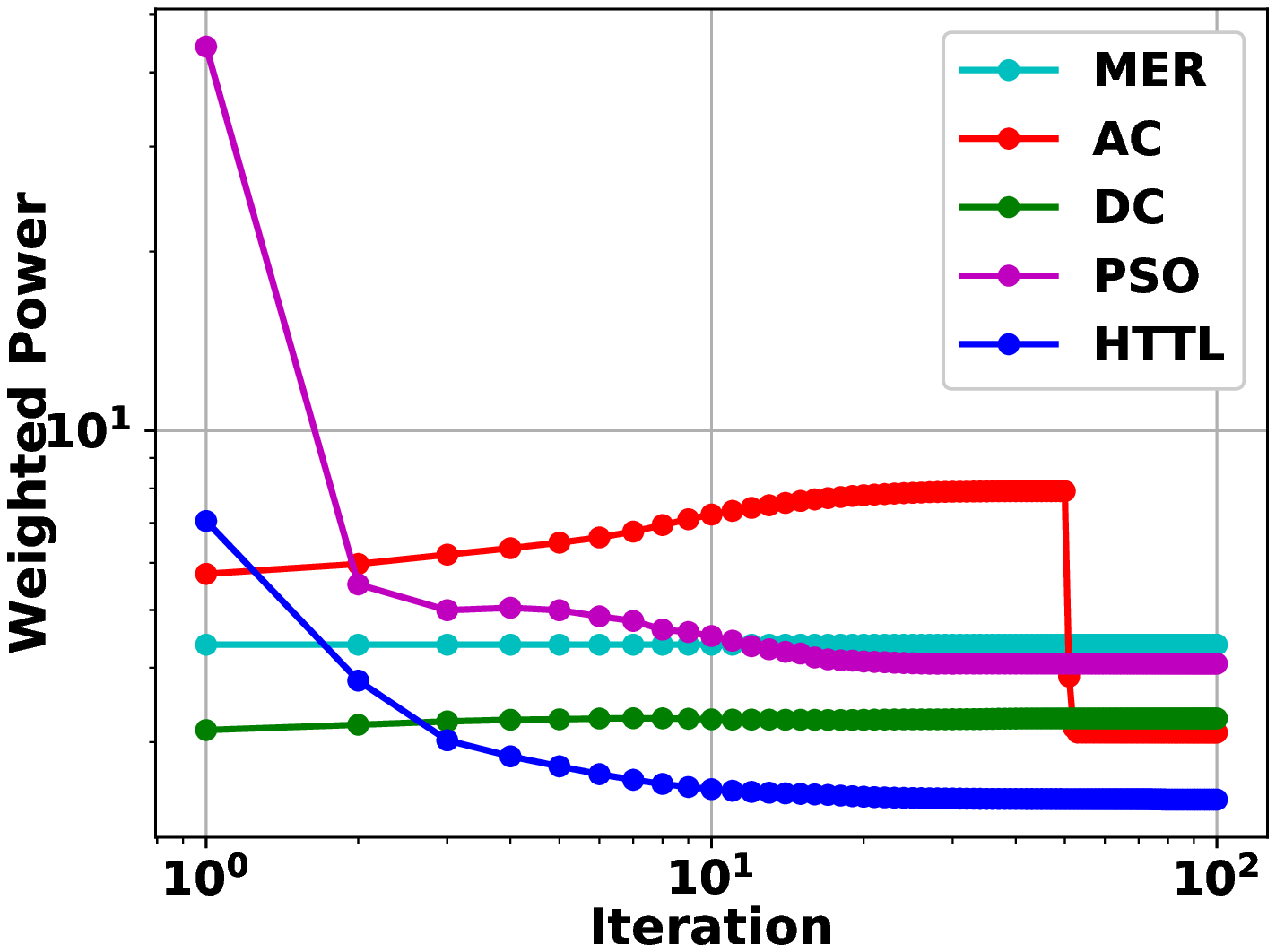}}
\label{distortion-vs-iteration-WSN2-Uniform}
\hspace{-3.65mm}
\subfloat[]{\includegraphics[width=42.0mm]{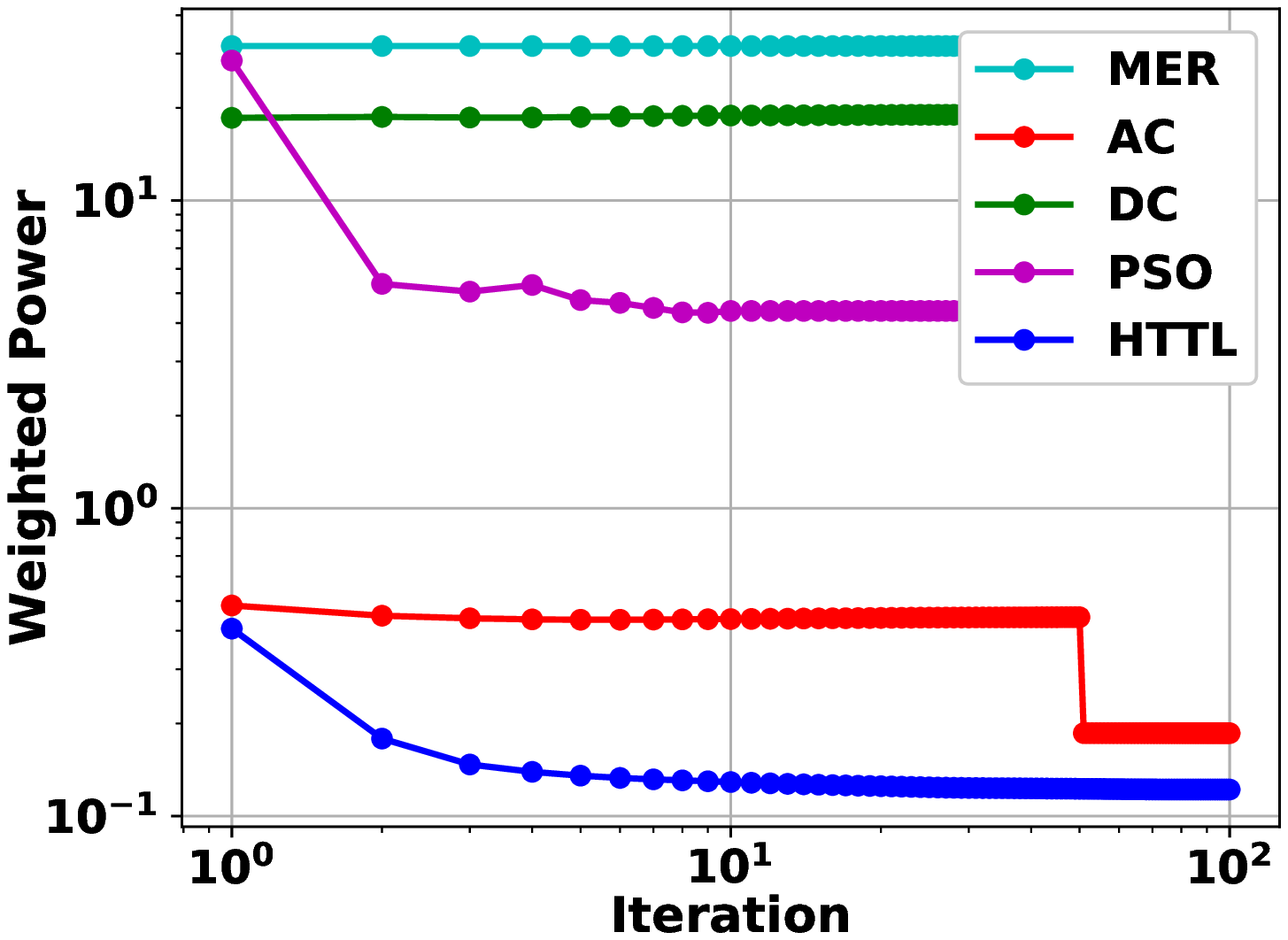}}
\label{distortion-vs-iteration-WSN1-NonUniform}
\hspace{-3.65mm}
\subfloat[]{\includegraphics[width=42.0mm]{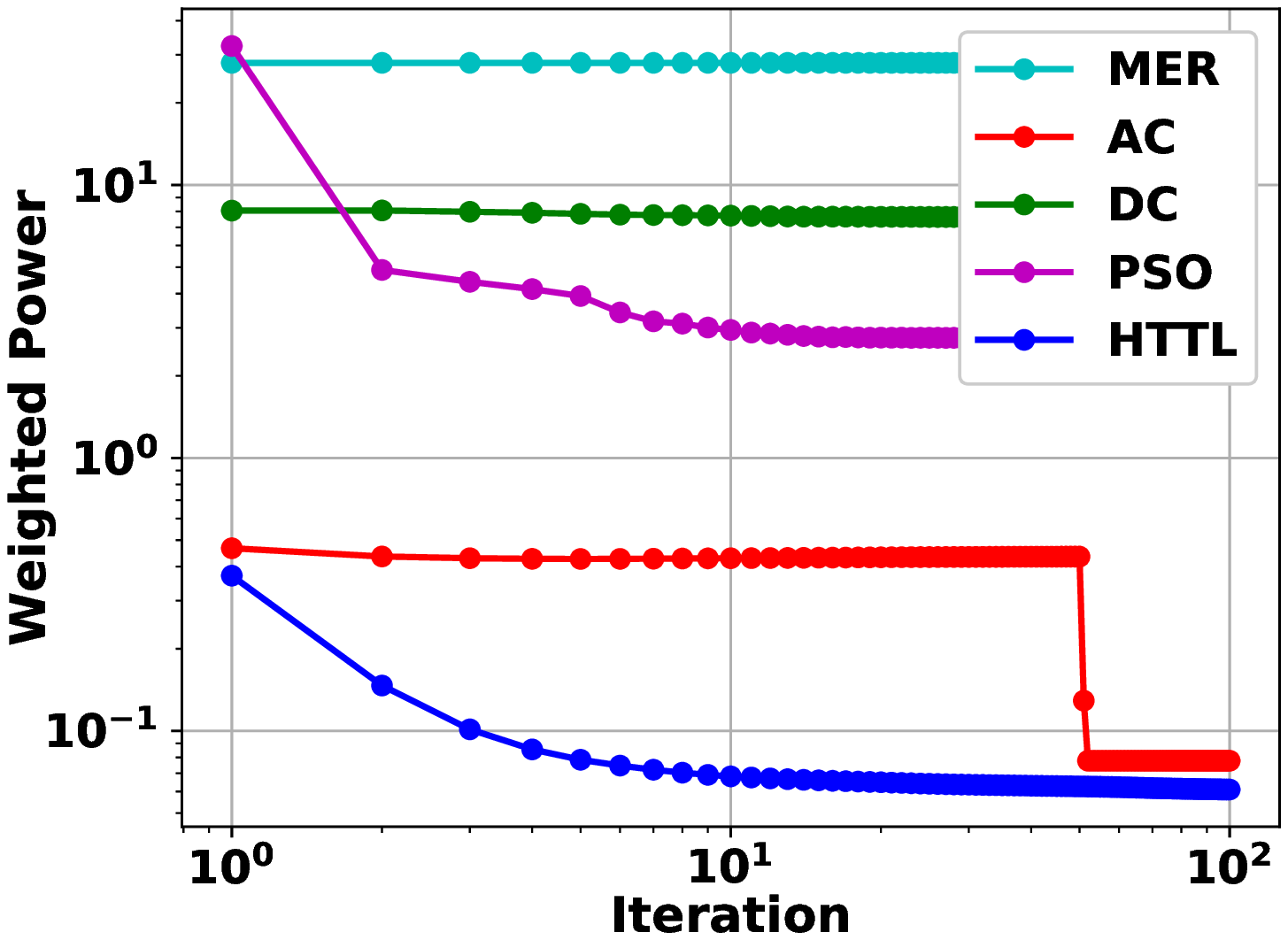}}
\label{distortion-vs-iteration-WSN2-NonUniform}
\vspace{-3mm}
\captionsetup{justification=justified}
\caption{\small{Weighted power versus iteration for different algorithms ($\beta=0.25$). (a) WSN1/Uniform pdf, (b) WSN2/Uniform pdf, (c) WSN1/Mixture of Gaussian pdf, (d) WSN2/Mixture of Gaussian pdf.}}
\label{distortion-vs-iteration}
\end{figure}

\begin{figure}[!htb]
\centering
\subfloat[]{\includegraphics[width=42.1mm]{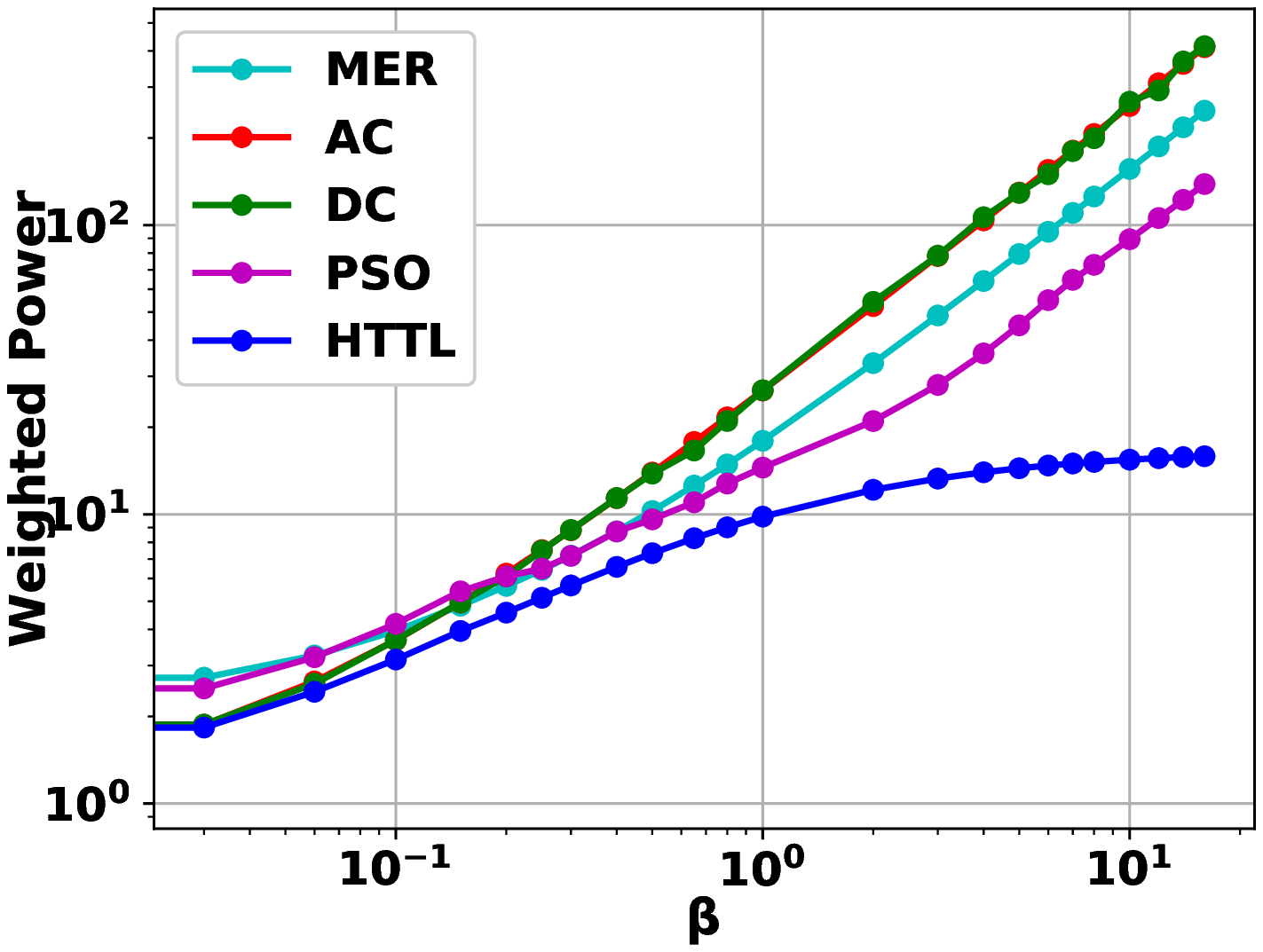}
\label{DistortionWSN1_Uniform}}
\hspace{-4.9mm}
\subfloat[]{\includegraphics[width=42.1mm]{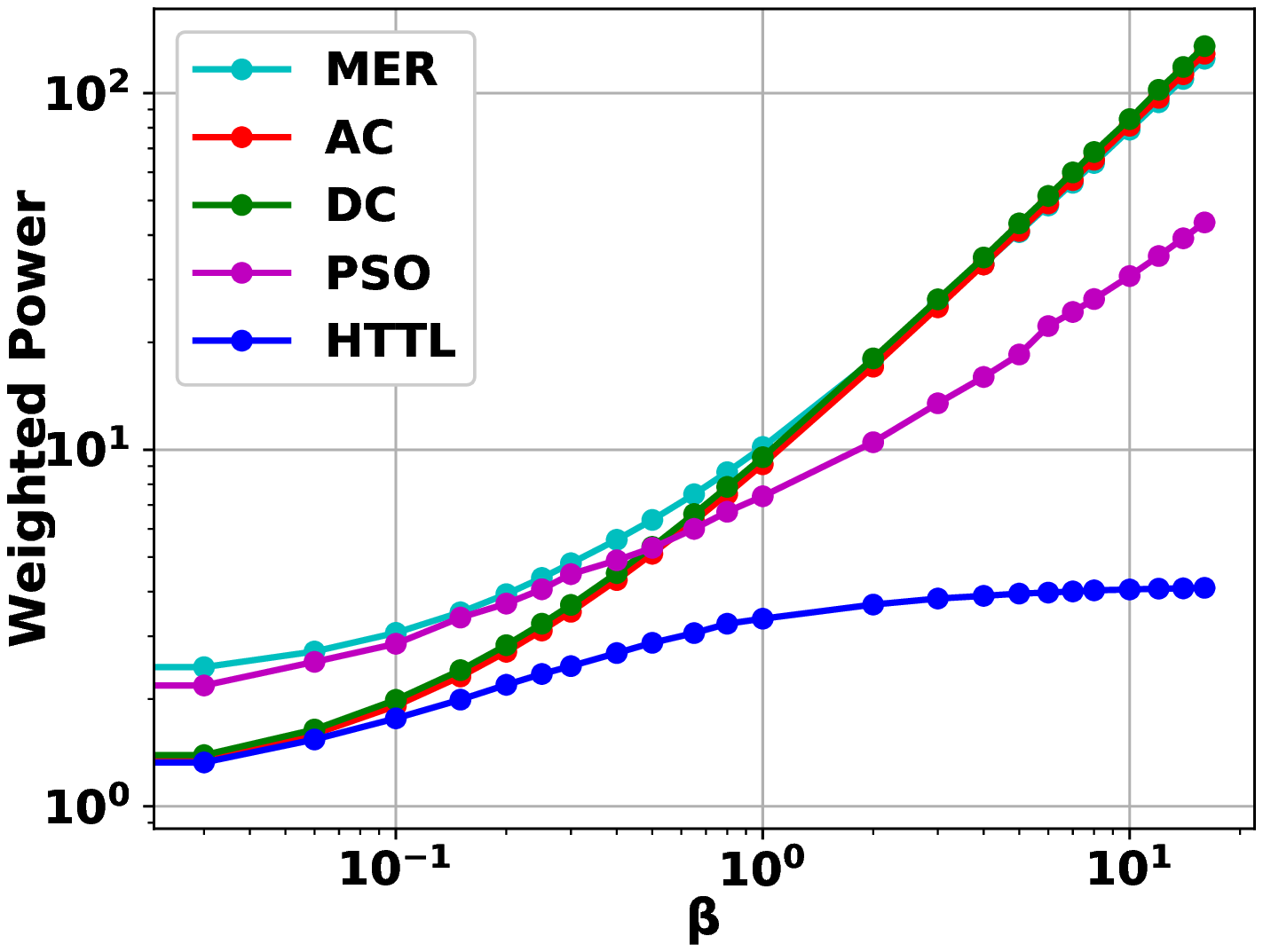}
\label{DistortionWSN2_Uniform}}
\hspace{-4.3mm}
\subfloat[]{\includegraphics[width=42.1mm]{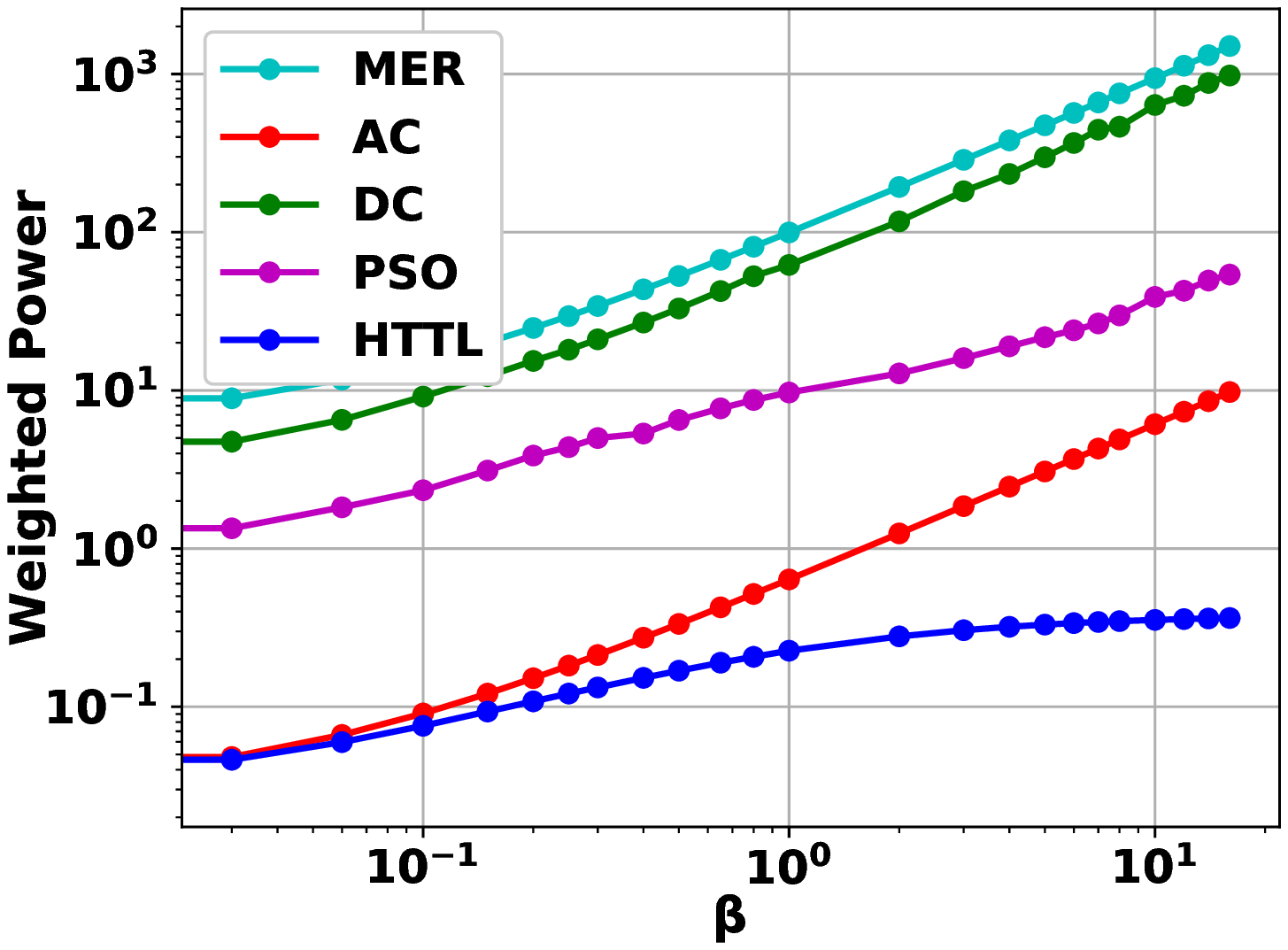}
\label{DistortionWSN1_Gaussian}}
\hspace{-4.2mm}
\subfloat[]{\includegraphics[width=42.1mm]{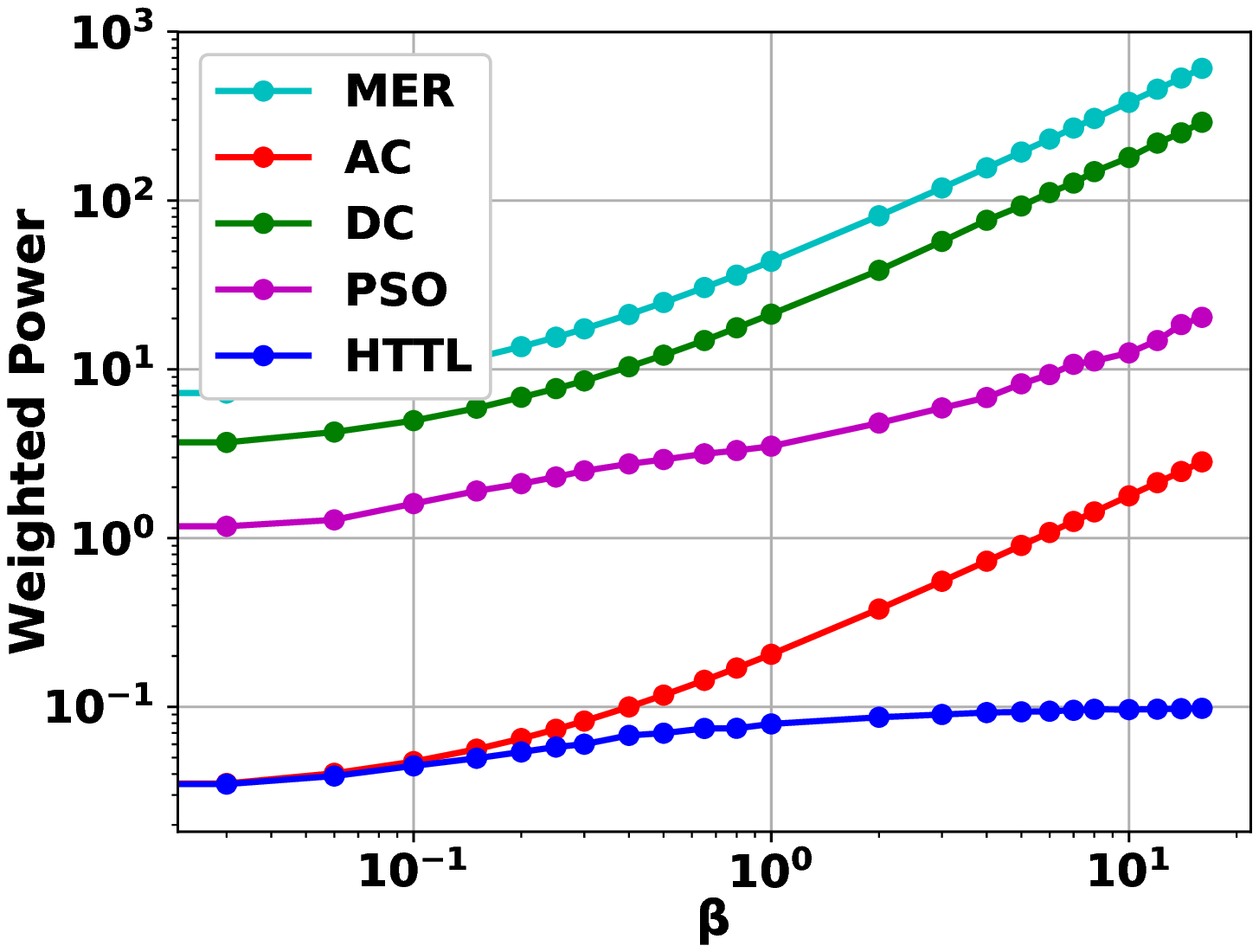}
\label{DistortionWSN2_Gaussian}}
\captionsetup{justification=justified}
\vspace{-3mm}
\caption{\small{Power consumption of different two-tier WSNs/sensor density functions. (a) WSN1/Uniform pdf, (b) WSN2/Uniform pdf, (c) WSN1/Mixture of Gaussian pdf, (d) WSN2/Mixture of Gaussian pdf.}}
\label{DistortionComparsion}
\end{figure}

\begin{figure}[!htb]
\centering
\subfloat[]{\includegraphics[width=35.5mm]{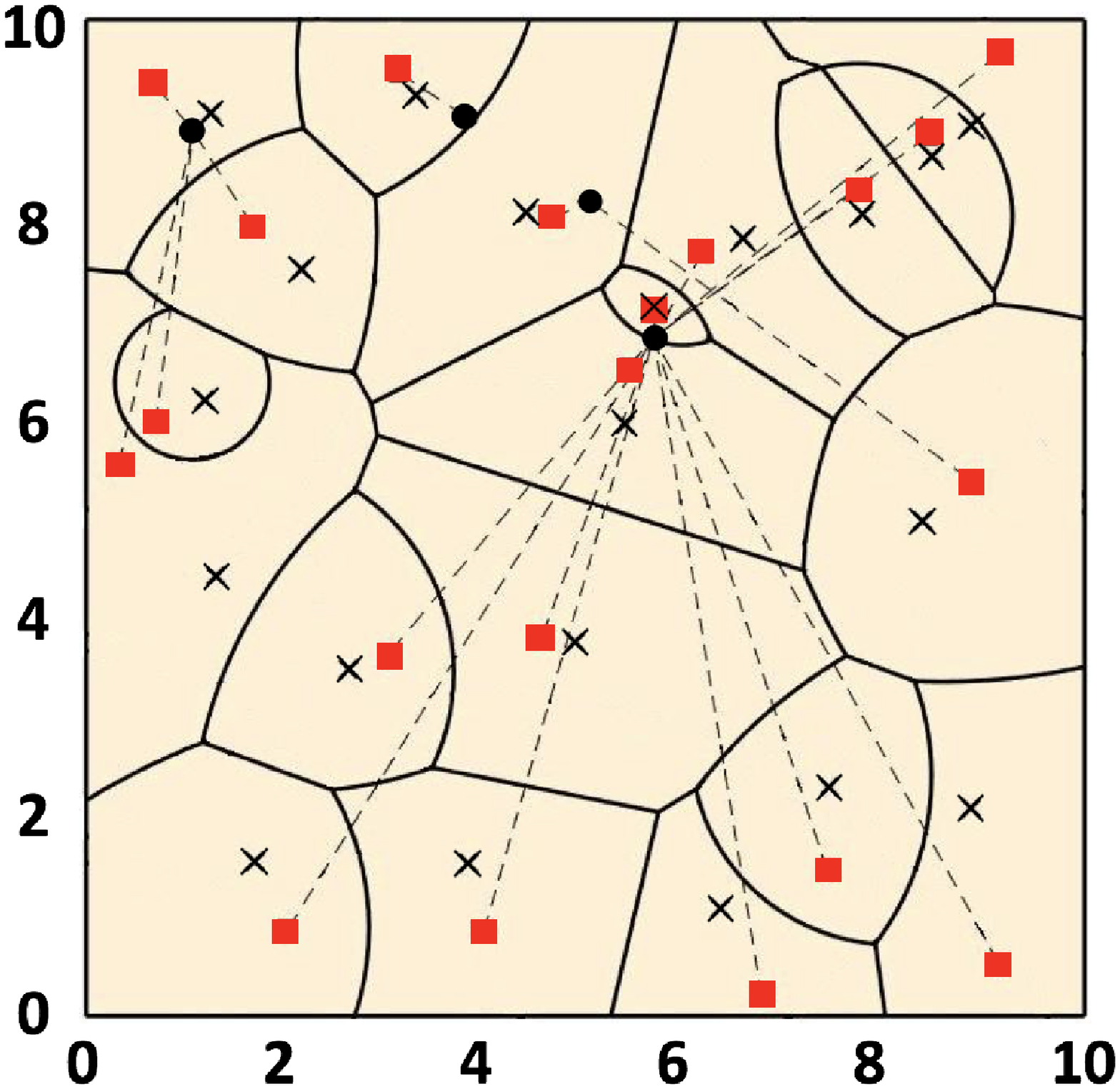}}
\label{MER-WSN2-NonUniform-Deployment}
\hspace{-6.5mm}
\subfloat[]{\includegraphics[width=35.5mm]{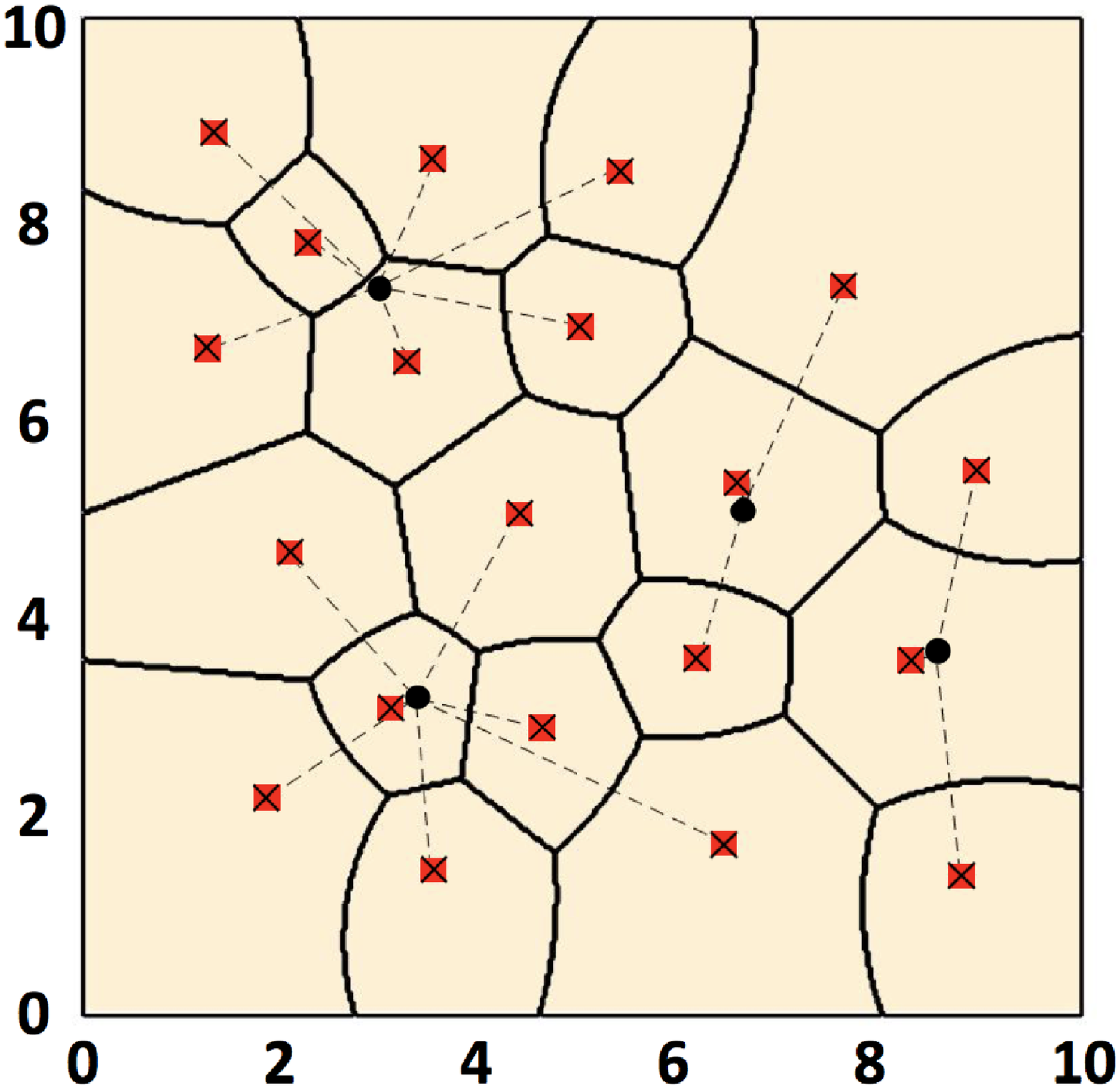}}
\label{AC-WSN2-NonUniform-Deployment}
\hspace{-6.5mm}
\subfloat[]{\includegraphics[width=35.5mm]{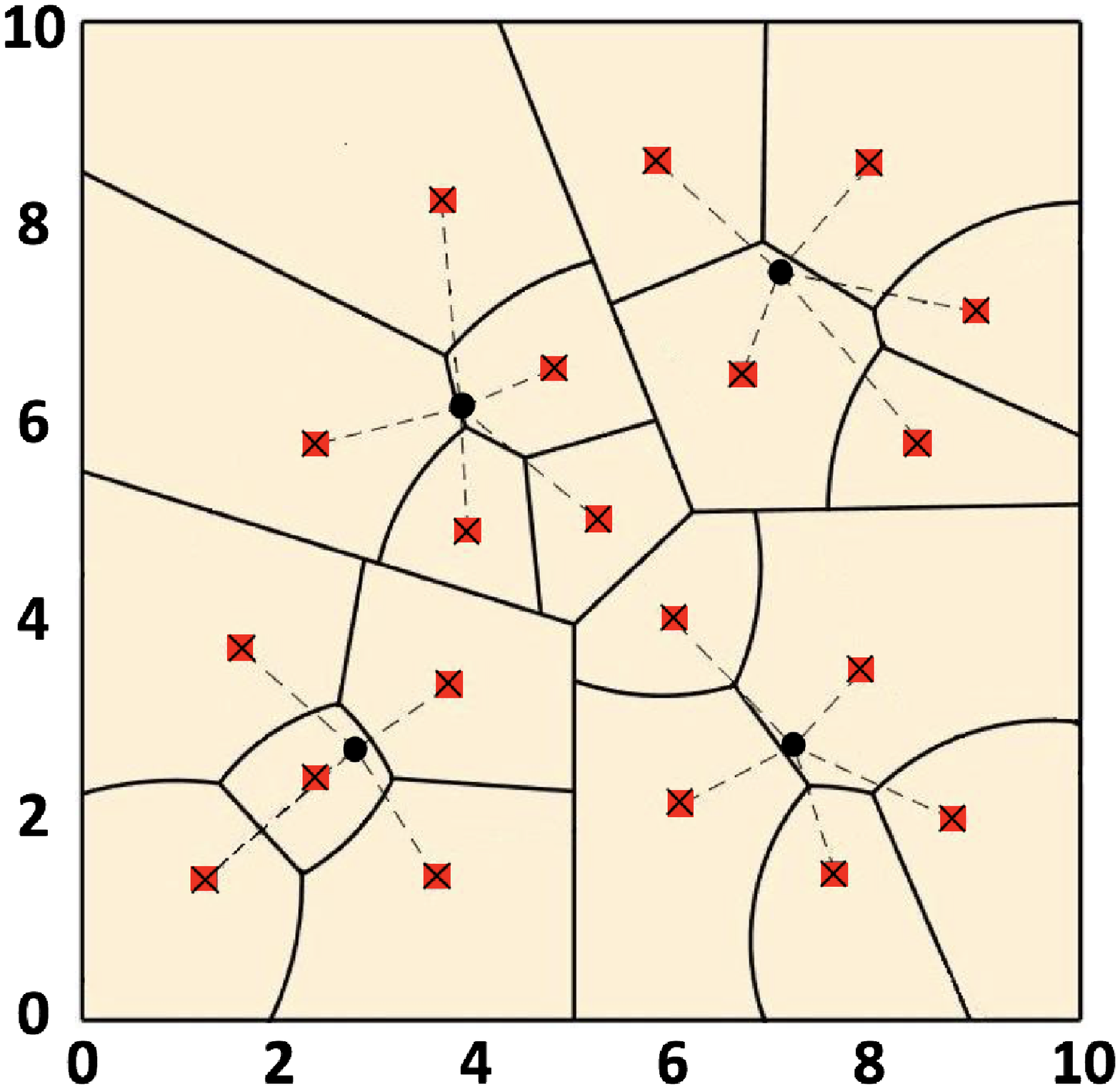}}
\label{DC-WSN2-NonUniform-Deployment}
\hspace{-6.5mm}
\subfloat[]{\includegraphics[width=35.5mm]{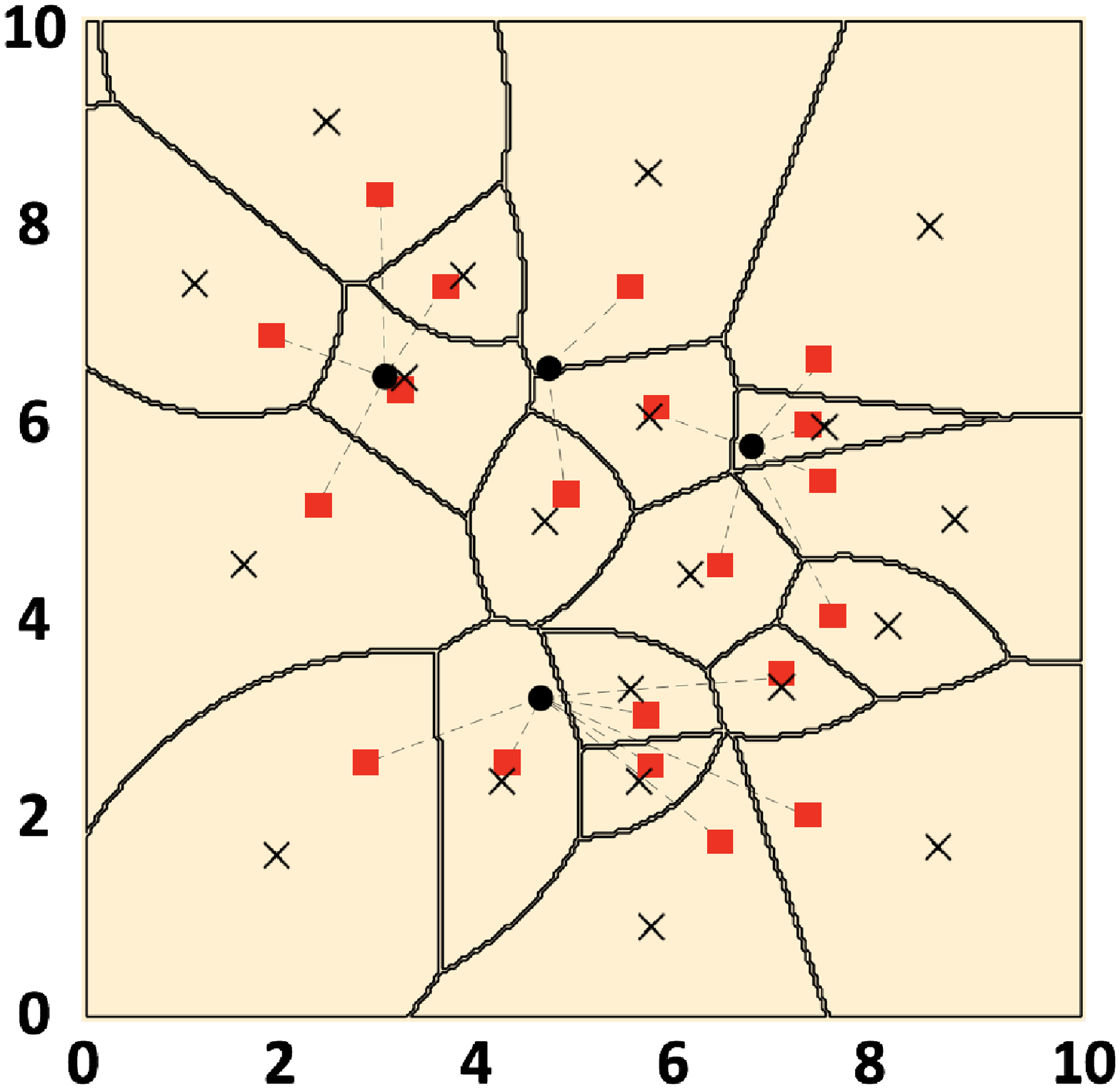}}
\label{PSO-WSN2-NonUniform-Deployment}
\hspace{-6.5mm}
\subfloat[]{\includegraphics[width=35.5mm]{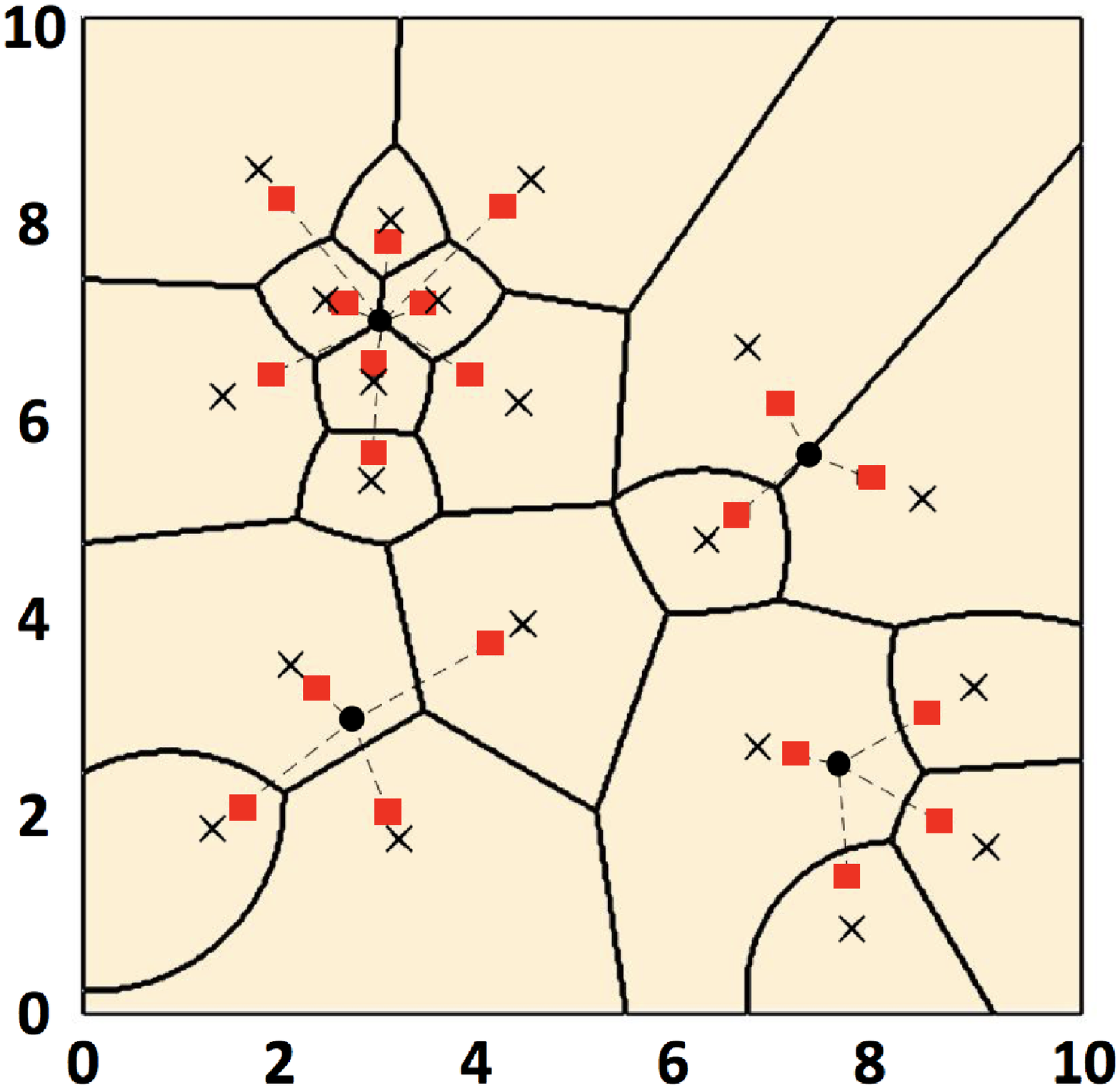}}
\label{HTTL-WSN2-NonUniform-Deployment}
\vspace{-3mm}
\captionsetup{justification=justified}
\caption{\small{Node deployment for different algorithms with $\beta=0.25$ in WSN2 and the mixture of Gaussian sensor density function. (a) MER (b) AC (c) DC (d) PSO (e) HTTL.}}
\label{WSN2-NonUniform-Deployment}
\end{figure}

The weighted power consumption over the iterations of MER, AC, DC, PSO and HTTL algorithms in WSN1 and WSN2 for $\beta=0.25$ are shown in Figs. \ref{distortion-vs-iteration}a and \ref{distortion-vs-iteration}b for uniform sensor density function, and in Figs. \ref{distortion-vs-iteration}c and \ref{distortion-vs-iteration}d for the Gaussian mixture given in (\ref{Mixture_of_Gaussian_Distribution}).
Weighted power consumption of MER, AC, DC, PSO and HTTL algorithms in WSN1 and WSN2 are illustrated in Figs. \ref{DistortionWSN1_Uniform} and \ref{DistortionWSN2_Uniform} for uniform sensor density function, and in Figs. \ref{DistortionWSN1_Gaussian} and \ref{DistortionWSN2_Gaussian} for the Gaussian mixture given in (\ref{Mixture_of_Gaussian_Distribution}). Obviously, our proposed algorithm, HTTL, outperforms the other four algorithms in both WSN1 and WSN2. For instance, HTTL Algorithm yields the power consumption of $2.351$ for WSN2, $\beta = 0.25$ and uniform distribution, which is lower than the values $4.371$, $3.113$, $3.253$  and $4.063$ obtained from MER, AC, DC and PSO algorithms, respectively. Similarly, for the case of WSN2 and mixture of Gaussian, HTTL Algorithm yields the power consumption of $0.058$ which is lower than the values $15.484$, $0.074$, $7.677$ and $2.301$ obtained from MER, AC, DC and PSO algorithms, respectively. Unlike other methods, HTTL Algorithm exploits the trade-off between Sensor and AP power consumptions; hence, the energy consumption gap between HTTL and other  algorithms increases as the  AP energy consumption becomes more important ($\beta$ increases). For $\beta=0.25$, the final node deployment for WSN2 and the mixture of Gaussian sensor density function given in (\ref{Mixture_of_Gaussian_Distribution}) is shown in Fig. \ref{WSN2-NonUniform-Deployment} where APs, FCs and centroid of regions are denoted via red squares, black circles and crosses, respectively.

\begin{figure}[!htb]
\centering
\subfloat[]{\includegraphics[width=40.5mm]{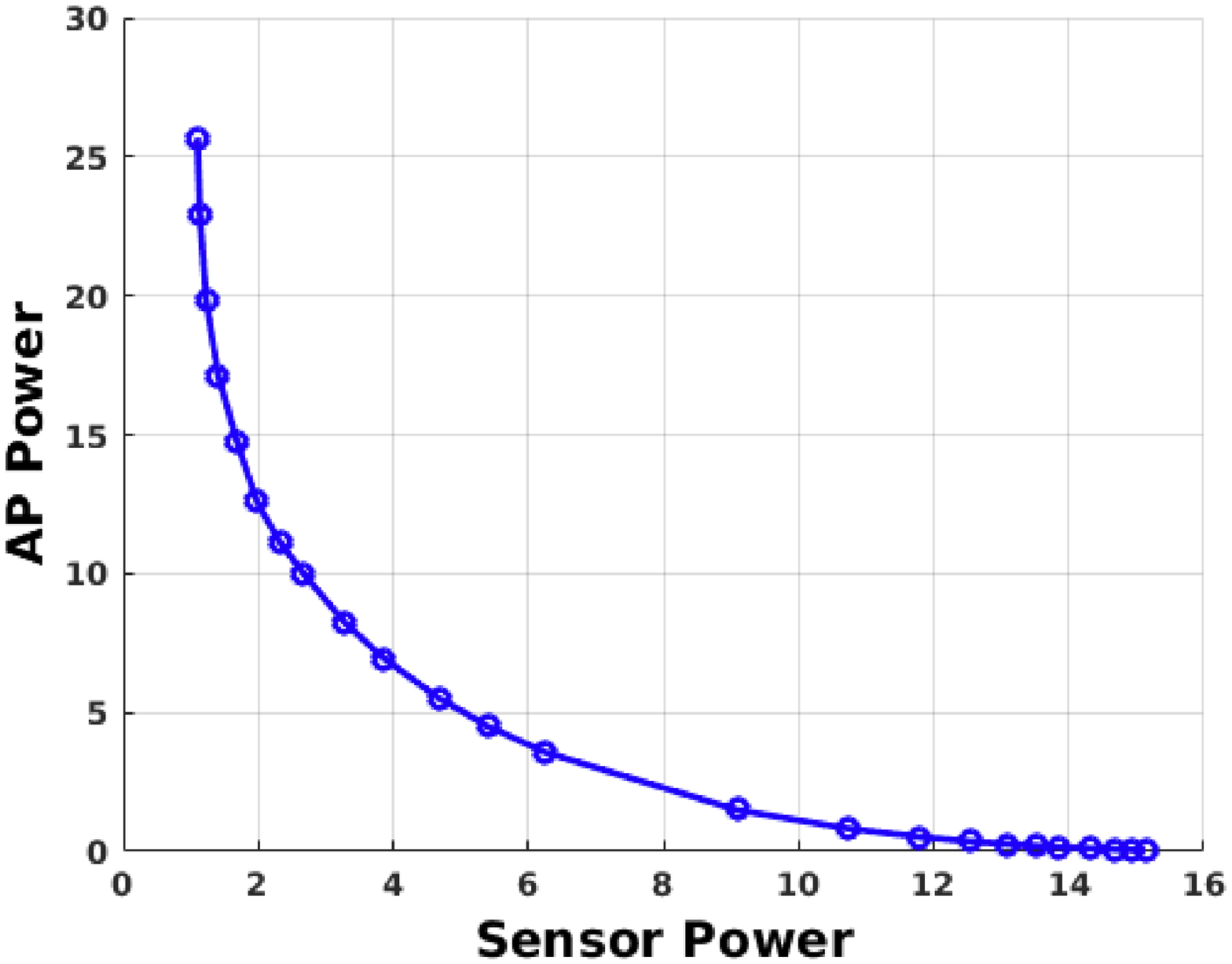}
\label{Power_Tradeoff_WSN1_Uniform}}
\hspace{-2.3mm}
\subfloat[]{\includegraphics[width=40.5mm]{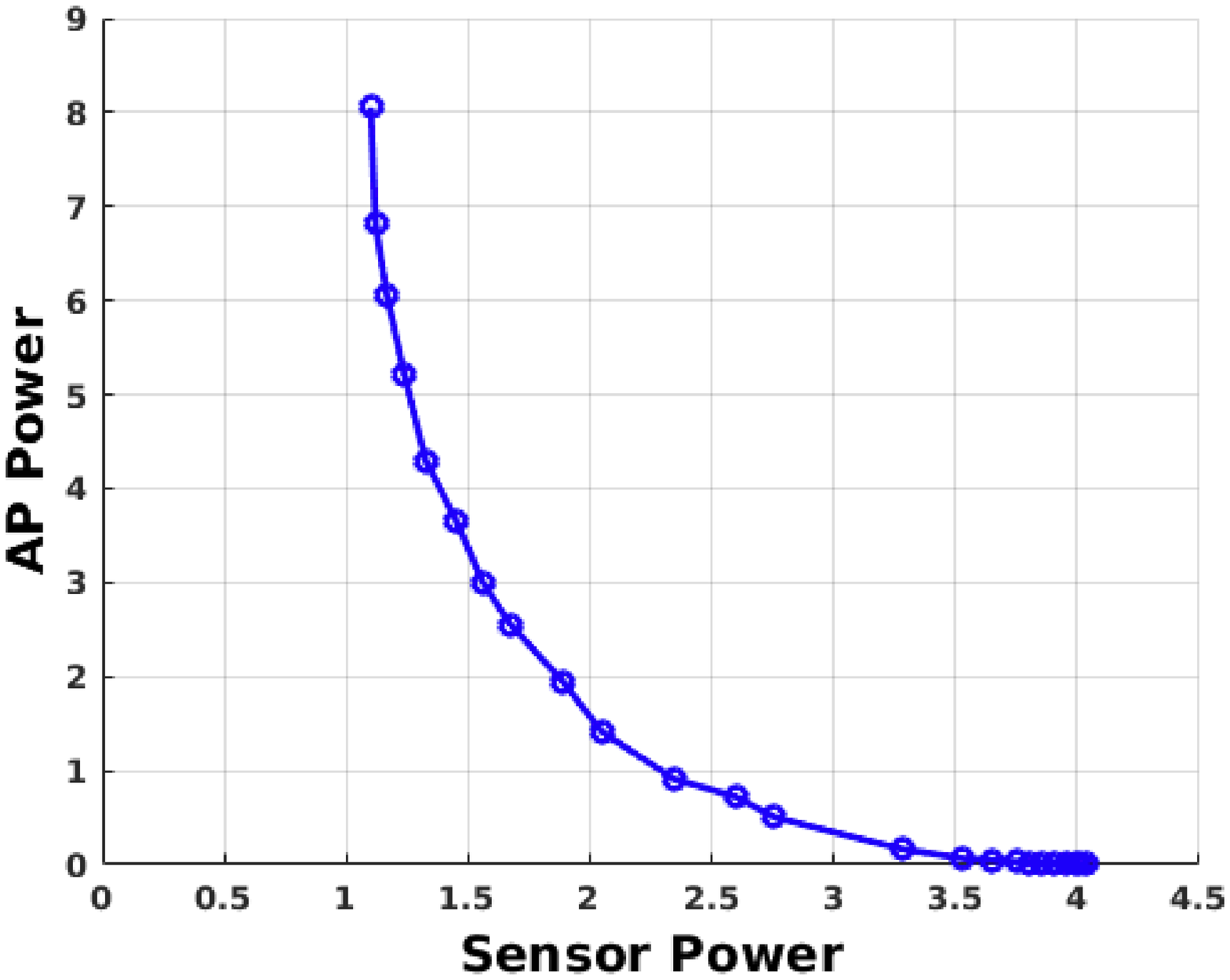}
\label{Power_Tradeoff_WSN2_Uniform}}
\hspace{-2.3mm}
\subfloat[]{\includegraphics[width=40.5mm]{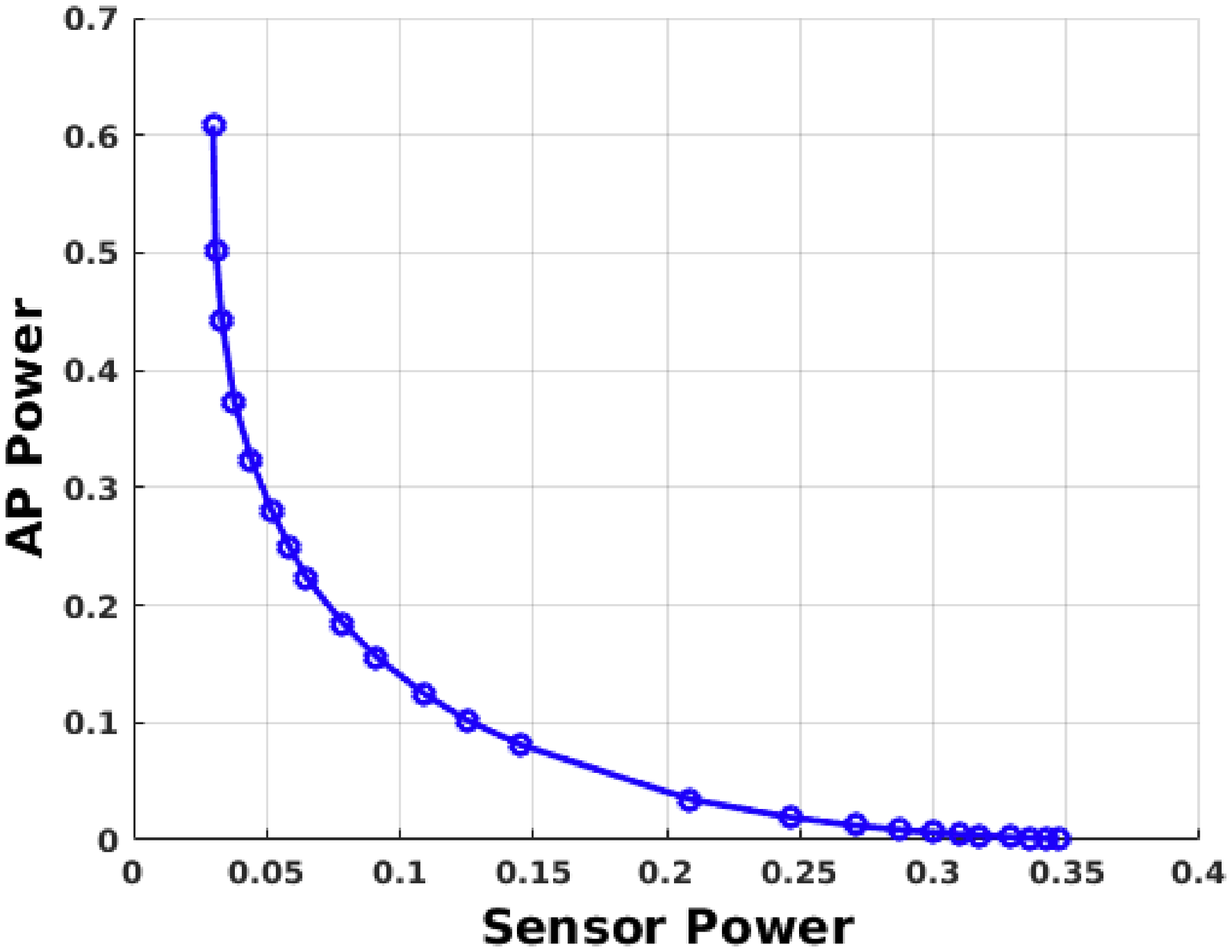}
\label{Power_Tradeoff_WSN1_Gaussian}}
\hspace{-2.3mm}
\subfloat[]{\includegraphics[width=40.5mm]{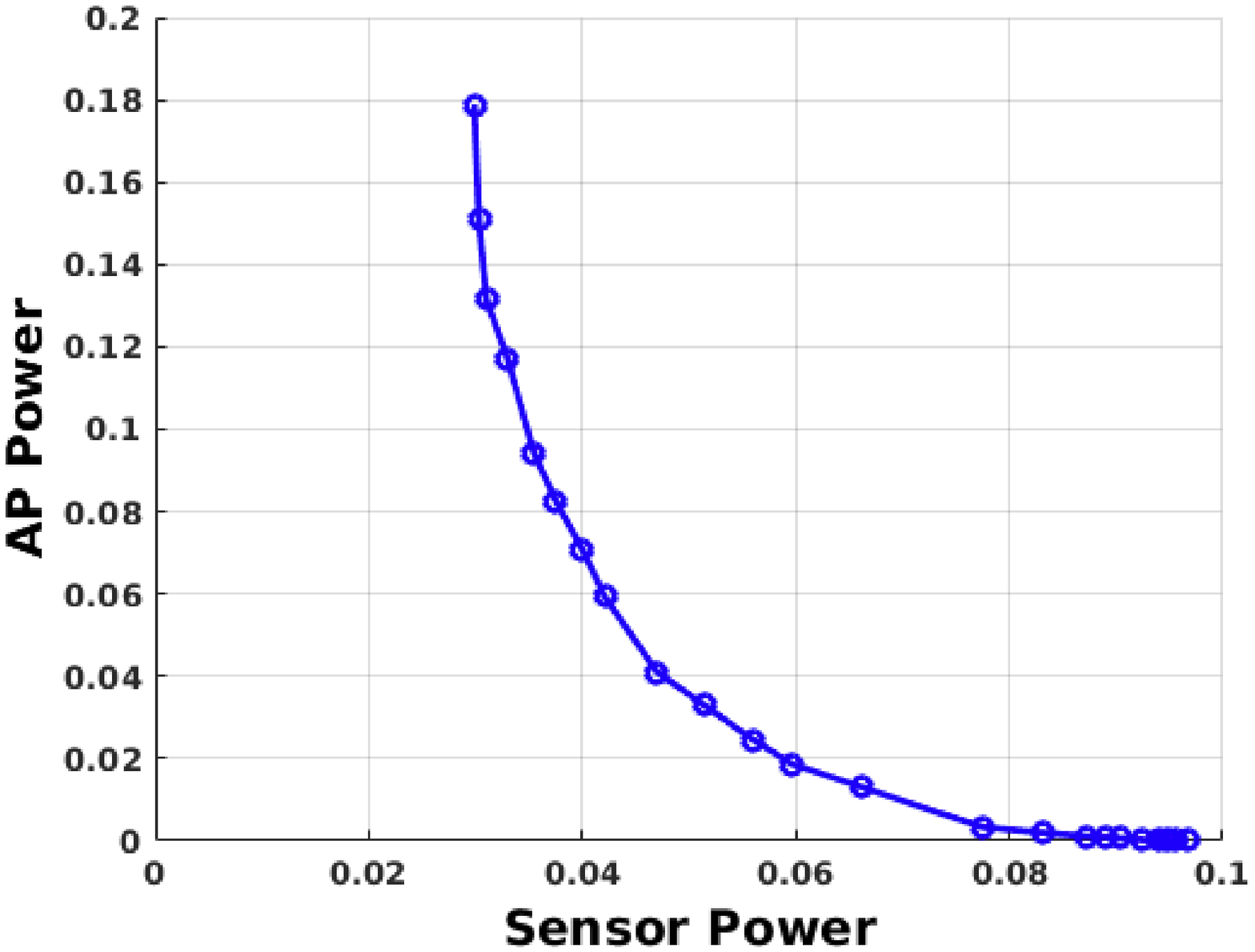}
\label{Power_Tradeoff_WSN2_Gaussian}}
\vspace{-4mm}
\captionsetup{justification=justified}
\caption{\small{AP-Sensor power trade-off for HTTL Algorithm (a) WSN1/Uniform pdf, (b) WSN2/Uniform pdf, (c) WSN1/Mixture of Gaussian pdf, (d) WSN2/Mixture of Gaussian pdf.}}
\label{Power Tradeoffs}
\end{figure}

Note that the two-tier power consumption defined in (\ref{eq4}) represents a trade-off between the Sensor power $\overline{\mathcal{P}}^{\mathcal{S}}$ and AP power $\overline{\mathcal{P}}^{\mathcal{A}}$, and this trade-off is illustrated as the AP-Sensor power functions for WSN1 and WSN2 in Figs. \ref{Power_Tradeoff_WSN1_Uniform} and \ref{Power_Tradeoff_WSN2_Uniform} for uniform distribution, and in Figs. \ref{Power_Tradeoff_WSN1_Gaussian} and \ref{Power_Tradeoff_WSN2_Gaussian} for the mixture of Gaussian sensor density function, respectively. For small values of $\beta$, sensor power contributes to the two-tier power consumption more than AP power; hence, the optimal deployment tends to minimize $\overline{\mathcal{P}}^{\mathcal{S}}$, while $\overline{\mathcal{P}}^{\mathcal{A}}$ tends to be minimized in an optimal node placement for large values of $\beta$. Intuitively, moving APs towards the FCs, usually, will increase the average distance between sensors and APs, resulting in the increase of the sensor power. On the other hand, moving APs toward geometric centroids of their corresponding regions, usually, will increase their distances to the FCs, which leads to an increase in the AP power. This is shown in Fig. \ref{Power Tradeoffs} where the AP-Sensor power function $A(s)$ decreases as $s$ increases. Lemma \ref{AP_Sensor_power_function_is_non_increasing} indicates that $A(s)$ is non-zero on the intervals $\left[D_{20}, D_1 \right)$ and $\left[D_{20}, D_4 \right)$ for WSN1 and WSN2, respectively. Simulations show that AP-sensor power function is a piece-wise continuous convex function, as we demonstrated earlier for the setting in Lemma \ref{special_case_closed_form_A(s)}.

\begin{table}[!bth]
\setlength\abovecaptionskip{0pt}
\setlength\belowcaptionskip{0pt}
\centering
\caption{Power Constraint Parameters}
\begin{tabular}{|c|c|c|c|c|c|c|c|}
\hline
Parameters & \!\!$\quad\sigma^2\quad$\!\! & \!\!$\quad\sigma^2_{1:4}\quad$\!\! & \!\!$\quad\sigma^2_{5:10}\quad$\!\! & \!\!$\quad\sigma^2_{11:20}\quad$\!\! \\
\hline
Values & 4 & 25 & 16 & 9\\
\hline
\end{tabular}
\label{Power_Const_Table}
\end{table}

Next, we consider a transmission power constraint on sensors and APs. The value of parameters $\sigma^2$ and $\sigma^2_n, n\in \mathcal{I_A}$ in (\ref{w_to_Pn_and_Qm_power_constraint}) are provided in Table \ref{Power_Const_Table}. 
According to Table \ref{Power_Const_Table}, strong APs ($n\in \{1,\ldots,10 \}$) also tend to have more available power than weak APs ($n\in \{11,\ldots, 20\}$).

\begin{table*}[t]
  \centering
  \caption{{ Coverage and power comparison for uniform sensor density function.}}
  \vspace{-3mm}
\begin{tabular}{cc|c|c|c|c|c|c}
 \toprule
          &              &{\bf MER}&{\bf AC} &{\bf DC} &{\bf RNDWSN}&{\bf IRNP}&{\bf Limited-HTTL}\\\hline\hline
{\bf WSN1}&{\bf Power}   &$1.1287$ &$2.1812$ &$1.3972$ &$4.0105$    &$4.4258$  &      $3.2151$    \\ 
          &{\bf Coverage}&$33.90\%$&$53.01\%$&$40.31\%$&$74.13\%$   &$80.04\%$ &      $78.26\%$   \\\hline
{\bf WSN2}&{\bf Power}   &$0.8843$ &$2.3309$ &$2.6340$ &$3.9463$    &$4.7733$  &      $2.1305$    \\ 
          &{\bf Coverage}&$38.55\%$&$82.26\%$&$91.79\%$&$81.48\%$   &$95.09\%$ &      $94.66\%$   \\
 \bottomrule
\end{tabular}
  \label{power_coverage_comparison_uniform}
\end{table*}

The two-tier power consumption and coverage of different algorithms for $\beta = 0.25$ and uniform sensor density function are summarized in Table \ref{power_coverage_comparison_uniform}. IRNP Algorithm yields the maximum coverage in WSN1; however, the $1.78\%$ improvement in the coverage over our proposed Limited-HTTL Algorithm comes at the cost of $38\%$ increase in power consumption. Our algorithm also outperforms RNDWSN Algorithm in terms of both power and coverage. Similarly, although IRNP Algorithm results in less than $1\%$ improvement in coverage compare to Limited-HTTL Algorithm in WSN2, it consumes more than twice power used by our proposed algorithm. Limited-HTTL Algorithm also outperforms the other algorithms in terms of both coverage and power consumption in WSN2.



\begin{figure}[!htb]
\centering
\subfloat[]{\includegraphics[width=27mm]{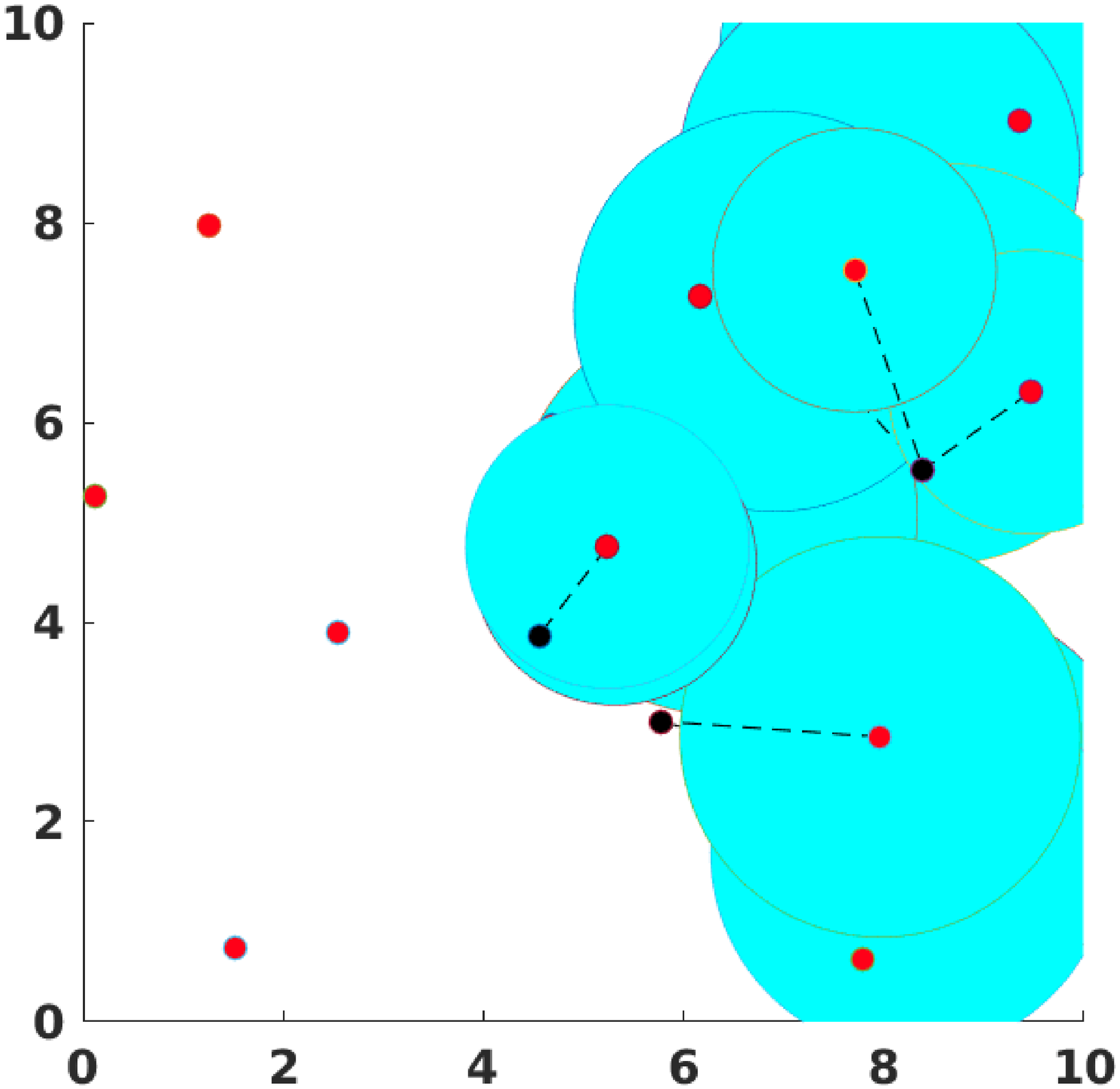}}
\label{Gaussian_WSN2_Covered_Area_MER}
\hspace{-2mm}
\subfloat[]{\includegraphics[width=27mm]{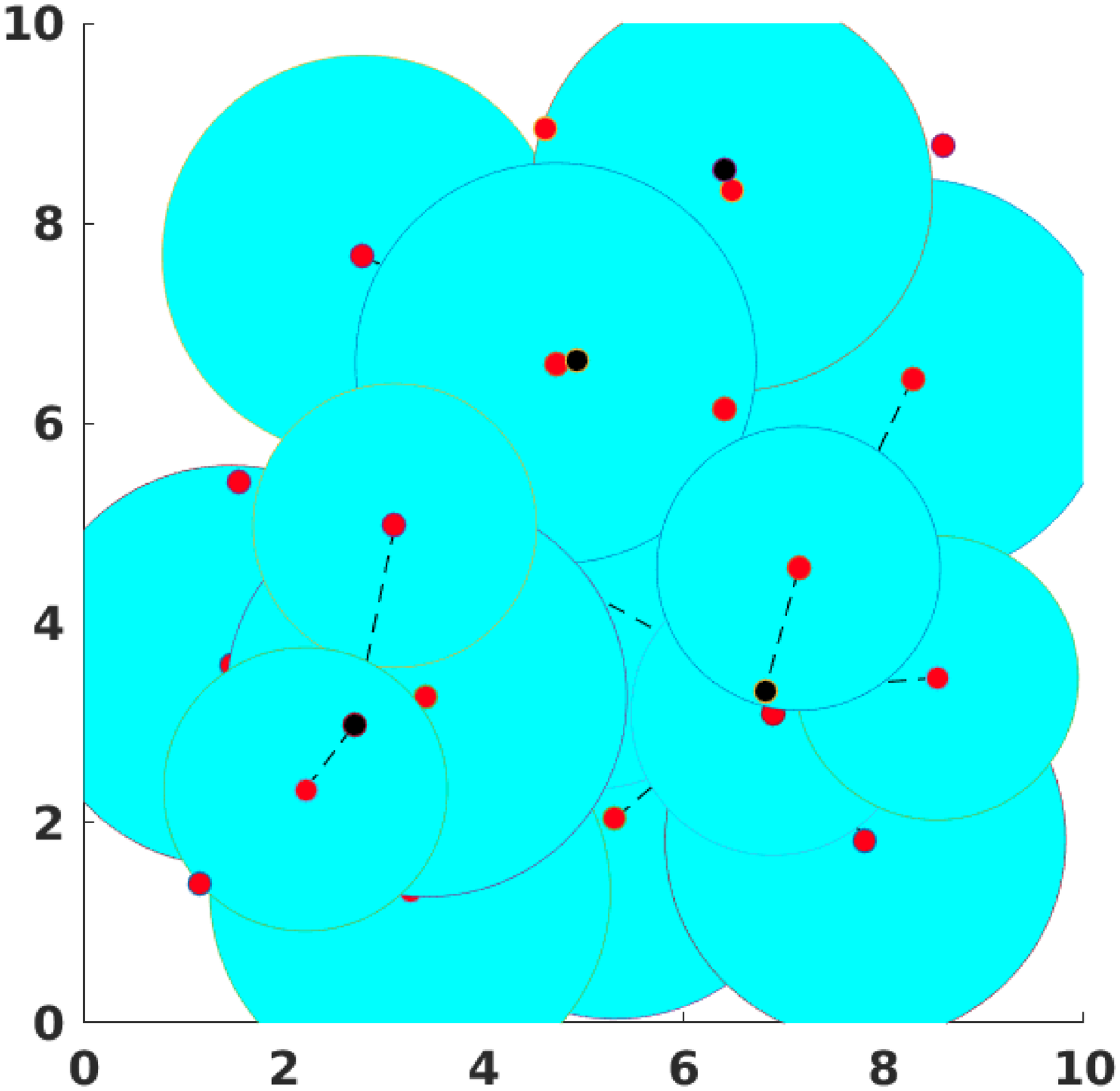}}
\label{Gaussian_WSN2_Covered_Area_AC}
\hspace{-2mm}
\subfloat[]{\includegraphics[width=27mm]{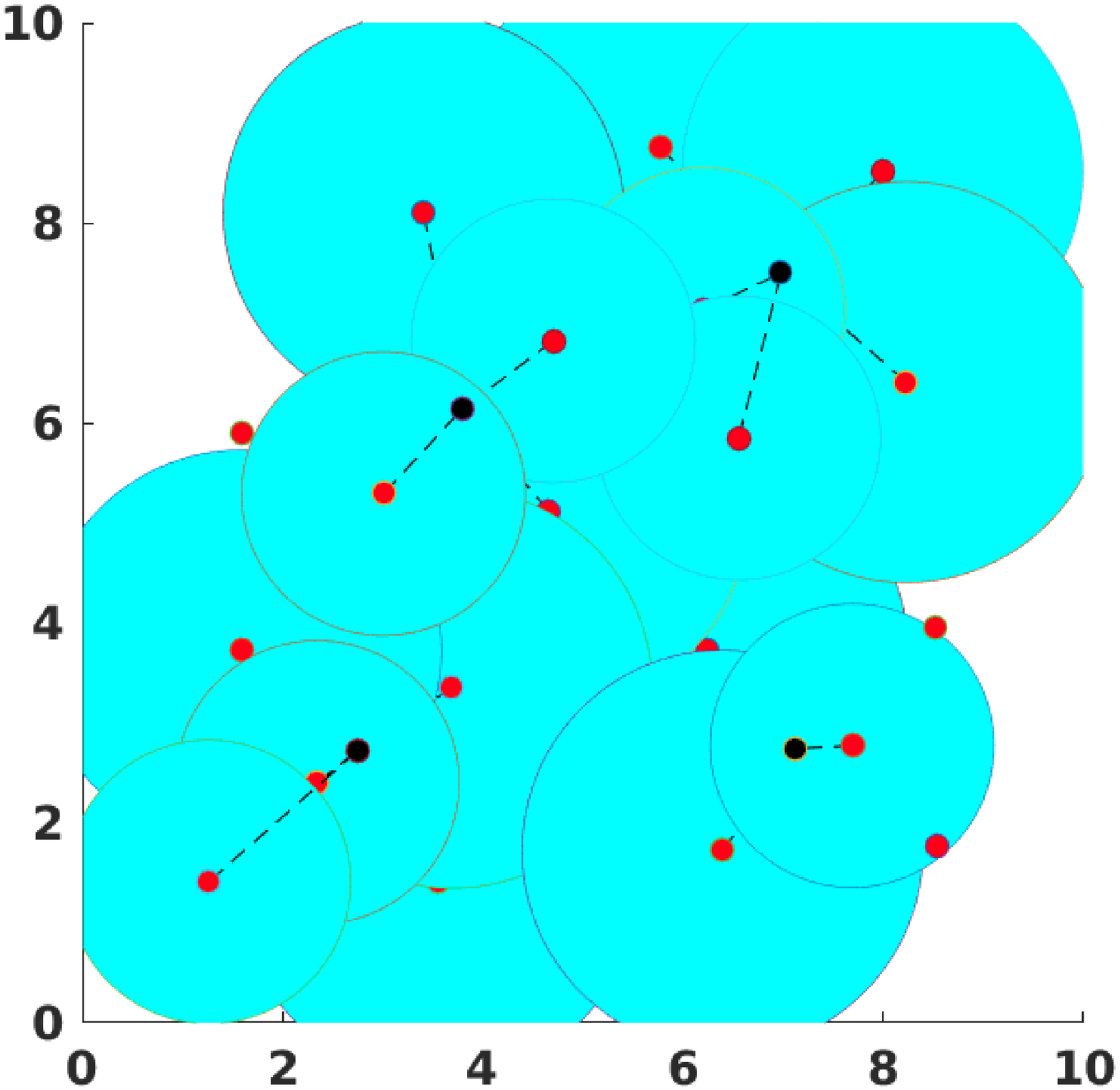}}
\label{Gaussian_WSN2_Covered_Area_DC}
\hspace{-2mm}
\subfloat[]{\includegraphics[width=27mm]{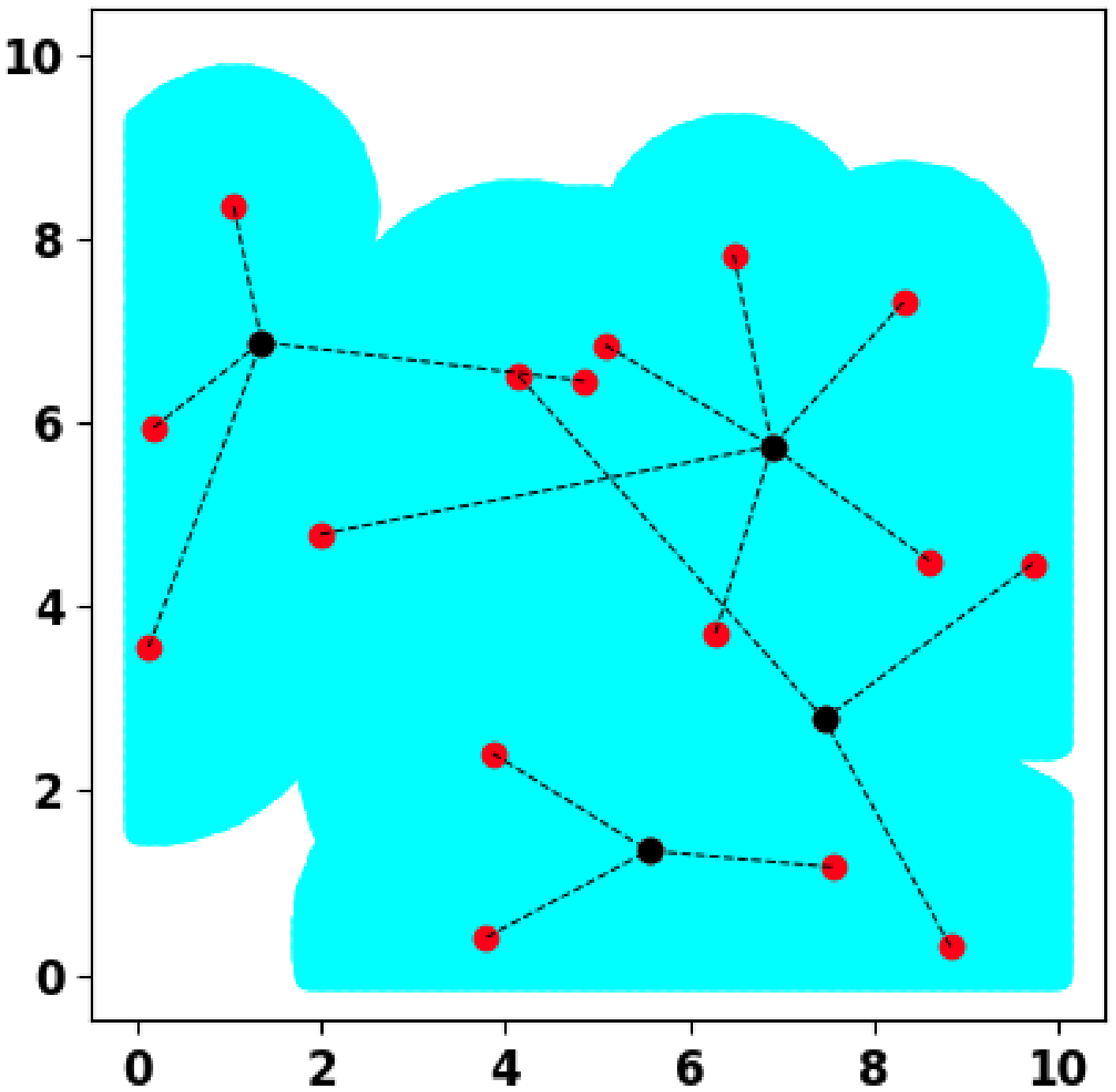}}
\label{Gaussian_WSN2_Covered_Area_RNDWSN}
\hspace{-2mm}
\subfloat[]{\includegraphics[width=27mm]{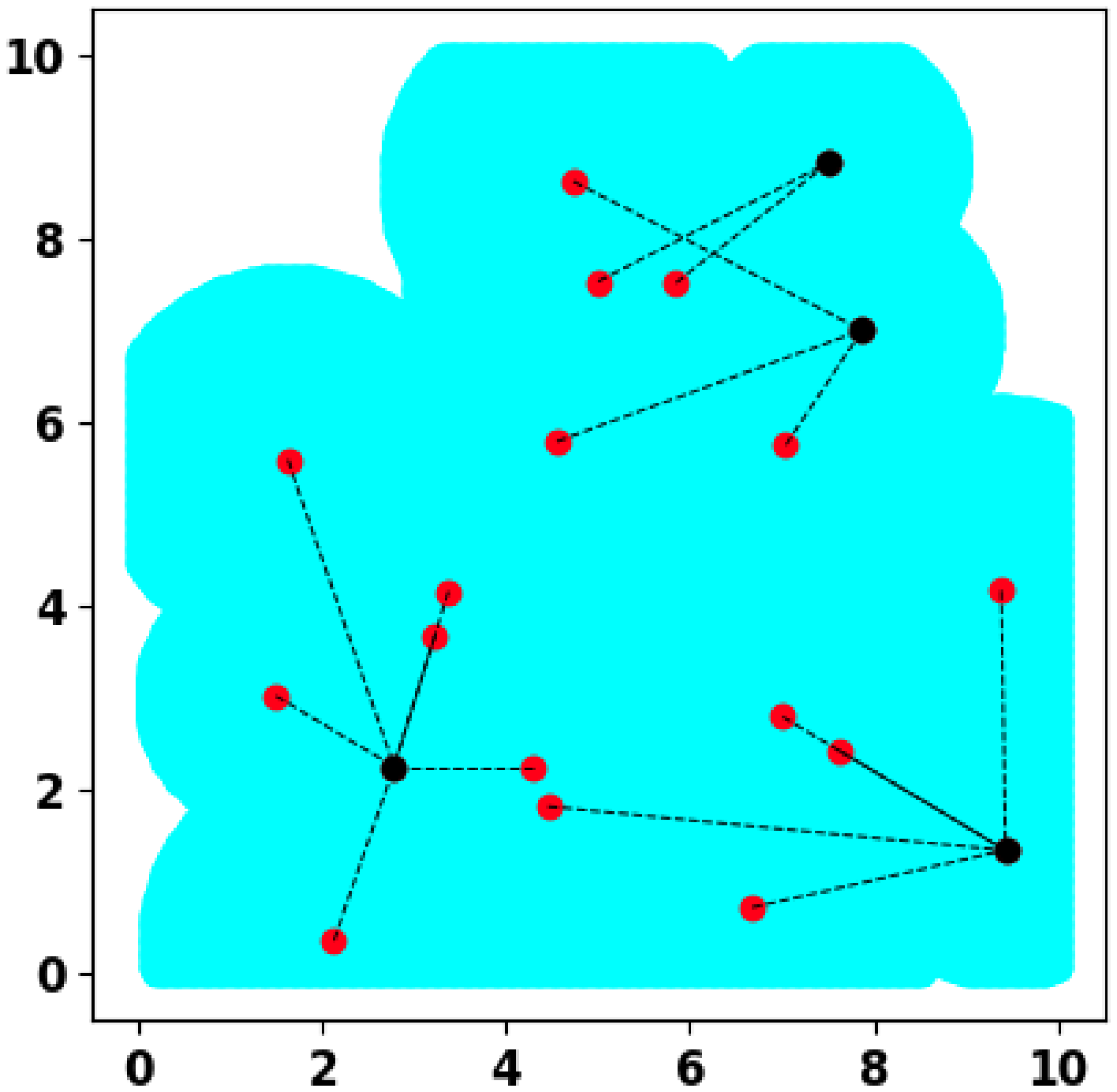}}
\label{Gaussian_WSN2_Covered_Area_IRNP}
\hspace{-2mm}
\subfloat[]{\includegraphics[width=27mm]{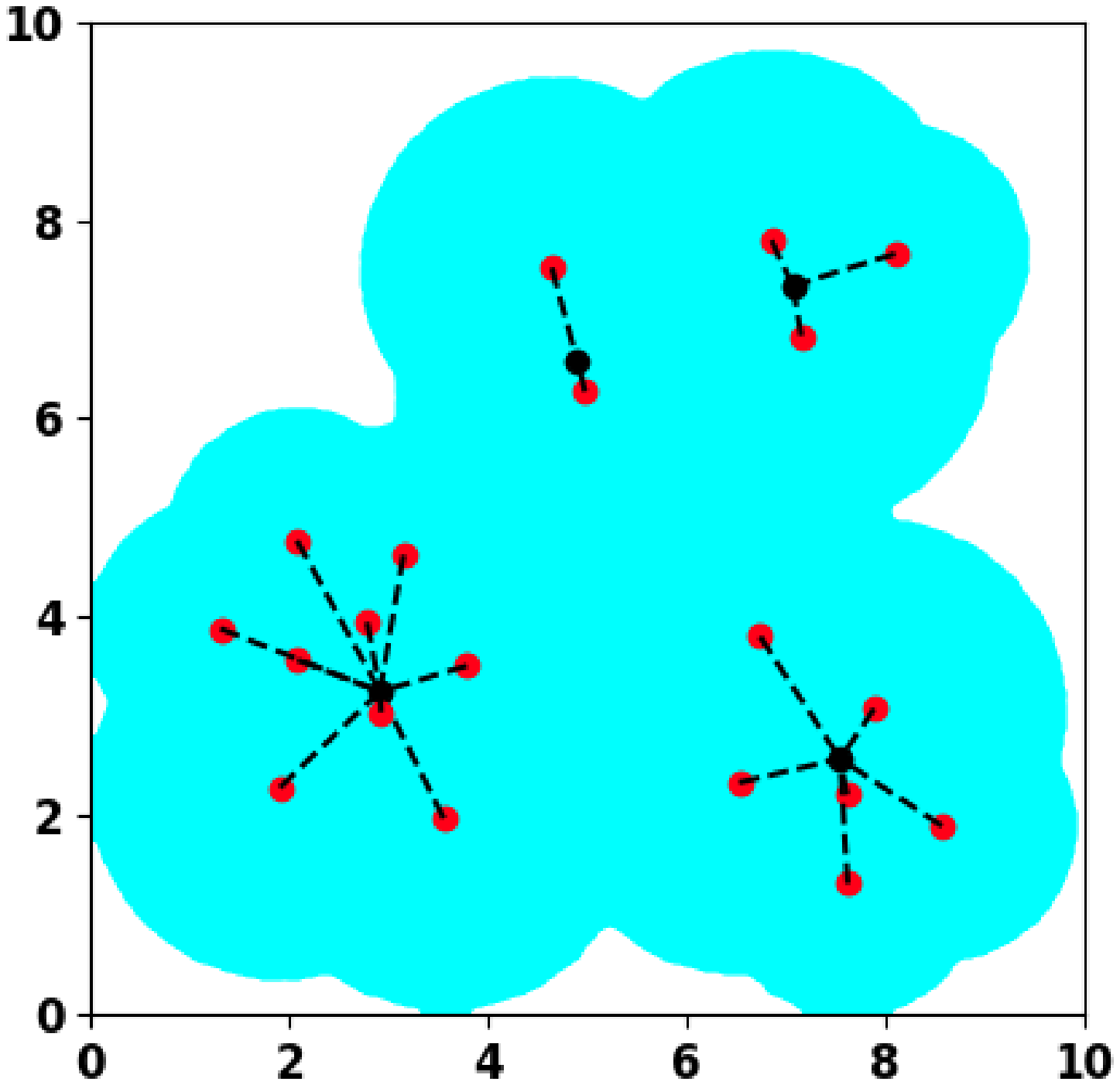}}
\label{Gaussian_WSN2_Covered_Area_HTTL}

\vspace{-3mm}
\captionsetup{justification=justified}
\caption{\small{Node deployment for different algorithms with $\beta=0.25$ and the mixture of Gaussian sensor density function in WSN2. (a) MER (b) AC (c) DC (d) RNDWSN (e) IRNP (f) Limited-HTTL.}}
\label{Gaussian_WSN2_Covered_Area}
\end{figure}

The two-tier power consumption and coverage of different methods for $\beta = 0.25$ and Gaussian mixture sensor density function given in (\ref{Mixture_of_Gaussian_Distribution}) are summarized in Table \ref{power_coverage_comparison_gaussian}. RNDWSN and IRNP algorithms result in less than $1\%$ improvement in coverage compare to Limited-HTTL Algorithm in WSN1; however, their power consumption is about twice that of our proposed algorithm. Similar results for AC and IRNP algorithms in WSN2 show that about $1\%$ increase in the coverage obtained by Limited-HTTL Algorithm  leads to $42\%$ and $210\%$ increase in power consumption, respectively. Finally, our proposed algorithm outperforms DC and RNDWSN methods in terms of both coverage and power consumption in WSN2. Note that when communication range is limited, MER Algorithm usually yields poor performance since many APs fall outside the communication range of their corresponding FC, and they cannot transfer their collected data from sensors to fusion centers. Fig. \ref{Gaussian_WSN2_Covered_Area} shows the optimal node deployment and covered area for different algorithms in WSN2 with $\beta = 0.25$ and mixture of Gaussian sensor density function.

\begin{table*}[t]
  \centering
  \caption{{Coverage and power comparison for the Gaussian mixture sensor density function.}}
  \vspace{-3mm}
\begin{tabular}{cc|c|c|c|c|c|c}
 \toprule
          &              &{\bf MER}&{\bf AC} &{\bf DC} &{\bf RNDWSN}&{\bf IRNP}&{\bf Limited-HTTL}\\\hline\hline
{\bf WSN1}&{\bf Power}   &$1.6810$ &$2.3428$ &$1.5385$ &$4.9187$    &$4.4630$  &     $2.2659$     \\ 
          &{\bf Coverage}&$43.24\%$&$75.72\%$&$63.04\%$&$92.45\%$   &$92.12\%$ &     $91.68\%$    \\\hline
{\bf WSN2}&{\bf Power}   &$1.5285$ &$1.6436$ &$1.6676$ &$4.0627$    &$3.5923$  &     $1.1565$     \\ 
          &{\bf Coverage}&$46.43\%$&$98.64\%$&$97.13\%$&$95.34\%$   &$99.32\%$ &     $98.11\%$    \\
 \bottomrule
\end{tabular}
  \label{power_coverage_comparison_gaussian}
\end{table*}

To evaluate the performance of our method in real world applications, we conduct experiments on the daily weather data of the Colorado state, i.e. precipitation, relative humidity, temperature etc. Sensory data is obtained with the same rate from $286$ locations that form a $13\times 22$ grid across Colorado. 
We consider a heterogeneous WSN with $40$ APs and $8$ FCs. The power constraints and other parameter values are provided in Table \ref{SWAT_DATA} \cite{Heinzelman}.

\begin{table}[t]
\centering
  \caption{Simulation Parameters}
  \vspace{-3mm}
\begin{tabular}{ccccccccccc}
 \toprule
&&\multicolumn{2}{c}{{\bf Parameters} {\bf$\left(\textrm{pWatt/m}^2\right)$}}  &        &         &&  &\multicolumn{2}{c}{{\bf Power Constraints} {\bf $\left(\textrm{milliWatt}\right)$}} &\\\cline{1-6}\cline{8-11}
$\mathbf{a_{1:20}}$&$\mathbf{a_{21:40}}$&$\mathbf{b_{1:8,1:4}}$&$\mathbf{b_{1:8,5:8}}$& $\mathbf{b_{9:40,1:4}}$ &$\mathbf{b_{9:40,5:8}}$&&$\mathbf{\sigma^2}$&$\mathbf{\sigma^2_{1:8}}$&$\mathbf{\sigma^2_{9:20}}$&$\mathbf{\sigma^2_{21:40}}$\\
$\mathbf{1}$       &   $\mathbf{2}$     &   $\mathbf{1}$       &      $\mathbf{2}$    &    $\mathbf{2}$         &     $\mathbf{4}$     && $\mathbf{6.4}$   &     $\mathbf{19.6}$       &      $\mathbf{14.4}$           &      $\mathbf{10.0}$        \\ 
 \bottomrule
\end{tabular}
  \label{SWAT_DATA}
\end{table}

Table \ref{real_world_power_coverage_comparison_uniform} summarizes the two-tier power consumption and coverage of different methods. Our method outperforms AC, RNDWSN and IRNP algorithms in terms of both total coverage and power consumption. While providing lower power, MER Algorithm yields poor performance since many sensory locations fall outside the communication range of their nearby APs. Finally, DC Algorithm yields $3\%$ improvement in power consumption although it provides a  significantly lower coverage value compared with our algorithm. 

\begin{table*}[t]
  \centering
  \caption{Coverage and power (Watt) comparison for the climate data.}
  \vspace{-3mm}
\begin{tabular}{c|c|c|c|c|c|c}
 \toprule
                        &{\bf MER}&{\bf AC} &{\bf DC} &{\bf RNDWSN}&{\bf IRNP}&{\bf Limited-HTTL}\\\hline\hline
          {\bf Power}   &$0.7052$ &$1.0169$ &$0.8846$ &$1.1978$    &$1.3788$  &      $0.9151$    \\ 
          {\bf Coverage}&$60.14\%$&$82.17\%$&$78.32\%$&$76.57\%$   &$89.51\%$ &      $96.15\%$   \\
 \bottomrule
\end{tabular}
  \label{real_world_power_coverage_comparison_uniform}
\end{table*}

\section{Conclusion}\label{sec:conclusion}

A heterogeneous two-tier network which collects data from a large-scale wireless sensor to heterogeneous fusion centers through heterogeneous access points is discussed. We studied the minimum power that ensures reliable communication on such two-tier networks and modeled it as an optimization problem. Different from the homogeneous two-tier networks, a novel Voronoi Diagram is proposed to provide the best cell partition for the heterogeneous network. The necessary conditions of optimal node deployment imply that every access point should be placed between its connected fusion center and the geometric center of its cell partition. By defining an appropriate power consumption measure, we proposed a heterogeneous two-tier Lloyd Algorithm (HTTL) to minimize the power consumption. Simulation results show that HTTL Algorithm greatly saves the weighted power or energy in a heterogeneous two-tier network. When communication range is limited, our novel Limited-HTTL Algorithm ensures that all APs are active. Simulation results show that our algorithms provide superior results, in terms of both power consumption and network coverage.

\appendices

\section{}\label{Appendix_A}

Proof of Proposition 1: For $U=\left(S_1,S_2,...,S_N \right)$, the left-hand side of (\ref{eq7}) can be written as:
\begin{align}\label{appendixA_eq1}
& \overline{\mathcal{P}}(P,Q,U,T) = \sum_{n=1}^N\int_{S_n}(a_n\|p_n-w\|^2 \nonumber + \beta b_{n,T(n)}\|p_n - q_{T(n)}\|^2 )f(w)dw \nonumber  
           \\& \geq  \sum_{n=1}^N\int_{S_n}\min_j (a_j\|p_j-w\|^2 \nonumber + \beta b_{j,T(j)}\|p_j - q_{T(j)}\|^2 )f(w)dw \nonumber 
            = \int_{\Omega}\min_j (a_j\|p_j-w\|^2 \nonumber + \\& \beta b_{j,T(j)}\|p_j - q_{T(j)}\|^2 )f(w)dw \nonumber 
           = \sum_{n=1}^N\int_{V_n}\min_j (a_j\|p_j-w\|^2 \nonumber  + \beta b_{j,T(j)}\|p_j - q_{T(j)}\|^2 )f(w)dw \nonumber 
           \\& = \sum_{n=1}^N\int_{V_n}(a_n\|p_n-w\|^2 \nonumber + \beta b_{n,T(n)}\|p_n - q_{T(n)}\|^2 )f(w)dw \nonumber 
           = \overline{\mathcal{P}}(P,Q, \mathbf{V}, T).
\end{align}
Hence, the generalized Voronoi diagram is the optimal partition for any deployment $(P,Q,T)$.$\hfill\blacksquare$

\section{}\label{Appendix_B}

Proof of Lemma 1: Given $N$ APs and $M$ FCs $(M<N)$, first we demonstrate that there exists an optimal node deployment such as $\left(\widehat{P},\widehat{Q},\widehat{\mathbf{R}},\widehat{T} \right)$ in which each FC has at most one connected AP at the same location, i.e., for each $m\in \mathcal{I_B}$, the cardinality of the set $\{n|\widehat{T}(n)=m, \widehat{p}_n=\widehat{q}_m\}$ is less than or equal to 1. For this purpose, we consider an optimal node deployment $\left(P^*,Q^*,\mathbf{R}^*,T^* \right)$ and assume that there exist at least two distinct indices $n_1,n_2\in \mathcal{I_A}$ and an index $m\in \mathcal{I_B}$ such that $T^*(n_1)=T^*(n_2)=m$, and $p^*_{n_1}=p^*_{n_2}=q^*_m$. We have:
\begin{equation}
     \overline{\mathcal{P}}_{n_1} =\int_{R^*_{n_1}}(a_{n_1}\|p^*_{n_1}-w\|^2   + \beta b_{n_1,m}\|p^*_{n_1} - q^*_m\|^2  )f(w)dw  = \int_{R^*_{n_1}}a_{n_1}\|p^*_{n_1}-w\|^2 f(w)dw,
\end{equation}
\begin{equation}
    \overline{\mathcal{P}}_{n_2}=\int_{R^*_{n_2}}(a_{n_2}\|p^*_{n_2}-w\|^2   + \beta b_{n_2,m}\|p^*_{n_2} - q^*_m\|^2  )f(w)dw  = \int_{R^*_{n_2}}a_{n_2}\|p^*_{n_2}-w\|^2 f(w)dw.
\end{equation}
Without loss of generality, we can assume that $a_{n_1}\leq a_{n_2}$. Hence, we have:
\begin{multline}\label{eq15}
    \overline{\mathcal{P}}_{n_1} + \overline{\mathcal{P}}_{n_2} = \nonumber \int_{R^*_{n_1}}a_{n_1}\|p^*_{n_1}-w\|^2 f(w)dw  + \int_{R^*_{n_2}}a_{n_2}\|p^*_{n_2}-w\|^2 f(w)dw  \\
 \geq \int_{R^*_{n_1}}a_{n_1}\|p^*_{n_1}-w\|^2  f(w)dw  + \int_{R^*_{n_2}}a_{n_1}\|p^*_{n_1}-w\|^2f(w)dw \nonumber 
=\nonumber \int_{R^*_{n_1}\bigcup R^*_{n_2}}a_{n_1}\|p^*_{n_1}-w\|^2f(w)dw, 
\end{multline}
which implies that if we update the cell partition for AP $n_1$ to be $R^*_{n_1}\bigcup R^*_{n_2}$, and place the AP $n_2$ to an arbitrary location different from $q^*_m$ with a corresponding zero volume cell partition, the resulting power consumption will not increase, and the obtained node deployment is also optimal. Note that in this newly obtained optimal power consumption, AP $n_2$ is not in the same location as FC $m$ anymore. This procedure is continued until we reach an optimal deployment, denoted via $\left(\widehat{P}, \widehat{Q},\widehat{\mathbf{R}}, \widehat{T}  \right)$, in which each FC has at most one connected AP upon it.

Since $M<N$ and each FC has at most one AP upon it, there exists an index $k\in \mathcal{I_A}$ such that $\widehat{p}_k\neq  \widehat{q}_{\widehat{T}(k)}$. In order to show that the optimal two-tier power consumption with $N$ APs and $M+1$ FCs is less than that of $N$ APs and $M$ FCs, it is sufficient to construct a node deployment with $N$ APs and $M+1$ FCs such as $\left(P'',Q'',\mathbf{R}'',T'' \right)$ that achieves lower power consumption than $\overline{\mathcal{P}}\left(\widehat{P}, \widehat{Q},\widehat{\mathbf{R}}, \widehat{T}  \right)$. For each $n\in \mathcal{I_A}$, let $\widehat{v}_n=\int_{\widehat{R}_n}f(w)dw$ denote the volume of the region $\widehat{R}_n$. We consider two cases: (i) If $\widehat{v}_k > 0$, then we set $P''=\widehat{P}$, $Q''=\left(\widehat{q}_1,\widehat{q}_2,...,\widehat{q}_M , q''_{M+1}=\widehat{p}_k  \right)$, $\mathbf{R}''=\widehat{\mathbf{R}}$ and $T''(n)=\widehat{T}(n)$ for $n\neq k$ and $T''(k) = M+1$. Note that
\begin{align}
    &{ }\int_{\widehat{R}_k}\left(a_k\|\widehat{p}_k - w\|^2 + \beta b_{k,\widehat{T}(k)} \|\widehat{p}_k - \widehat{q}_{\widehat{T}(k)}\|^2   \right)f(w)dw \nonumber 
    \\&> \int_{\widehat{R}_k} \left(a_k\|\widehat{p}_k - w\|^2 \right)f(w)dw = \int_{\widehat{R}_k} \left(a_k\|\widehat{p}_k - w\|^2 + \beta b_{k,M+1}\|\widehat{p}_k - q''_{M+1}\|^2    \right)f(w)dw
\end{align}
implies that in the deployment $\left(P'',Q'',\mathbf{R}'',T'' \right)$, the contribution of the AP $k$ to the total power consumption has decreased. Since the contribution of other APs to the power consumption has not changed, we have $\overline{\mathcal{P}}\left(P'', Q'', \mathbf{R}'', T''  \right) < \overline{\mathcal{P}}\left(\widehat{P}, \widehat{Q},\widehat{\mathbf{R}}, \widehat{T}  \right)$ and the proof is complete. (ii) If $\widehat{v}_k = 0$, then AP $k$ does not contribute to the optimal power consumption $\overline{\mathcal{P}}\left(\widehat{P},\widehat{Q},\widehat{\mathbf{R}},\widehat{T}  \right)$, and it can be placed anywhere within the target region $\Omega$. Since the set $\left\{\widehat{p}_1,...,\widehat{p}_N,\widehat{q}_1,...,\widehat{q}_M  \right\}$ has zero measure, there exists a point $x\in \Omega$ and a threshold $\delta \in \mathbb{R}^+$ such that $\mathcal{B}\left(x,\delta \right)=\left\{w\in \Omega | \|x-w\|\leq \delta \right\}$ does not include any point from the set $\left\{\widehat{p}_1,...,\widehat{p}_N,\widehat{q}_1,...,\widehat{q}_M  \right\}$. Since $f(.)$ is positive, continuous and differentiable over $\Omega$, for each $0 < \epsilon < \delta$ the region $\mathcal{B}(x,\epsilon)=\left \{w\in \Omega \big| \|w-x\|\leq \epsilon   \right\}$ has positive volume, i.e., $\int_{\mathcal{B}(x,\epsilon)}f(w)dw > 0$. Given $0< \epsilon < \delta$, assume that:
\begin{equation}\label{eq18.5}
    \mathcal{B}(x,\epsilon)\subset \widehat{R}_n,
\end{equation}
for some $n\in \mathcal{I_A}$; therefore, the contribution of the region $\mathcal{B}(x,\epsilon)$ to the total power consumption $\overline{\mathcal{P}}\left(\widehat{P},\widehat{Q},\widehat{\mathbf{R}},\widehat{T} \right)$ is equal to:
\begin{equation}\label{eq19}
    \int_{\mathcal{B}(x,\epsilon)}\left(a_n\|\widehat{p}_n - w\|^2 + \beta b_{n,\widehat{T}(n)}\|\widehat{p}_n - \widehat{q}_{\widehat{T}(n)}\|^2        \right)f(w)dw.
\end{equation}
As $\epsilon \longrightarrow 0$, (\ref{eq19}) can be approximated as:
\begin{equation}\label{eq20}
  \Delta_n \times   \int_{\mathcal{B}(x,\epsilon)}f(w)dw,
\end{equation}
where $\Delta_n = \left(a_n\|\widehat{p}_n - x\|^2 + \beta b_{n,\widehat{T}(n)}\|\widehat{p}_n - \widehat{q}_{\widehat{T}(n)}\|^2        \right)$. If we set $p''_k=q''_{M+1}=x$ and $R''_k = \mathcal{B}(x,\epsilon)$ and $T''(k)=M+1$, then the contribution of the region $\mathcal{B}(x,\epsilon)$ to the total power consumption $\overline{\mathcal{P}}\left(P'',Q'',\mathbf{R}'',T'' \right)$ is equal to:
\begin{equation}\label{eq21}
\int_{\mathcal{B}(x,\epsilon)}\left(a_k\|p''_k-w\|^2 + \beta b_{k,M+1}\|p''_k - q''_{M+1}\|^2 \right)f(w)dw  = a_k \int_{\mathcal{B}(x,\epsilon)}\left(\|x - w\|^2 \right)f(w)dw.
\end{equation}
The below equation for the ratio of power consumption in (\ref{eq20}) and (\ref{eq21})
\begin{equation}\label{eq23}
    \lim_{\epsilon \longrightarrow 0}\frac{a_k \int_{\mathcal{B}(x,\epsilon)}\left(\|x - w\|^2 \right)f(w)dw}{\Delta_n \times   \int_{\mathcal{B}(x,\epsilon)}f(w)dw} = 0
\end{equation}
implies that there exists an $\epsilon^*\in (0,\delta)$ such that the contribution of the region $\mathcal{B}(x,\epsilon^*)$ to the total power in $\overline{\mathcal{P}}\left( P'', Q'', \mathbf{R}'', T'' \right)$ will be less than that of $\overline{\mathcal{P}}\left(\widehat{P}, \widehat{Q}, \widehat{\mathbf{R}}, \widehat{T}\right)$. Hence, we set $P''=\left(p''_1,p''_2,...,p''_N \right)$ where $p''_i = \widehat{p}_i$ for $i\neq k$, and $p''_k = x$. Also, we set $Q''=\left(\widehat{q}_1,\widehat{q}_2,...,\widehat{q}_M,q''_{M+1} = x   \right)$. The partitioning $\mathbf{R}''=\left(R''_1,...,R''_N   \right)$ is defined as $R''_i=\widehat{R}_i$ for $i\neq k$ and $i\neq n$, $R''_k = \mathcal{B}(x,\epsilon^*)$ and $R''_n = \widehat{R}_n - \mathcal{B}(x,\epsilon^*)$. Finally, we set $T''(i) = \widehat{T}(i)$ for $i\neq k$ and $T''(k) = M+1$. As mentioned earlier, the two-tier power consumption $\overline{\mathcal{P}}\left(P'',Q'',\mathbf{R}'',T''   \right)$ is less than $\overline{\mathcal{P}}\left (\widehat{P}, \widehat{Q}, \widehat{\mathbf{R}}, \widehat{T}    \right)$. Note that if the region $\mathcal{B}(x,\epsilon)$ is a subset of more than one region, (\ref{eq18.5}) to (\ref{eq20}) and (\ref{eq23}) can be modified accordingly and a similar argument shows that the resulting power consumption will be improved in the new deployment, and the proof is complete.$\hfill\blacksquare$

\section{}\label{Appendix_C}

Proof of Corollary 1: Assume that there exists an index $m\in \mathcal{I_B}$ in the optimal node deployment $\left(P^*, Q^*, \mathbf{R}^*, T^* \right)$ such that $\bigcup_{n:T^*(n)=m}R^*_n$ has zero volume. Consider the node deployment $\left(P',Q',\mathbf{R}',T' \right)$ where $P'=P^*$, $Q' =  \left(q^*_1,...,q^*_{m-1},q^*_{m+1},...,q^*_M\right)$, $\mathbf{R}' = \mathbf{R}^*$ and $T'(i)=T^*(i)$ for indices $i\in \mathcal{I_A}$ such that $T^*(i)\neq m$. Note that for indices $i\in \mathcal{I_A}$ such that $T^*(i)=m$, we can define $T'(i)$ arbitrarily because the corresponding regions $R'_i$ have zero volume. Since $\bigcup_{n:T^*(n)=m}R^*_n$ has zero volume, we have $    \overline{\mathcal{P}}\left (  P',Q',\mathbf{R}',T'  \right) = \overline{\mathcal{P}}\left (  P^*,Q^*,\mathbf{R}^*,T^*  \right)$ which is in contradiction with Lemma 1 since the optimal node deployment $\left(P^*, Q^*, \mathbf{R}^*, T^* \right)$ for $N$ APs and $M$ FCs has not improved the node deployment $\left(P',Q',\mathbf{R}',T' \right)$ for $N$ APs and $M-1$ FCs in terms of power consumption.$\hfill\blacksquare$

\section{}\label{Appendix_D}
Proof of Proposition 2: First, we study the shape of the Voronoi regions in (\ref{eq5}). Let $\mathcal{B}(c,r)=\{\omega|\|\omega-c\|\leq r\}$ be a disk centered at $c$ with radius $r$ in two-dimensional space. In particular, $\mathcal{B}(c,r)=\emptyset$ when $r\leq 0$.
Let $\mathcal{HS}(A, B)=\left\{\omega|A\omega+B\leq0\right\}$ be a half space, where $A\in\mathbb{R}^2$ is a vector and $B\in\mathbb{R}$ is a constant. 
For $i,j \in \mathcal{I_A}$, we define 
\begin{equation}
    V_{ij}(P,Q,T)\triangleq\{\omega|a_i\|p_i-w\|^2 + \beta b_{i, T(i)}\|p_i-q_{T(i)}\|^2 \leq  a_j\|p_j-w\|^2 + \beta b_{j, T(j)}\|p_j-q_{T(j)}\|^2 \}
\label{Vij}
\end{equation}
to be the pairwise Voronoi region of AP $i$ where only AP $i$ and $j$ are considered. Then, AP $i$'s Voronoi region can be represented as $V_i(P, Q) = \left[\bigcap_{j\neq i}V_{ij}(P,Q)\right]\bigcap\Omega$. By expanding (\ref{Vij}) and straightforward algebraic calculations, the pairwise Voronoi region $V_{ij}$ is derived as:
\begin{align}
    V_{ij}=\Omega\cap\begin{cases}
         \mathcal{HS}\left(A_{ij}, B_{ij}\right)& \quad ,  a_i=a_j  \\
        \mathcal{B}\left(c_{ij}, r_{ij}\right) &\quad , a_i>a_j , L_{ij}\geq 0\\
        \varnothing & \quad ,  a_i>a_j , L_{ij}< 0\\
        \mathcal{B}^c\left(c_{ij}, r_{ij}\right) &\quad , a_i<a_j , L_{ij} \geq 0 \\
        \mathbb{R}^2 & \quad , a_i<a_j , L_{ij}< 0
        \end{cases},
\end{align}
where $A_{ij} = a_jp_j-a_ip_i$, $B_{ij}=\frac{\left(a_i\|p_i\|^2-a_j\|p_j\|^2+\beta b_{i, T(i)}\|p_i-q_{T(i)}\|^2-\beta b_{j, T(j)}\|p_j-q_{T(j)}\|^2\right)}{2}$, $c_{ij}=\frac{a_ip_i-a_jp_j}{a_i-a_j}$, $L_{ij} = \frac{a_ia_j\|p_i-p_j\|^2}{\left(a_i-a_j\right)^2} -\beta \times \frac{ b_{i, T(i)}\|p_i\!-\!q_{T(i)}\|^2- b_{j, T(j)}\|p_j\!-\!q_{T(j)}\|^2}{(a_i-a_j)}$, $r_{ij} = \sqrt{\max\left(L_{ij}, 0\right)}$, and $\mathcal{B}^c(c_{ij},r_{ij})$ is the complementary of  $\mathcal{B}(c_{ij},r_{ij})$. Note that for two distinct indices such as $i,j\in \mathcal{I_A}$, if $a_i>a_j$ and $L_{ij}<0$, then two regions $\Omega\cap\mathcal{B}(c_{ij},r_{ij})$ and $\varnothing$ differ only in the point $c_{ij}$. Similarly, for $a_i<a_j$ and $L_{ij}<0$, two regions $\Omega\cap\mathcal{B}^c(c_{ij},r_{ij})$ and $\Omega$ differ only in the point $c_{ij}$. If we define:
\begin{equation}\label{voronoi_v_n}
    \overline{V}_k = \left [ \bigcap_{i:a_k>a_i} \mathcal{B}(c_{ki},r_{ki})  \right] \bigcap \left[\bigcap_{i:a_k=a_i} \mathcal{HS}(A_{ki},B_{ki}) \right] \bigcap \left[ \bigcap_{i:a_k<a_i} \mathcal{B}^c(c_{ki},r_{ki})  \right] \bigcap \Omega,
\end{equation}
then two regions $\overline{V}_k$ and $V_k$ differ only in finite number of points. As a result, integrals over both $\overline{V}_k$ and $V_k$ have the same value since the density function $f$ is continuous and differentiable, and removing finite number of points from the integral region does not change the integral value. Note that if $V_k$ is empty, the Proposition 1 in \cite{JG2} holds since the integral over an empty region is zero. If $V_k$ is not empty, the same arguments as in Appendix A of \cite{JG2} can be replicated since $\overline{V}_k$ in (\ref{voronoi_v_n}) is similar to (31) in \cite{JG2}.

Using parallel axis theorem \cite{parallel_reference}, the two-tier power consumption can be written as:
\begin{align}\label{distortion_parallel_axis_theorem}
    \overline{\mathcal{P}}\left(P,Q,\mathbf{V},T \right) &= \sum_{n=1}^{N}\int_{V_n}\bigg(a_n\|p_n \! - \! w\|^2  + \beta b_{n,T(n)}\|p_n \! - \! q_{T(n)}\|^2  \bigg)f(w)dw  \\
    &= \sum_{n=1}^{N}\bigg( \int_{V_n} \! a_n\|c_n \! - \! w\|^2f(w)dw + a_n\|p_n \! -\! c_n\|^2v_n +\beta b_{n,T(n)}\|p_n \! - \! q_{T(n)}\|^2v_n \bigg).\nonumber
\end{align}
Using Proposition 1 in \cite{JG2}, since the optimal deployment $\left(P^*,Q^*  \right)$ satisfies zero gradient, we take the partial derivatives of (\ref{distortion_parallel_axis_theorem}) as follows:
\begin{align}\label{partial_derivative}
    \frac{\partial \overline{\mathcal{P}}}{\partial p^*_n} & = 2\left[ a_n(p^*_n - c^*_n) + \beta b_{n,T^*(n)}(p^*_n - q^*_{T^*(n)}) \right]v^*_n  =0, \nonumber \\
    \frac{\partial \overline{\mathcal{P}}}{\partial q^*_m}& = 2\sum_{n:T^*(n)=m} \beta b_{n,m}(q^*_m - p^*_n)v^*_n = 0.
\end{align}
By solving (\ref{partial_derivative}), we have the following necessary conditions:
\begin{equation}\label{necesasry_condition}
    p_n^* = \frac{a_nc^*_n + \beta b_{n,T^*(n)}q^*_{T^*(n)}}{a_n + \beta b_{n,T^*(n)}} \qquad , \qquad
    q^*_m = \frac{\sum_{n:T^*(n)=m}b_{n,m}p^*_nv^*_n}{\sum_{n:T^*(n)=m}b_{n,m}v^*_n},
\end{equation}
and the proof is complete.$\hfill\blacksquare$

\section{}\label{Appendix_F}

Proof of Lemma 2: Using Lemma 3 in \cite{JPH}, it can be easily shown that the optimal quantization regions are two closed intervals. Without loss of generality, let $\mathbf{R} = \{R_1,R_2\}$, where $R_1 = \left[0,r \right]$ and $R_2 = \left[r, 1 \right]$ be the optimal partitioning. Thus, we have $c_1 = \frac{r}{2}$ and $c_2 = \frac{1+r}{2}$. Using (\ref{eq9}), we have:
\begin{equation}\label{appendix_I_eq_1}
        p_1 =\frac{a_1c_1 + \beta b_{1,1}q}{a_1+\beta b_{1,1}} =\frac{r+2\beta'q}{2\left(1+\beta'\right)} \qquad , \qquad
        p_2 =\frac{a_2c_2 + \beta b_{2,1}q}{a_2+\beta b_{2,1}}=\frac{1+r+2\beta'q}{2\left(1+\beta' \right)},
\end{equation}
where $\beta'=\beta \times \kappa$. Therefore, the two-tier power in the regions $R_1$ and $R_2$ are given by:
\begin{align}\label{appendix_I_eq_2}
    \overline{\mathcal{P}}_1 &= a_1\int_{0}^{r}\left[ \left(\frac{r+2\beta'q}{2\left(1+\beta'\right)}-w\right)^2 +\beta' \frac{\left(r-2q\right)^2 }{4\left(1+\beta' \right)^2} \right] dw, \\
    \overline{\mathcal{P}}_2 &= a_2\int_{r}^{1}\left[ \left(\frac{1+r+2\beta'q}{2\left(1+\beta'\right)}-w\right)^2 + \beta' \frac{\left(1+r-2q\right)^2}{4\left(1+\beta' \right)^2} \right] dw, \nonumber
\end{align}
and $\overline{\mathcal{P}}(r,q)=\overline{\mathcal{P}}_1+\overline{\mathcal{P}}_2$ is the total two-tier power consumption. Simplifying (\ref{appendix_I_eq_2}) yields:
\begin{align}\label{appendix_I_eq_3}
    \overline{\mathcal{P}}_1 &= \frac{a_1 r}{4\left(1+\beta' \right)^2}\times \Bigg (\beta' \left(r-2q \right)^2 +\frac{1}{3}\times \bigg [\left(r+2\beta' q\right)^2 \nonumber \\ & + \left(r+2\beta' q \right)\left(2\beta' (q -  r) -r \right) + \left(2\beta' (q - r) -r \right)^2 \bigg ] \Bigg ), \nonumber \\
    \overline{\mathcal{P}}_2&= \frac{a_2(1-r)}{4\left(1+\beta' \right)^2}\times \Bigg (\beta'\left(1+r-2q \right)^2 + \frac{1}{3}\times \bigg[\left((1-r)+2\beta' (q-r) \right)^2 \nonumber \\ 
    &+ \left((1-r)+2\beta' (q-r) \right)\left((r-1) + 2\beta' (q-1) \right) + \left((r-1) + 2\beta' (q-1) \right)^2 \bigg ]   \Bigg ).
\end{align}
Since both $\overline{\mathcal{P}}_1(r,q)$ and $\overline{\mathcal{P}}_2(r,q)$ are continuous and differentiable functions of $r$ and $q$, the minimum occurs either at zero gradients, given by:
\begin{equation}\label{appendix_I_eq_4}
    \frac{\partial \overline{\mathcal{P}}}{\partial q} = 0 \qquad , \qquad \frac{\partial \overline{\mathcal{P}}}{\partial r} = 0.
\end{equation}
or at the boundaries, i.e., $q,r\in\{0,1\}$. First, we focus on the zero gradient equations. Simplifying (\ref{appendix_I_eq_4}) yields the following:
\begin{equation}\label{appendix_I_eq_5}
    q = \frac{a_1 r^2 + a_2 (1-r^2)}{2\left(a_1 r + a_2(1-r) \right)},
\end{equation}
\begin{equation}\label{appendix_I_eq_6}
     3\left(4\beta' + 1 \right)\left(a_1 - a_2 \right)r^2 + 12\beta' \left(a_1-a_2\right)q^2  
    -24\beta' \left(a_1-a_2 \right)qr + 3a_2(2r-1) =0.
\end{equation}
If $a_1=a_2$, then the unique solution to (\ref{appendix_I_eq_5}) and (\ref{appendix_I_eq_6}) is $q=r=\frac{1}{2}$; otherwise, by substituting (\ref{appendix_I_eq_5}) in (\ref{appendix_I_eq_6}) we have the following fourth order polynomial equation:
\begin{multline}\label{appendix_I_eq_7}
    \left(a_1 \! - \! a_2 \right)^3 \left(\beta' \! + \! 1\right)r^4 + 4a_2\left(a_1 \! -\! a_2 \right)^2\left(\beta' \! +\! 1 \right)r^3 
    + \big[\left(4\beta' \! +\! 5\right)a_2^2\left(a_1 \! - \! a_2\right) - \left(2\beta' \! + \! 1 \right)a_2 \times \\ \left(a_1 \! - \! a_2\right)^2    \big]r^2  + 2a_2^2\left[a_2 - \left(2\beta' \! + \! 1\right)\left(a_1 \! - \! a_2\right) \right]r 
    + a_2^2 \left[\beta' \left(a_1 \! - \! a_2 \right) -a_2  \right] =0.
\end{multline}
Solving (\ref{appendix_I_eq_7}) and substituting the roots into (\ref{appendix_I_eq_5}) gives the following pairs of solutions to (\ref{appendix_I_eq_4}):
\begin{align}\label{appendix_I_eq_8}
    r_1 = \frac{1}{1+\sqrt{\frac{a_1}{a_2}}} \quad &, \quad q_1 = \frac{1}{1+ \sqrt{\frac{a_1}{a_2}}}, \nonumber \\
    r_2 = \frac{1}{1-\sqrt{\frac{a_1}{a_2}}} \quad &, \quad q_2 = \frac{1}{1-\sqrt{\frac{a_1}{a_2}}}, \nonumber \\
    r_3 = \frac{1-\sqrt{\frac{\beta'}{\beta' +1}}\sqrt{\frac{a_1}{a_2}}}{1-\frac{a_1}{a_2}} \textrm{ } &,\textrm{ } q_3 = \frac{1 - \left(\frac{\sqrt{\frac{\beta'}{\beta' +1}} + \sqrt{\frac{\beta' + 1}{\beta'}} }{2}  \right)\sqrt{\frac{a_1}{a_2}}}{1-\frac{a_1}{a_2}}, \nonumber \\
        r_4 = \frac{1+\sqrt{\frac{\beta'}{\beta' +1}}\sqrt{\frac{a_1}{a_2}}}{1-\frac{a_1}{a_2}} \textrm{ } &,\textrm{ } q_4 = \frac{1 + \left(\frac{\sqrt{\frac{\beta'}{\beta' +1}} + \sqrt{\frac{\beta' + 1}{\beta'}} }{2}  \right)\sqrt{\frac{a_1}{a_2}}}{1-\frac{a_1}{a_2}},
\end{align}
which in turn, leads to the four possible power consumption values $\overline{\mathcal{P}}(r_i, q_i)$ for $i\in \{1,2,3,4\}$. By comparing all four feasible powers, it can be shown via straightforward algebraic calculations that $\overline{\mathcal{P}}(r_1,q_1)$ is always the minimum among the four candidate solutions. Therefore, the optimal FC location and partitioning are given by $q_1$ and  $\mathbf{R}=\{ R_1 = [0,r_1], R_2=[r_1,1]\}$, respectively. Using (\ref{appendix_I_eq_1}), the optimal AP locations can be calculated accordingly. Now, we consider the boundary case of $q,r\in \{0,1\}$. Note that $r\in \{0,1\}$ means that one of the regions is empty, i.e., the whole target region $\Omega=[0,1]$ sends its data to the stronger AP. As a result, we can achieve the optimal power consumption of $\frac{\min\left(a_1,a_2\right)}{12}$ by placing the stronger AP and the FC at the centroid of $\Omega$. The weaker AP will be used only if:
\begin{equation}\label{appendix_I_eq_9}
        \overline{\mathcal{P}}(r_1,q_1) < \frac{a_1}{12} \quad , \quad
        \overline{\mathcal{P}}(r_1,q_1) < \frac{a_2}{12}.
\end{equation}
Solving (\ref{appendix_I_eq_9}) yields the necessary and sufficient condition given in (\ref{necesssary_sufficient_condition_for_usefulness}). Therefore, if the condition in (\ref{necesssary_sufficient_condition_for_usefulness}) holds, both APs are useful and the optimal power consumption is given by $\overline{\mathcal{P}}(r_1,q_1)$ as it is given in (\ref{optimal_two_tier_distortion_useful_APs}); otherwise, using only the stronger AP yields a lower power consumption value given in (\ref{optimal_two_tier_distortion_useless_APs}) and the proof is complete.$\hfill\blacksquare$

\section{}\label{Appendix_E}

Proof of Proposition 3: In what follows, we demonstrate that none of the four steps in the HTTL algorithm will increase the two-tier power consumption. Given $P$, $Q$ and $\mathbf{R}$, updating the index map $T$ according to (\ref{eq8}) minimizes the total power consumption, i.e., the two-tier power consumption will not increase by the first step. Moreover, given $P$, $Q$ and $T$, Proposition \ref{allactive} indicates that updating $\mathbf{R}$ according to (\ref{eq5}) and (\ref{eq6}) provides the best partitioning; thus, the second step of the HTTL algorithm will not increase the power consumption either. 
We need the following equality, which can be derived from simple algebra, to continue the proof. 
\begin{equation}\label{appendix_F_1_eq}
\sum_{n:T(n)=m} b_{n,m}v_n\|p_n - q_m\|^2 =  \sum_{n:T(n)=m}b_{n,m}v_n(\|p_n - q'_m\|^2  + \|q_m - q'_m\|^2 ),
\end{equation}
where $q'_m=\frac{\sum_{n:T(n)=m}b_{n,m}p_nv_n}{\sum_{n:T(n)=m}b_{n,m}v_n}$. Now, the contribution of FC $m$ to the total power consumption can then be rewritten as:
\begin{align}\label{appendix_F_2}
    &\sum_{n:T(n)=m}\int_{R_n}\left(a_n\|p_n-w\|^2 + \beta b_{n,m}\|p_n-q_m\|^2\right)f(w)dw =
    \sum_{n:T(n)=m}\int_{R_n}a_n\|p_n-w\|^2f(w)dw \nonumber \\&+ \beta \left(\sum_{n:T(n)=m}b_{n,m}v_n  \right)\|q_m - q'_m\|^2 +
    \beta \left(\sum_{n:T(n)=m}b_{n,m}v_n\|p_n - q'_m\|^2 \right).
\end{align}
Now, given $P$, $\mathbf{R}$ and $T$, the first and third terms in the right hand side of (\ref{appendix_F_2}) are constant and moving $q_m$ toward $q'_m$ will not increase the power consumption in (\ref{appendix_F_2}). Therefore, the third step of the HTTL algorithm will not increase the total two-tier power consumption as well. We use the following equality to simplify the calculation:
\begin{equation}\label{appendix_F_3_eq}
    a_n\|p_n  -  w\|^2 \! + \! \beta b_{n,m}\|p_n - q_m\|^2 \! = \! (a_n \! +\! \beta b_{n,m})\! \left| \! \left|p_n \! - \! \frac{(a_nw \! + \! \beta b_{n,m}q_m)}{a_n \! + \! \beta b_{n,m}}\right| \! \right|^2 \! + \frac{\beta a_n b_{n,m}}{a_n \!\! + \!\! \beta b_{n,m}}\|w -  q_m\|^2.
\end{equation}
Using (\ref{appendix_F_3_eq}), for each index $n\in \mathcal{I_A}$ and the corresponding index $m=T(n)$, we can rewrite the contribution of AP $n$ to the total power consumption as:
\begin{align}\label{appendix_F_4}
    & \int_{R_n}(a_n\|p_n \! - \! w\|^2 \! + \! \beta b_{n,m}\|p_n \! - \! q_m\|^2  )f(w)dw  
    \stackrel{(a)}{=} \int_{R_n}\bigg[(a_n \! + \! \beta b_{n,m})\bigg| \! \bigg|p_n \! - \! \nonumber  \frac{(a_nw \! + \! \beta b_{n,m}q_m)}{a_n \! + \! \beta b_{n,m}}\bigg| \! \bigg|^2 \\& \! + \! \frac{\beta a_n b_{n,m}}{a_n \! + \! \beta b_{n,m}} \! \times \! \|w \! - \! q_m\|^2 \bigg]f(w)dw \stackrel{(b)}{=}
    \int_{R_n}\bigg[ \frac{a_n^2}{a_n \! + \! \beta b_{n,m}} \! \times \! \nonumber  \bigg| \! \bigg| \frac{( a_n \! + \! \beta b_{n,m} )p_n \! - \! \beta b_{n,m}q_m}{a_n} \! - \! w \bigg| \! \bigg|^2 \\& \! + \! \frac{\beta a_nb_{n,m}}{a_n \! + \! \beta b_{n,m}}\|w \! - \! q_m\|^2 \bigg]f(w)dw   \stackrel{(c)}{=}
     \int_{R_n}\bigg[ \frac{a_n^2}{a_n \! + \! \beta b_{n,m}} \! \times \! \nonumber  \bigg(\bigg| \! \bigg| \frac{( a_n \! + \! \beta b_{n,m} )p_n \! - \! \beta b_{n,m}q_m}{a_n} \! - \! c_n \bigg| \! \bigg|^2 \\& \! + \! \|c_n \! - \! w\|^2\bigg) \! + \!  \frac{\beta a_nb_{n,m}}{a_n \! + \! \beta b_{n,m}}\|w \! - \! q_m\|^2 \bigg]f(w)dw \nonumber \stackrel{(d)}{=}
      \int_{R_n}\bigg[\frac{a_n^2}{a_n \! + \! \beta b_{n,m}}\|c_n \! - \! w\|^2 \! + \! (a_n \! + \! \beta b_{n,m})\! \times \! \\& \bigg| \! \bigg|p_n \! - \! \frac{a_nc_n \! + \! \beta b_{n,m}q_m}{a_n \! + \! \beta b_{n,m}}   \bigg| \! \bigg|^2 \nonumber \! + \! \frac{\beta a_nb_{n,m}}{a_n \! + \! \beta b_{n,m}}\|w \! - \! q_m\|^2 \bigg] f(w)dw  \stackrel{(e)}{=} 
      \frac{a_n^2}{a_n \! + \! \beta b_{n,m}}\int_{R_n}\|c_n \! - \! w\|^2f(w)dw \nonumber \\& \! + \! (a_n \! + \! \beta b_{n,m})\|p_n \! - \! p'_n   \|^2 v_n \! + \! \frac{\beta a_nb_{n,m}}{a_n \! + \! \beta b_{n,m}}\int_{R_n}\|w\! - \! q_m\|^2 f(w)dw,
\end{align}
where $p'_n = \frac{a_nc_n + \beta b_{n,m}q_m}{a_n + \beta b_{n,m}}$. Note that Equality $(a)$ in (\ref{appendix_F_4}) comes from (\ref{appendix_F_3_eq}), and Equality $(c)$ follows from the parallel axis theorem. Now, given $Q$, $\mathbf{R}$ and $T$, the first and third terms in the right hand side of Equality $(e)$ in (\ref{appendix_F_4}) are constants and moving $p_n$ toward $p'_n$ will not increase the second term in (\ref{appendix_F_4}). Hence, the fourth step of the HTTL algorithm will not increase the total power consumption either. So, the HTTL algorithm generates a sequence of positive non-increasing power consumption values and thus, it converges. Note that if power consumption remains the same after an iteration of the algorithm, it means that none of the four steps has decreased the power consumption and the algorithm has already reached an optimal deployment.$\hfill\blacksquare$

\section{}\label{Appendix_H}

Proof of Lemma 3: Note that $\overline{\mathcal{P}}^{\mathcal{S}}(P,\mathbf{R})$ defined in (\ref{eq2}) is the distortion of a one-tier quantizer with parameters $a_1,\ldots, a_N$, node positioning $P=\left(p_1,\ldots, p_N \right)$ and partitioning $\mathbf{R}=\left(R_1,\ldots, R_N \right)$; thus, the minimum value that $\overline{\mathcal{P}}^{\mathcal{S}}(P,\mathbf{R})$ can achieve is $D_N$ given in (\ref{D_N}), i.e., $\overline{\mathcal{P}}^{\mathcal{S}}\in \left[D_N, +\infty \right)$ which is the domain of the function $A(s)$. 

Let $\mathcal{F}(s)$ be the set of all feasible solutions for the power pair $\left(s,A(s)\right)$. We can rewrite (\ref{AP_Sensor_power_function}) as:
\begin{equation}\label{AP_Sensor_power_function_feasible_solutions}
    A(s) = \inf_{\left(P,Q,\mathbf{R},T  \right)\in \mathcal{F}(s)}\overline{\mathcal{P}}^{\mathcal{A}}\left(P,Q,\mathbf{R},T \right).
\end{equation}
It is self-evident that for two values of $s_1$ and $s_2$ such that $D_N\leq s_1 < s_2$, we have $\mathcal{F}(s_1) \subseteq \mathcal{F}(s_2)$, which implies that $A\left(s_1\right) \geq A\left(s_2 \right)$, i.e., $A(s)$ is a non-increasing function.

Without loss of generality, we assume that $a_1\leq a_2\leq \ldots \leq a_N$. If $s\in \left[D_M, +\infty \right)$, then $A(s)=0$ since if $X^*=\left(x_1^*,\ldots, x_M^* \right)$ and $\mathbf{R}^* = \left(R_1^*, \ldots, R_M^* \right)$ is the optimal deployment that achieves $D_M$ in (\ref{D_N}), then the deployment $\left(P,Q,\mathbf{R},T \right)$ where $P=\left(x_1^*, \ldots, x_M^*, x_1^*,x_1^*,\ldots, x_1^* \right)$, $Q = \left(x_1^*, \ldots, x_M^*  \right)$, $\mathbf{R}=\left(R_1^*,..., R_M^*, \varnothing, \varnothing , \ldots, \varnothing  \right)$ and $T^*(i)=i$ for each $i \in \mathcal{I_B}$ and $T^*(i) = 1$ for each $i\in \mathcal{I_A} - \mathcal{I_B}$ is a feasible solution for which $\overline{\mathcal{P}}^{\mathcal{S}}(P,\mathbf{R}) = D_M \leq s$ and $A(s)=0$. If $s\in \left[D_N, D_M \right)$, then the inequality $\overline{\mathcal{P}}^{\mathcal{S}}(P,\mathbf{R}) \leq s$ implies that $\overline{\mathcal{P}}^{\mathcal{S}}(P,\mathbf{R}) < D_M$, i.e., optimal APs should have at least $M+1$ different positions; therefore, the optimal AP power cannot be zero and the proof is complete. $\hfill\blacksquare$

\section{}\label{Appendix_I}

Proof of Lemma 4: Note that the pair $\left(s,\mathbf{R} \right)$ belongs to the domain of $A\left(s,\mathbf{R} \right)$ if and only if there exists a node positioning $P$ such that $\overline{\mathcal{P}}^{\mathcal{S}}(P,\mathbf{R}) \leq s$. Since we have $\overline{\mathcal{P}}^{\mathcal{S}}(P,\mathbf{R}) \geq \mathcal{H}(\mathbf{R})$ for any fixed partitioning $\mathbf{R}$, the domain of the function $A\left(s,\mathbf{R}\right)$ is $\left\{(s,\mathbf{R}) \big | s\geq \mathcal{H}(\mathbf{R})  \right\}$.

First, we show that $\mathcal{J}(\mathbf{R})$ is the minimum value of the quantity $\sum_{n=1}^{N}\int_{R_n}a_n\|x-w \|^2f(w)dw$ for a fixed $\mathbf{R}$. Using parallel axis theorem, we have:
\begin{equation}\label{all_APs_in_a_location_distortion}
    \sum_{n=1}^{N}\int_{R_n}a_n\|x-w \|^2f(w)dw = \sum_{n=1}^{N}a_n\|x - c_n \|^2v_n  + \sum_{n=1}^{N}\int_{R_n}a_n\|c_n - w\|^2f(w)dw,
\end{equation}
where $c_n$ is the centroid of the region $R_n$. Taking the derivative of (\ref{all_APs_in_a_location_distortion}) yields:
\begin{equation}\label{optimal_q^*}
\frac{\partial}{\partial x}\sum_{n=1}^{N}\int_{R_n}a_n\|x-w \|^2f(w)dw = \sum_{n=1}^{N}2a_n(x-c_n)v_n = 0,
\end{equation}
i.e., $x^* = \frac{\sum_{n=1}^{N}a_n v_n c_n}{\sum_{n=1}^{N}a_n v_n}$ where $v_n$ is the volume of $R_n$. Substituting $x^*$ into (\ref{all_APs_in_a_location_distortion}) yields:
\begin{equation}
    \mathcal{J}(\mathbf{R}) = \min_{x}\sum_{n=1}^{N}\int_{R_n}a_n\|x-w \|^2f(w)dw.
\end{equation}
If $s\in \left[\mathcal{J}(\mathbf{R}), +\infty \right)$ then $A(s,\mathbf{R}) = 0$ because for the deployment $P=\left(p_1,\ldots,p_N \right) = \left(x^*,\ldots,x^*   \right)$ and $Q=(q)=(x^*)$, we have $\overline{\mathcal{P}}^{\mathcal{S}}(P,\mathbf{R}) = \mathcal{J}(\mathbf{R}) \leq s$ and $\overline{\mathcal{P}}^{\mathcal{A}}\left(P,Q,\mathbf{R},T \right)=0$. Now, we determine the value of $A(s,\mathbf{R})$ for $s\in \left[\mathcal{H}(\mathbf{R}), \mathcal{J}(\mathbf{R}) \right)$. We have:
\begin{align}\label{rewrite_AP_power_function}
     \overline{\mathcal{P}}^{\mathcal{A}}\! \left(P,Q,\mathbf{R},T \right) &= \sum_{n=1}^{N}\int_{R_n} \! b_{n,1}\|p_n - q \|^2f(w)dw = \sum_{n=1}^{N}b_{n,1}\|p_n - q \|^2v_n = \kappa\sum_{n=1}^{N}a_n\|p_n - q \|^2v_n \nonumber \\
     &= \kappa\sum_{n=1}^{N}\|p_n\sqrt{a_nv_n} - q\sqrt{a_nv_n} \|^2 = \kappa\times \|\Tilde{\mathbf{p}} - \Tilde{\mathbf{q}} \|^2,
\end{align}
where $\Tilde{\mathbf{p}} = \left(p_1\sqrt{a_1v_1}, \ldots, p_N\sqrt{a_Nv_N}   \right)$ and $\Tilde{\mathbf{q}} = \left(q\sqrt{a_1v_1}, \ldots, q\sqrt{a_Nv_N}   \right)$. Similarly, we can rewrite the Sensor-power function as:
\begin{equation}\label{rewrite_Sensor_power_function}
    \begin{split}
        \overline{\mathcal{P}}^{\mathcal{S}}(P,\mathbf{R}) &= \sum_{n=1}^{N}\int_{R_n}a_n\|p_n - w\|^2 f(w) dw = \sum_{n=1}^{N}a_n\|p_n-c_n\|^2v_n + \mathcal{H}(\mathbf{R}) \\
        &= \sum_{n=1}^{N}\|p_n\sqrt{a_nv_n} - c_n\sqrt{a_nv_n} \|^2 + \mathcal{H}(\mathbf{R}) = \|\Tilde{\mathbf{p}} - \Tilde{\mathbf{c}} \|^2 + \mathcal{H}(\mathbf{R}),
    \end{split}
\end{equation}
where $\Tilde{\mathbf{c}} = \left(c_1\sqrt{a_1v_1}, \ldots, c_N\sqrt{a_Nv_N} \right)$. Note that $\mathcal{H}(\mathbf{R})$ is a constant since $\mathbf{R}$ is fixed. Therefore, we have:
\begin{equation}\label{alternative_definition_of_A(s,R)}
    A(s, \mathbf{R}) = \inf_{\left(\Tilde{\mathbf{p}}, \Tilde{\mathbf{q}} \right): \|\Tilde{\mathbf{p}} - \Tilde{\mathbf{c}}\|^2 \leq \left(s - \mathcal{H}(\mathbf{R}) \right) }\kappa\times \|\Tilde{\mathbf{p}}-\Tilde{\mathbf{q}} \|^2.
\end{equation}
Note that for any fixed value of $\Tilde{\mathbf{q}}$, (\ref{alternative_definition_of_A(s,R)}) implies that we want to minimize the distance from the point $\Tilde{\mathbf{p}}$ to $\Tilde{\mathbf{q}}$ while it remains within a radius of $\sqrt{s-\mathcal{H}(\mathbf{R})}$ of the point $\Tilde{\mathbf{c}}$. By using a simple geometric reasoning, it can be shown that $\Tilde{\mathbf{p}}$ lies on the segment connecting $\Tilde{\mathbf{c}}$ to $\Tilde{\mathbf{q}}$, i.e., there exists a coefficient $\lambda \geq 0$ for which we have:
\begin{equation}\label{p_lies_between_c_and_q}
    \Tilde{\mathbf{p}} = \frac{\Tilde{\mathbf{q}} + \lambda \Tilde{\mathbf{c}}}{1 + \lambda},
\end{equation}
i.e., for any $\Tilde{\mathbf{q}}$, the constraint in (\ref{alternative_definition_of_A(s,R)}) is equivalent to:
\begin{equation}\label{constraint_equation}
    \lambda:(1+\lambda)^2\geq \frac{\|\Tilde{\mathbf{q}} - \Tilde{\mathbf{c}} \|^2}{s - \mathcal{H}(\mathbf{R})}.
\end{equation}
Therefore, (\ref{alternative_definition_of_A(s,R)}) can be rewritten as:
\begin{equation}\label{A(s,R)_based_on_lambda}
        A(s,\mathbf{R}) = \inf_{Q}\inf_{\lambda:(1+\lambda)^2\geq \frac{\|\Tilde{\mathbf{q}} - \Tilde{\mathbf{c}} \|^2}{s - \mathcal{H}(\mathbf{R})}}\kappa \times \bigg| \! \bigg|\frac{\Tilde{\mathbf{q}} + \lambda \Tilde{\mathbf{c}}}{1 + \lambda}- \Tilde{\mathbf{q}} \bigg| \! \bigg|^2 
        = \inf_{\lambda:(1+\lambda)^2\geq \frac{\sum_{n=1}^{N}a_n\|q-c_n \|^2v_n}{s - \mathcal{H}(\mathbf{R})}}\inf_{q} G(q, \lambda),
\end{equation}
where:
\begin{equation}\label{G(q)}
        G(q,\lambda) =\kappa\times \sum_{n=1}^{N}a_n\bigg| \! \bigg|\frac{q + \lambda c_n}{1 + \lambda}-q \bigg| \! \bigg|^2v_n 
        =\kappa\times \left(\frac{\lambda}{1+\lambda} \right)^2 \sum_{n=1}^{N}a_n\|c_n - q\|^2v_n.
\end{equation}
Taking the derivative of $G(q)$ w.r.t. the FC location $q$ yields:
\begin{equation}\label{derivative_of_G_wrt_q}
    \begin{split}
        \frac{\partial G(q, \lambda)}{\partial q} &=\kappa\times \left(\frac{\lambda}{1+\lambda} \right)^2\sum_{n=1}^{N}2a_nv_n\left(q-c_n \right)=0,
    \end{split}
\end{equation}
i.e., $q^* = \frac{\sum_{n=1}^{N}a_nv_nc_n}{\sum_{n=1}^{N}a_nv_n}$. By substituting $q^*$ into (\ref{A(s,R)_based_on_lambda}), we have:
\begin{equation}\label{A(s,R)_just_function_of_lambda}
    A(s,\mathbf{R}) = \inf_{\lambda:\lambda\geq \sqrt{\frac{\sum_{n=1}^{N}a_n\|q^*-c_n \|^2v_n}{s - \mathcal{H}(\mathbf{R})}} - 1}G(q^*,\lambda).
\end{equation}
Since $G(q^*,\lambda)$ depends on $\lambda$ through the coefficient $\left(\frac{\lambda}{1 + \lambda} \right)^2$ that increases with $\lambda$, the infimum in (\ref{A(s,R)_just_function_of_lambda}) occurs for:
\begin{equation}\label{optimal_lambda}
    \lambda^* = \sqrt{\frac{\sum_{n=1}^{N}a_n\|q^*-c_n \|^2v_n}{s - \mathcal{H}(\mathbf{R})}} - 1 
    = \sqrt{\frac{\mathcal{J}(\mathbf{R}) - \mathcal{H}(\mathbf{R})}{s-\mathcal{H}(\mathbf{R})}} - 1,
\end{equation}
where the second equality follows from the parallel axis theorem. Substituting $\lambda^*$ into (\ref{A(s,R)_just_function_of_lambda}) yields the formula in (\ref{closed_form_formula}) for $A(s,\mathbf{R})$ and the proof is complete.$\hfill\blacksquare$

\section{}\label{Appendix_J}

Proof of Lemma 5: Note that the constrained optimization in (\ref{AP_Sensor_power_function}) is equivalent to the unconstrained optimization in (\ref{eq4}). As we showed earlier in Appendix \ref{Appendix_F}, if the condition in (\ref{necesssary_sufficient_condition_for_usefulness}) holds, the optimal partitioning is two closed intervals $\left[0,r^* \right]$ and $\left[r^*, 1 \right]$ where the FC is located at $r^*=q^* = \frac{1}{1+\sqrt{\frac{a_1}{a_2}}}$, in which case we have:
\begin{equation}\label{appendix_V_eq1}
    \mathcal{J}(\mathbf{R}) - \mathcal{H}(\mathbf{R}) = \sum_{n = 1}^{2}a_n\|q^* - c_n\|^2v_n 
    =  \frac{1}{4} \times \left[a_1{q^*}^3 + a_2\left(1-q^*\right)^3 \right], 
\end{equation}
\begin{equation}\label{appendix_V_eq2}
    \begin{split}
        s - \mathcal{H}(\mathbf{R}) &= s - \int_{0}^{q^*}a_1\Big| \! \Big|\frac{q^*}{2}-w\Big| \! \Big|^2f(w)dw - \int_{q^*}^{1}a_2 \Big| \! \Big| \frac{1+q^*}{2} - w \Big| \! \Big|^2 f(w)dw \\
        & = s - \frac{1}{12}\left[a_1{q^*}^3 + a_2\left(1-q^*\right)^3 \right].
    \end{split}
\end{equation}
Substituting (\ref{appendix_V_eq1}) and (\ref{appendix_V_eq2}) into (\ref{closed_form_formula}) yields (\ref{special_case_closed_form_formula_for_A(s)}) for $\frac{1}{12}\left(\frac{\sqrt{a_1a_2}}{\sqrt{a_1} + \sqrt{a_2}  }  \right)^2 \leq s \leq \frac{1}{3}\left(\frac{\sqrt{a_1a_2}}{\sqrt{a_1} + \sqrt{a_2}  }  \right)^2$. However, if the condition in (\ref{necesssary_sufficient_condition_for_usefulness}) does not hold, the optimal partitioning is when the region corresponding to the weaker AP is empty, and both FC $q$ and the stronger AP are located at the centroid of the target space; hence, $A(s)=0$ and $\overline{\mathcal{P}}^{\mathcal{S}}\left(P,\mathbf{R} \right) = \frac{\min\left(a_1,a_2\right)}{12}$. Since $\frac{\min\left(a_1,a_2\right)}{12}\leq \frac{1}{3}\left(\frac{\sqrt{a_1a_2}}{\sqrt{a_1} + \sqrt{a_2}  }  \right)^2$ with equality if and only if $a_1=a_2$, (\ref{special_case_closed_form_formula_for_A(s)}) is only valid for $\frac{1}{12}\left(\frac{\sqrt{a_1a_2}}{\sqrt{a_1} + \sqrt{a_2}  }  \right)^2 \leq s < \frac{\min\left(a_1,a_2\right)}{12}$, and $A(s)=0$ for $s\geq \frac{\min\left(a_1,a_2\right)}{12}$.$\hfill\blacksquare$

\section{}\label{Appendix_K}

Proof of Proposition 4: In what follows, we prove that none of the four steps in the Limited-HTTL algorthim will increase the two-tier power consumption. Note that APs in the set $\{ n\big | T(n)=-1  \}$ are neither used for target region partitioning, nor they contribute to the total power consumption; hence, given $P$, $Q$ and $R$, updating the index map $T$ according to (\ref{limited_com_range_index_map}) will not increase the power consumption. Furthermore, partitioning the target region according to the generalized Voronoi diagram is the best partitioning according to Proposition \ref{allactive}, and the two-tier power consumption will not be increased by the second stage of Limited-HTTL Algorithm.

Next, for a given $P$, $R$ and $T$, (\ref{appendix_F_2}) indicates that decreasing the distance between $q_m$ and $q'_m = \frac{\sum_{n:T(n)=m}b_{n,m}p_nv_n}{\sum_{n:T(n)=m}b_{n,m}v_n}$ will decrease the two-tier power consumption. Note that moving FC $m$ to $\widehat{q}_m$ will not increase the power consumption since $\|\widehat{q}_m - q'_m \| \leq \|q_m - q'_m \|$, and $\widehat{q}_m$ is still in the communication range of APs associated to FC $m$. Finally, (\ref{appendix_F_4}) implies that decreasing the distance between $p_n$ and $p'_n = \frac{a_nc_n + \beta b_{n,m}q_m}{a_n + \beta b_{n,m}}$ will decrease the two-tier power consumption. Note that moving AP $n$ to $\widehat{p}_n$ will not increase the power consumption since $\|\widehat{p}_n - p'_n \| \leq \|p_n - p'_n \|$, and $\widehat{p}_n$ is still in the communication range of the FC $q_{T{(n)}}$. Since none of the above four stages will increase the power consumption, Limited-HTTL Algorithm generates a sequence of positive non-increasing power consumption values and thus, it converges.$\hfill\blacksquare$

\ifCLASSOPTIONcaptionsoff
  \newpage
\fi

\end{document}